\documentclass[12pt]{article}
\usepackage{latexsym}
\usepackage{amsmath}
\usepackage{amssymb}
\usepackage{epsfig,graphics}
\usepackage{graphicx}
\usepackage{booktabs}
\usepackage{multirow}
\usepackage{caption}

\usepackage{subcaption}
\usepackage{tikz}
\usetikzlibrary{matrix,snakes,arrows,shapes,decorations.pathmorphing,decorations.markings,calc}

\newcommand{\be}{\begin{equation}}
\newcommand{\ee}{\end{equation}}

\begin{document}

\begin{titlepage}

\vspace*{0.6in}
 
\begin{center}
{\large\bf SO(N) gauge theories in 2+1 dimensions:\\
glueball spectra and confinement}\\
\vspace*{0.85in}
{Richard Lau and Michael Teper\\
\vspace*{.2in}
Rudolf Peierls Centre for Theoretical Physics, University of Oxford,\\
1 Keble Road, Oxford OX1 3NP, UK
}
\end{center}

\vspace*{1.2in}

\begin{center}
{\bf Abstract}
\end{center}

We calculate the spectrum of light glueballs and the string tension in a number of $SO(N)$ lattice
gauge theories in $2+1$ dimensions, with $N$ in the range $3\leq N\leq 16$. After extrapolating 
to the continuum limit and then to $N=\infty$ we compare to the spectrum and string tension of the
$SU(N\to\infty)$ gauge theory and find that the most reliably and precisely calculated physical
quantities are consistent in that limit. We also compare
the glueball spectra of those pairs of $SO(N)$ and $SU(N^\prime)$ theories that possess the
same Lie algebra, i.e. $SO(3)$ and $SU(2)$, $SO(4)$ and $SU(2) \times SU(2)$, $SO(6)$ and $SU(4)$,
and find that for the very lightest glueballs the spectra are consistent within each such pair,
as are the string tensions and the couplings. Where there are apparent discrepancies  
they are typically for heavier glueballs, where the  systematic errors are much harder to control.
We calculate the $SO(N)$ string tensions with a particular focus on the confining properties
of $SO(2N+1)$ theories which, unlike $SO(2N)$ theories, possess a trivial centre. We find
that for both the light glueballs and for the string tension $SO(2N)$ and $SO(2N+1)$ gauge theories
appear to form a single smooth sequence.

\vspace*{0.95in}

\leftline{{\it E-mail:} r.lau1@physics.ox.ac.uk, m.teper1@physics.ox.ac.uk}

\end{titlepage}

\setcounter{page}{1}
\newpage
\pagestyle{plain}

\tableofcontents

\section{Introduction}
\label{section_intro}

While the continuum physics of $SU(N)$ gauge theories has been extensively studied via 
lattice calculations in both 2+1 and 3+1 dimensions, much less is known about 
$SO(N)$ gauge theories. The fact that $SO(N)$ theories are different from $SU(N)$ and yet 
are `near neighbours', suggests that studying these theories is worthwhile.

$SU(N)$ and $SO(N)$ gauge theories are `near neighbours' in at least two ways. Firstly, 
the large $N$ limits of these theories are known to coincide
\cite{CL_N}
at the diagrammatic level (up to a factor of 2 in $g^2$).
Moreover the orbifold equivalence
\cite{Orb_equiv}
between $SO(2N)$ and $SU(N)$ 
gauge theories implies that they have the same physics in the common sector of states
when $N\to\infty$
\cite{Orb_phys}.
So it would be interesting to confirm these expectations with a non-perturbative 
lattice calculation of, for example, their common (positive charge conjugation)
mass spectra, and also to investigate how $SO(2N+1)$ gauge theories fit in with this. 
Secondly, certain low $N$ pairs of theories possess the same Lie algebras: 
$SU(2)$ and $SO(3)$; $SU(2)\times SU(2)$ and $SO(4)$; $SU(4)$ and $SO(6)$. Again it would be 
interesting to know if the spectra are the same or if the differences in the global properties 
of the groups, to which large fields may be sensitive, influence the spectrum.
We recall, for example, that the centre of $SU(N)$ is $Z_N$ while the centre of $SO(2N)$ is 
only $Z_2$ and that of $SO(2N+1)$ is trivial. In models of confinement based on dual disorder 
loops (centre vortices)
\cite{tHooft_ZN},
one might expect the differing centres to lead, for example, to differing string tensions. 
In the case of $SO(2N+1)$ theories, which have a trivial centre, do we in fact have linear
confinement? And what of the deconfining transition and its critical exponents? If the
spectra of the pairs of unitary and orthogonal theories that share a common Lie algebra
are in fact identical (as naively expected) how does this constrain their $N$-dependence,
both in $SU(N)$ and $SO(N)$ gauge theories?

These and other interesting questions about $SO(N)$ gauge theories can be addressed by 
lattice calculations in both 
$2+1$ and $3+1$ dimensions. The $D=3+1$ case is clearly of more direct physical interest, 
but standard lattice calculations encounter an obstacle in the existence of a first-order 
strong to weak coupling phase transition that, for small $N$, occurs at a small value of 
the lattice spacing
\cite{FBRLMT_SON}. 
This means that a lattice on the weak coupling side will need to be very 
large in lattice units if it is to have an adequately large physical volume.
While this obstacle should be surmountable,
it has led us to give priority to calculations in $D=2+1$ where the analogous transition 
is more manageable and where the interesting field theoretic questions can also be 
addressed with, moreover, greater precision. 

In this paper our aim is to calculate the masses of the lightest glueball states
of $D=2+1$ $SO(N)$ gauge theories for various values of $N$ with sufficient acccuracy
to be able to  extrapolate the lattice results to the continuum limit. We choose
the range of values of $N$, $N\in[3,16]$, to be large enough to allow a plausible 
extrapolation to the $N=\infty$ limit where we can compare to existing results for $SU(N)$ 
theories. En route we will compare the various pairs of theories with the same Lie 
algebras, and we will comment on what this might imply for the $N$-dependence of the 
physics. We will also calculate string tensions, both in $SO(2N)$ theories where
confinement can be linked to their non-trivial $Z_2$ centre and, more interestingly, 
in  $SO(2N+1)$ gauge theories, where the centre is trivial. We will address the 
question of  whether $SO(2N+1)$ and $SO(2N)$ gauge theories form two separate 
sequences that only converge (if at all) at $N=\infty$ or if the two sets do in 
fact form a single sequence.

To achieve a truly convincing comparison between $SO(N)$ and $SU(N)$ gauge theories
we would need to obtain accurate calculations of at least a few states in each
spin-parity ($J^P$) sector of $SO(N)$, just as one already has in $SU(N)$
\cite{AAMT_SUN}.
However it turns out that we are not able to achieve that ideal in the present work,
primarily because our lattice glueball operators have poorer overlaps onto the
glueball states than in the case of $SU(N)$, particularly so for small values of $N$.
Nonetheless we are able to calculate the lightest $J^P=0^+,2^\pm$ glueball masses
with the necessary precision, as well as the string tension (at least for $N> 3$).
More massive states are subject to increasing systematic errors as the mass
increases. To indicate the seriousness of these systematic errors in our Tables
of masses (where the errors shown are purely statistical) we provide a `reliability'
grade ranging from $\alpha$, indicating no significant systematic error, to
$\phi$, the opposite extreme. These are discussed in detail in Section~\ref{subsubsection_grades},
and should be taken into account in all the comparisons we show.

In the next section we begin by recalling the large $N$ behaviour of $SO(N)$ and $SU(N)$ 
gauge theories and we outline our expectations for the relationship between those
$SO(N)$ and $SU(N^\prime)$ theories that share a common Lie algebra. We only briefly
mention how this might constrain the $N$-dependence of the glueball spectra of $SO(N)$
and $SU(N)$ theories since this question is discussed in detail in a recent letter
\cite{AARLMT_SONSUN}
that makes use of some preliminary results of the present paper.
In Section~\ref{section_prelim} we describe how the lattice calculations are performed
and we discuss what we believe to be the principal
sources of systematic error. We present in  Section~\ref{section_confinement}
our evidence for linear confinement in $SO(2N)$ and, more interestingly, in $SO(2N+1)$.
We follow that with detailed calculations of the string tension in
Section~\ref{section_stringtension} and, in Section~\ref{section_glueballs},
of the glueball spectrum. Then in Section~\ref{section_other} we touch upon
three further issues: we look at how well our results confirm the usual expectations
of how one takes the large-$N$ limit; we compare mass ratios in strong and weak coupling;
and we demonstrate that the spectrum of $SO(4)$ gauge theories appears to know about
the spinorial excitations even though there are no explicit spinorial fields present.
Finally our concluding section summarises and discusses the results of this paper.

Another interesting physics question which we have addressed elsewhere in some
detail is the deconfining transition
\cite{RLMT_Tc}
and, as remarked above, in
\cite{AARLMT_SONSUN}
we showed how the fact that certain $SO(N)$ theories have the same Lie algebras
as certain $SU(N^\prime)$ theories, may impose strong constraints upon the $N$-dependence
of both  $SO(N)$ and $SU(N)$ gauge theories. In addition
the question of whether the $Z_N$ centre symmetry, or at least its manifestation in
$k$-strings, is recovered in some sense as one approaches $N=\infty$ will be
addressed elsewhere. 
Finally we remark that our initial exploratory calculations comparing $SO(N)$ and
$SU(N)$ gauge theories were presented some time ago in
\cite{FBRLMT_SON},
and some of the results of the present paper have appeared in
\cite{RL_thesis}.

\section{Relations between SU(N) and SO(N)}
\label{section_sunsonconstraints}

\subsection{large N}
\label{subsection_largeN}

In $SU(N)$ gauge theories all-order diagrammatic arguments
\cite{tHooft_N},
supported by lattice calculations
\cite{BLMTUW_SUN,AAMT_SUN},
indicate that mass ratios equal their $N=\infty$ values up to $O(1/N^2)$ corrections:
\begin{equation}
\frac{M_i}{M_j}
\stackrel{N\to\infty}{=}
{\tilde{r}}_{ij}
+\frac{\tilde{c}_{1,ij}}{N^2}
+\frac{\tilde{c}_{2,ij}}{N^4}
+ ... \qquad : \quad SU(N)
\label{eqn_suN}
\end{equation}
where ${\tilde{r}}_{ij}$ is the value of the mass ratio in the $SU(\infty)$ theory.
In $SO(N)$ gauge theories a similar diagrammatic analysis
\cite{CL_N}
suggests that
\begin{equation}
\frac{M_i}{M_j}
\stackrel{N\to\infty}{=}
{{r}}_{ij}
+\frac{{c}_{1,ij}}{N}
+\frac{{c}_{2,ij}}{N^2}
+ ... \qquad : \quad SO(N)
\label{eqn_soN}
\end{equation}
Moreover the leading  planar diagrams are the same in both cases
\cite{CL_N}, 
so we might expect an identical $N=\infty$ spectrum, i.e.
\begin{equation}
{\tilde{r}}_{ij} = {{r}}_{ij}
\label{eqn_soNsuN}
\end{equation}
in the common $C=+$ sector of the two theories. One of the purposes of
our calculations in this paper is to test this expectation. We also
recall that a planar $N=\infty$ limit
requires that we hold $g^2N$ fixed as $N\to\infty$, and to obtain the same 
limit in $SO(N)$ and $SU(N)$ we need to match the couplings as
\cite{CL_N} 
\begin{equation}
\left.g^2\right|_{SO(N)}
\stackrel{N\to\infty}{=}
2 \times \left.g^2\right|_{SU(N)}
\label{eqn_g}
\end{equation}
or, equivalently, we need to match $SO(2N)$ and $SU(N)$ theories at the same  
coupling. That this holds beyond perturbation theory is something we will
test in this paper. Finally we remark that there exists a large-$N$ orbifold 
equivalence between $SO(2N)$ and $SU(N)$ gauge theories
\cite{Orb_equiv}
which has been shown
\cite{Orb_phys}
to imply that at $N=\infty$ the theories have the same physics, in
their common sector of states.

We recall that $g^2$ has dimensions of mass in $D=2+1$ and so we can compare
dimensionless ratios $M_i/g^2N$ in the different theories, which we will
do later on in this paper. In addition, one of the masses can be the square root
of the confining string tension $\sigma$.

An important aside is that although $SO(2N)$ and $SU(N)$ gauge theories have
identical planar limits, the former has just a $Z_2$ centre while the latter's
centre is $Z_N$, and the centre of $SO(2N+1)$ is trivial. This raises
interesting questions about the confining properties of these theories,
especially for odd $N$.

\subsection{Lie algebra equivalences}
\label{subsection_liealgebra}

Where an  SO($N$) and SU($N^\prime$) gauge group share the same Lie algebra
the naive expectation is that corresponding physical quantities should be equal.  
For colour singlet quantities such as $C=+$ `glueball' masses, what
the `corresponding physical quantities' are is obvious. For flux tubes and 
string tensions this is less obvious, because these carry flux in a certain 
representation of the group, and one needs to establish the `corresponding'
representations. One needs to be equally careful with couplings, since these
can be defined in various ways. In this section we briefly summarise
what the correspondences are for the three such pairs of groups.

Testing these expectations is necessary because, for example, $SU(2)$ and $SO(3)$ 
have different topology, and so large field fluctuations will not be identical.
This may affect the glueball spectrum and/or the string tension. (It will
certainly do so at strong coupling.)  One of the main purposes of
our calculations is to test the validity of these naive expectations.

\subsubsection{SU(2) and SO(3)}
\label{subsubsection_su2andso3}

As is well known, $SO(3)$ is locally equivalent to $SU(2)$
in the adjoint representation. Since the fundamental of SO(3) is a 
$\underline{3}$ and this is just the $J=1$ (adjoint) representation of 
$SU(2)$, fundamental $SO(3)$ flux tubes will correspond to $SU(2)$ flux 
tubes carrying adjoint flux. These adjoint flux tubes are expected to be
unstable (they can be broken by gluon pairs from the vacuum)
and can decay to the vacuum and, once they are long and massive enough, into 
glueballs. This instability is consistent with the fact that $SO(3)$ does not
possess a non-trivial center which would prevent the mixing of a winding flux 
tube operator with the contractible operators that project onto glueball states.
Thus the $\sigma$ extracted in $SO(3)$ corresponds to the adjoint string 
tension in $SU(2)$. This is not straightforward to test since there are no
really precise calculations of the $SU(2)$ adjoint string tension. Existing
calculations of the latter
\cite{AAMT_SUN,AAMT_string}
do however support the approximate validity of Casimir scaling
\cite{CS}, 
which asserts that the ratio of string tensions corresponding to flux tubes
carrying flux in representations $R$ and $R^\prime$ is given by
\begin{equation}
\frac{\sigma_R}{\sigma_{R^\prime}}
=
\frac{C_2(R)}{C_2(R^\prime)}
\label{eqn_CS}
\end{equation}
where $C_2(R)$ is the quadratic Casimir of $R$. In our case this predicts
\begin{equation}
\frac{\sigma_{adj}}{\sigma_f}
=
\frac{8}{3} \qquad ; \qquad SU(2).
\label{eqn_CSsu2}
\end{equation}
In practice we shall use the values obtained in
\cite{AAMT_SUN,AAMT_string}
rather than this very rough estimate, since those calculations use the same methods
as used in this paper.

To compare $SU(2)$ and $SO(3)$ couplings it is useful to recall the study
of mixed fundamental-adjoint $SU(2)$ actions, where the fundamental and adjoint 
lattice actions have $1/g^2_f$ and $1/g^2_a$ factors respectively
\cite{FA_SU2}.
Using the standard plaquette action (for notation see Section~\ref{subsection_lattice})
we can write this mixed lattice action, for $SU(N)$, as
\begin{equation}
S = \beta_f \sum_p \left\{1-\frac{1}{N_f}ReTr_f U_p\right\}
+
\beta_a \sum_p \left\{1-\frac{1}{N_a}Tr_a U_p\right\}
\label{eqn_SmixedSUN}
\end{equation}
where $U_p$ is the ordered product of link matrices around the boundary
of the plaquette $p$.
Here the first term is in the fundamental and the second is in the adjoint,
with $f,a$ denoting fundamental and adjoint respectively. For $SU(N)$, 
we have $N_f=N$, $N_a=N^2-1$, $\beta_f=2N_f/ag_f^2=2N/ag_f^2$ and 
$\beta_a=2N_a/ag_a^2=2(N^2-1)/ag_a^2$. Using $Tr_aU=|Tr_f U|^2-1$ we expand
the group elements in terms of Lie algebra potentials, and obtain the 
standard continuum action with a prefactor of $1/g^2$ if we choose
\begin{equation}
\frac{1}{g^2} = \frac{1}{g_f^2} + \frac{2N}{g_a^2}.
\label{eqn_gfgaSUN}
\end{equation}
So if we use actions that are entirely adjoint or entirely fundamental,
then we will get the same physics if we use
\begin{equation}
g_a^2 = 2Ng_f^2 \stackrel{SU2}{=} 4g^2_f
\label{eqn_gfgaSU2}
\end{equation}
and this is one of the relations we will wish to test via the $SO(3)$ 
lattice calculations in this paper.

\subsubsection{SU(4) and SO(6)}
\label{subsubsection_su4andso6}

Since $SU(4)$ and $SO(6)$ also have the same Lie algebra, we might
expect that their $C=+$ glueball spectra will be identical. For the 
string tensions, we need to know what the fundamental of $SO(6)$ 
corresponds to in $SU(4)$. Now we recall that in $SU(4)$
\begin{equation}
\underline{4} \otimes \underline{4}  =  \underline{6} \oplus \underline{10}
\label{eqn_44to6}
\end{equation}
where the $\underline{6}$ corresponds to the $k=2$ antisymmetric
representation (which indeed is $C=+$ for SU(4)) and this will map 
to the fundamental $\underline{6}$ of $SO(6)$. Thus in testing the
equivalence of $SU(4)$ and $SO(6)$ we should compare the
$SO(6)$ fundamental string tension to the  $k=2A$ string tension in $SU(4)$ 
which, in terms of the fundamental $SU(4)$ string tension, has the value
\cite{AAMT_SUN}
(see also
\cite{BBMT_kstring})
\begin{equation}
\frac{\sigma_{2A}}{\sigma_f}
= 
1.357 \pm 0.003 \qquad ; \quad SU(4).
\label{eqn_2AtoFsu4}
\end{equation}
This implies that when we calculate glueball masses in units of the 
string tension, the relevant comparison to make is
\begin{equation}
\left.\frac{M_G}{\surd\sigma}\right|_{so6}
\stackrel{?}{=}
\left.\frac{M_G}{\surd\sigma_{2A}}\right|_{su4}
\label{eqn_Kso6su4}
\end{equation}

To determine the relationship between the $SU(4)$ and $SO(6)$ couplings
we use the same argument as for $SU(2)$and $SO(3)$. Working with the $SO(6)$ 
action should be equivalent to working in $SU(4)$ with the fields in the
$k=2A$ representation. So one can think of using a mixed $SU(N)$ lattice
plaquette action 
\begin{equation}
S = \beta_f \sum_p \left\{1-\frac{1}{N_f}ReTr_f U_p\right\}
+
\beta_{2A} \sum_p \left\{1-\frac{1}{N_{2A}}Tr_{2A} U_p\right\}
\label{eqn_Smixedsu4}
\end{equation}
where 
\begin{equation}
\beta_f = 2N_f/g^2_f \quad ; \quad \beta_{2A}=2N_{2A}/g^2_{2A}
\label{eqn_beta}
\end{equation}
analogous to the mixed fundamental-adjoint action in eqn(\ref{eqn_SmixedSUN}). 
For $SU(4)$, the sizes of the representations are $N_f=4, \ N_{2A}=6$. Using 
\begin{equation}
Tr_{2A} U_p = \frac{1}{2}\left\{ (Tr_fU_p)^2- Tr_fU^2_p \right\}
\label{eqn_trace2A}
\end{equation}
and performing a weak coupling expansion we find that
\begin{equation}
\left.g^2\right|_{so6} = \left.g^2_{2A}\right|_{su4} = \left.2 g^2_f\right|_{su4}
\label{eqn_gso6}
\end{equation}

It is interesting to compare this to the leading-order large-$N$ expectation
that we should  match fundamental couplings of $SO(2N)$ and $SU(N)$ theories, 
i.e. $g^2|_{so6} = g^2|_{su3}$ in the present case. For $SU(N)$ theories
the leading order expectation is that we keep the 't Hooft coupling
constant, i.e. $4  g^2|_{su4} = 3 g^2|_{su3}$. All this leads to the expectation
\begin{equation}
\left.g^2\right|_{so6} 
\stackrel{large \ N}{=} \left.g^2\right|_{su3} 
\stackrel{large \ N}{=} \frac{4}{3} \left.g^2\right|_{su4} 
\label{eqn_gso6largeN}
\end{equation}
which is quite different from eqn(\ref{eqn_gso6}). That is to say, the leading 
large-$N$ result is certainly a poor approximation when applied to $g^2$ at
small values of $N$.

\subsubsection{SU(2)$\times$SU(2) and SO(4)}
\label{subsubsection_su2andso4}

The Lie algebra equivalence of $SO(4)$ and $SU(2)\times SU(2)$ suggests that 
the glueball spectra of these two gauge theories might be the same. Now, the glueball 
spectrum of  $SU(2)\times SU(2)$ will consist of two sets of glueballs that do not 
interact with each other, each being identical to that of $SU(2)$. There will
be multi-glueball states consisting of some glueballs from one $SU(2)$ and
some from the other, so the spectrum of $SU(2)\times SU(2)$ is not identical
to that of $SU(2)$. (Especially in a finite volume.) However the single particle
spectrum should be the same, and that is what we are primarily interested in
calculating in this paper.

Along the same lines,  we would naively expect that the fundamental SO(4) flux tube 
contains fundamental flux of each of the two $SU(2)$ groups, and since these two 
fluxes do not interact, we should have 
\begin{equation}
\left.\sigma\right|_{so4} = 2\left.\sigma\right|_{su2}.
\label{eqn_Kso4}
\end{equation}
We will test this relation below.

As for the couplings, one would expect the $SU(2)$ and $SO(4)$ couplings to be the 
same, except for the fact that the $SU(2)$ group elements are $4\times 4$ matrices.
To compensate for the extra trace factors, we expect the actual relationship
between the couplings to be
\begin{equation}
\left.g^2\right|_{so4} = 2 \left.g^2_f\right|_{su2}
\label{eqn_gso4}
\end{equation}
\subsection{predicting $SO(N)$ from $SU(N)$}
\label{subsection_soNfromsuN}

Let us assume we know the glueball spectrum of $SU(N)$ theories. 
What does that tell us about the spectrum of $SO(N)$ gauge theories?
If we assume that (i) the single particle spectra of both $SO(3)$ and $SO(4)$ are 
the same as that of $SU(2)$, (ii) the spectrum of  $SO(6)$ is the same as 
that of $SU(4)$, (iii) the spectra of $SO(\infty)$ and $SU(\infty)$ are the same,
then this will strongly constrain the spectrum of $SO(N)$ gauge theories for all $N$.
In particular for any mass ratio where the expansion in eqn(\ref{eqn_soN}) 
needs only terms up to $O(1/N^3)$ to accurately describe the $N$-dependence 
for $N\geq 3$, these constraints means that the $N$-dependence is completely
determined by the values in $SU(N)$. 

More generally the $SO(N)$ and $SU(N)$ mass ratios are mutually constrained
through eqn(\ref{eqn_soN}) and eqn(\ref{eqn_suN}) by such Lie algebra and
large-$N$ equivalences. Obviously these constraints become stronger the fewer
the terms needed in eqns(\ref{eqn_suN},\ref{eqn_soN}) to describe the mass ratio
for $SO(N\geq 3)$ and $SU(N\geq 2)$. What makes these observations relevant is that 
in practice, as we shall see below, one only needs one or two corrections to describe 
the $N$-dependence of ratios of a number of the lightest masses in $SO(N)$ gauge 
theories. And this is also the case in $SU(N)$
\cite{AAMT_SUN}.
This interesting relation between  $SO(N)$ and $SU(N)$ gauge theories
is explored in detail in 
\cite{AARLMT_SONSUN}
and we refer the reader to that paper.

\section{Preliminaries}
\label{section_prelim}

\subsection{calculating on the lattice}
\label{subsection_lattice}

Our lattice field variables are SO($N$) matrices, $U_l$, residing on the links $l$
of the $L^2_s L_t$ or $L_xL_yL_t$ lattice whose spacing is $a$ and upon which we impose
periodic boundary conditions. The Euclidean path integral is 
$Z=\int {\cal{D}}U \exp\{-S[U]\}$ where ${\cal{D}}U$ is the Haar masure. We 
use the standard plaquette action,
\begin{equation}
S = \beta \sum_p \left\{1-\frac{1}{N} Tr U_p\right\}  \quad ; \quad \beta=\frac{2N}{ag^2}
\label{eqn_S}
\end{equation}
where $U_p$ is the ordered product of link matrices around the plaquette $p$.
As a convenient shorthand, we shall use $u_p \equiv \frac{1}{N} Tr U_p$.
We have written $\beta=2N/ag^2$, but strictly speaking $ag^2$ is just one 
possible definition of the dimensionless coupling on the length scale $a$,
so if we were to be punctilious we would write  $\beta=2N/ag_p^2$ with 
$ag^2_p = ag^2 + ca^2g^4 +... \to ag^2 $ as $a\to 0$.

We update the fields using a natural extension to $SO(N)$
\cite{FBRLMT_SON,RL_thesis}
of the standard $SU(N)$ 
Cabibbo-Marinari algorithm. Suppose we generate $N_z$ field 
configurations, $\{U_l\}^{I=1,...,N_Z}$, with the measure ${\cal{D}}U \exp\{-S[U]\}$.
Then we can estimate the expectation value of some functional $\Phi$ of the
fields (this may be a correlation function) by the simple average,
\begin{equation}
\langle \Phi[U] \rangle = \frac{1}{Z}\int {\cal{D}}U \Phi[U] \exp\{-S[U]\}
= \frac{1}{N_z} \sum_{I=1}^{N_z}  \Phi[U^I]
+O(1/\surd{N_z}),
\label{eqn_PhiMC}
\end{equation}
where the last term denotes the statistical error. In our calculations
we only use heat bath updates, although it would be desirable to
include over-relaxation so as to accelerate the exploration of
the theory's phase space.

\subsection{calculating energies}
\label{subsection_energies}

Our SO($N$) calculations closely parallel those in SU($N$), so we will be
brief here and refer the reader to other papers
\cite{BLMTUW_SUN,AAMT_SUN}
for details. It is of
course important to be aware when these calculations become less reliable,
and this will be addressed in Section~\ref{subsection_errors}.

\subsubsection{glueball masses}
\label{subsubsection_glueballs}

In two spatial dimensions the rotation group is Abelian and spins are
$J=0, \pm1,\pm2,...$ (ignoring more exotic possibilities such as anyons).
Parity, P, does not commute with rotations, $J \stackrel{P}{\to} -J$, 
so particle (`glueball') states can be labelled by parity $P=\pm$ and
$|J|$.  (Charge conjugation is necessarily positive for $SO(N)$.) For $J\neq 0$ 
the above implies that continuum $P=\pm$ states in an infinite volume 
are necessarily degenerate although this is not necessarily true for even $J$
at finite lattice spacing and/or in a finite volume
\cite{BLMTUW_SUN,HM_thesis}.
Thus a useful check that our volume is 
large enough and that our lattice is fine enough, is provided by 
comparing the $J^P=2^+$ and $2^-$ masses.

Ground state masses $M$ are calculated from the asymptotic time
dependence of correlators, i.e.
\begin{equation}
\langle \phi^\dagger(t) \phi(0) \rangle 
= \sum_n |\langle vac|\phi^\dagger|n \rangle|^2 e^{-E_n t} 
\stackrel{t\to\infty}{\propto} e^{-Mt} 
\label{eqn_M}
\end{equation}
where $M$ is the mass of the lightest state with the quantum numbers
of the operator $\phi$. The operator $\phi$ will be the product of
$SO(N)$ link matrices around some closed path, with the trace then 
taken. We will use zero momentum operators so that there is no
momentum integral on the right side of eqn(\ref{eqn_M}).
To calculate the excited states $E_n$ in eqn(\ref{eqn_M}), one
calculates (cross)correlators of several operators and uses these as a
basis for a systematic variational calculation in $e^{-Ht_1}$ where $H$
is the Hamiltonian (corresponding to our lattice transfer matrix) and
$t_1$ is some convenient distance. (Typically we choose $t_1=a$.) 
To have good overlaps onto the
desired states, so that one can evaluate masses at values of $t$
where the signal has not yet disappeared into the statistical noise,
one uses blocked and smeared operators. (For more details see e.g.
\cite{BLMTUW_SUN,AAMT_SUN}.)

An obvious remark. On a lattice $t=an_t$ and what we know is the number 
of lattice spacings $n_t$. So fitting an exponential as in eqn(\ref{eqn_M})
to some correlation function, will give us the value of $Mt=Man_t$. That
is to say, what we obtain is a value for $aM$, the mass in lattice units.

Another remark: in reality the  $t\to\infty$ limit is not accessible in a
numerical calculation. This is trivially so because our lattice has a finite
extent in time, but more importantly because the statistical errors
on pure glue correlators are (usually) roughly constant in $t$ so the
error to signal ratio grows roughly exponentially in $t$. In practice
what one does is to find the lowest value $t=t_0$ such that the
correlator $C(t)=\langle \phi^\dagger(t) \phi(0) \rangle$
can be fitted with a single exponential
$\propto \exp(-Mt)$ for $t\geq t_0$, and one takes $M$ as the estimate of
the true mass. This procedure works well if $aM$ is small enough that
many values of $C(t=an_t\geq t_0)$ have small errors. Otherwise the fact
of a `good' fit may not be very significant. This will be a problem
when the mass $aM$ is not small, but also when the overlap of the
glueball wave-functional onto our operator basis is not close to unity.
The latter turns out, unfortunately, to be the case in our $SO(N)$
calculations at small $N$, in contrast to $SU(N)$. The accompanying systematic errors
are discussed in more detail in Section~\ref{subsubsection_Eplateaux}

We are on a square lattice so we have exact rotational invariance only
under rotations of $\pi/2$. So for each operator $\phi_0$ we can construct
the rotated operators,  $\phi_{\pi/2}$,  $\phi_{\pi}$,  $\phi_{3\pi/2}$ and
we can use these to construct a spin $J$ operator 
\begin{equation}
\phi_J(t) = 
\sum_{n=0}^3 \phi_{n \pi/2}(t) \exp\{iJn\pi/2\}.
\label{eqn_phiJ}
\end{equation}
Since this sum is invariant under $J\to J+4$, such an operator will
in fact project onto the tower of spins $J+4n$ with $n$ any integer. 
For simplicity we will refer to the states as having a spin equal to 
the minimum of these spins. (Often, but not always, the lightest 
glueball in such a tower does indeed possess the minimal spin.) 
If we decide to use parity eigenstates, then we start with the
operator $\phi_0^{P=\pm} = (1\pm P)\phi_0$, construct its rotations 
$\phi_{n\pi/2}^{P=\pm}$, and sum these up as in eqn(\ref{eqn_phiJ})
to obtain $\phi_J^{P=\pm}(t)$. Note that in this sum the parity
inverse of the rotated operator $\phi_{\theta}$ is a rotation
by $-\theta$ of $P\phi_0$, i.e. the $P=\pm$ operators have definite
$|J|$ but not definite $J$. For notational simplicity we shall, nonetheless,
label our operators by the symbol $J$. Thus using $J=0,1,2$ and $P=\pm$
operators we can obtain states of all spins and parities,
$J^P=0^\pm, 1^\pm, 2^\pm, \ldots $.

We now briefly describe one of the two sets of operators that we use in our 
glueball calculations. We take products of $SO(N)$ matrices around
closed curves  $\mathcal{C}$. The simplest such curves are the $1\times1$ 
plaquette, and the $1\times2$, $1\times3,\ldots$ rectangles. 
The square plaquette is invariant under $\pi/2$ rotations and so will only 
project on to $J=0$ states while the rectangles are invariant under $\pi$ 
rotations and so will only project on to $J=0$ and $J=2$ states. 
These curves are also invariant under parity and so will only project on 
to $P=+$ states. Hence, we need to consider more complicated curves to 
project on to $J=1$ and $P=-$ states. To do this, we consider curves 
constructed from squares and rectangles that have no rotational or 
reflection symmetry. We show four such curves in Figure~\ref{fig_glueball:curves}.
We use twelve such curves to build, by rotations and reflections, a basis for 
each $J^P$ state of twelve operators as described above.
We also use two rectangle-based operators for $0^+$ and $2^+$ states and 
the plaquette for the $0^+$ state. This means that we have a basis of fifteen 
elementary operators for the $0^+$ state and a basis of fourteen elementary 
operators for the $2^+$ state. We then add operators based on the same curves
but with the links replaced by `blocked links'. We recall
\cite{BLMTUW_SUN}
that these blocked link matrices join pairs of sites that are $1,2,4,8,...$ lattice 
spacings apart: $2^{n_B-1}$ spacings apart at blocking level $n_B$. We typically
include all blocking levels such that the blocked link fits into the lattice
e.g. up to blocking level 6 (length=32) on a $44^2$ spatial lattice.
So, for example,  in this case the total number of operators is 
$6\times 15 = 80$ for the $0^+$ and the variational analysis will produce
for us $80$ supposed (approximate) eigenstates. It is not implausible that this
basis is large enough that we have a good enough overlap onto the lightest
few states in each $J^P$ channel, without missing any. As an explicit check 
of this we will later compare spectra obtained with this basis and the somewhat
different (and larger) basis that is being used in similar $SU(N)$ calculations
\cite{AAMT_SUN}.
We remark that this alternative basis is the one used in our $SO(3)$, $SO(5)$
and $SO(6)$ spectrum calculations.

\subsubsection{string tensions}
\label{subsubsection_strings}

To calculate the string tension $\sigma$ we calculate the ground state 
energy $E(l)$ of a flux tube that winds around a periodic spatial torus of 
length $l$.  If the length $l=aL_s$ of the torus is
large then $E(l) \simeq \sigma l$ where $\sigma$ is the string
tension. At finite $l$ there are corrections to this and we assume that
for our range of $l$ these are accurately incorporated in the simple
`Nambu-Goto' expression 
\begin{equation}
E(l)=\sigma l \left\{ 1 - \frac{\pi}{3\sigma l^2} \right\}^{1/2}
\label{eqn_NG}
\end{equation}
which is what we shall use to extract $\sigma$ from $E(l)$. This expression
incorporates all the known universal corrections to the flux tube energy,
when one expands $E(l)$ in powers of $1/l^2\sigma$
\cite{OA_string}.
(See also
\cite{MLPWJD_string}.)
It has also been shown to arise as a good approximation at small $l$ from
the near-integrability of the world-sheet action
\cite{SD_string}.
Moreover it has been checked numerically with some precision in $SU(N)$ for
flux tubes carrying both fundamental and higher representation fluxes.  (See e.g.
\cite{AAMT_string,MTstring_old}
and references therein.) We shall provide some checks for $SO(N)$ in 
Section~\ref{section_confinement}.

The flux tube energy can be calculated just like a glueball mass, except
that instead of taking products of our matrices around contractible loops,
we do so about the non-contractible loops that wind once around the torus. 
The simplest such operator is the Polyakov loop $l_p$, i.e. the trace of the 
product of link matrices along a minimal length curve that closes around the
spatial torus,
\begin{equation}
l_p(n_t) =  \sum_{n_y} \mathrm{Tr} 
\left\{\prod^{L_x}_{n_x=1} U_x(n_x,n_y,n_t)\right\} \,,
\label{eqn_poly}
\end{equation}
where we have taken the product of the link matrices in the $x$-direction 
around the $x$-torus of length $l=aL_x$, with
$(x,y,t)=(an_x,an_y,an_t)$, and we sum over $n_y$ to produce
an operator with zero transverse momentum, $p_\perp = p_y = 0$.
We also use blocked/smeared versions of this. Most of our 
calculations of the string tension use this set of operators. However
some calculations, in particular most of the $SO(3)$, $SO(5)$ and $SO(6)$ ones, use 
a much larger basis of operators, which incorporates a variety of non-minimal 
winding curves. Such a large basis is essential for obtaining excited 
flux tube states, but makes little difference to the ground state
calculations that are of interest in this paper.

For even $N$, the theory has a $Z_2$ symmetry that ensures that $\langle l_p \rangle = 0$ as 
long as the symmetry is not spontaneously broken, and indeed that 
$\langle l_p \phi_G \rangle \equiv 0$ where $\phi_G$ is any contractible loop (such as one
uses for glueball operators). That is to say we have a stable flux tube state that winds 
around the torus. (Of course, linear confinement only arises if the flux
does not spread out arbitrarily far, which is an additional  dynamical question.)
For $N$ odd we have no $Z_2$ symmetry to invoke and
whether such theories are linearly confining is an interesting question
that we shall address later on in this paper.

\subsection{continuum limit}
\label{subsection_continuum}

Given a value of $\beta$ in eqn(\ref{eqn_S}) we can calculate some masses (or energies) $a m_i$ 
and the string tension, $a^2\sigma$, in lattice units, as described above. 
However what we want is the spectrum in some physical units in the
continuum limit $\beta\to\infty$. We can take ratios of masses so that the 
lattice units cancel $am(a)/a\mu(a) =m(a)/\mu(a)$, but this ratio will still depend on 
the discretisation $a$. However the fact that the theory becomes
free at short distances, since $g^2$ has dimensions of $[m]$ so that 
the dimensionless expansion parameter for physics on the length scale $l$
will be $g^2l \stackrel{l\to 0}{\rightarrow} 0$, allows us to control the 
expansion of the lattice action in terms of continuum fields as $a\to 0$.
For our plaquette action it is known that the leading correction is $O(a^2)$,
i.e. 
\begin{equation}
\frac{am_i(a)}{am_j(a)}=\frac{m_i(a)}{m_j(a)} 
=\frac{m_i(0)}{m_j(0)} + c_{ijk} a^2 m^2_k(a) + O(a^4).
\label{eqn_mimjcont}
\end{equation}

One can use any (sensible) mass $m_k$ to set the scale of the corrections;
different choices, as well as the $a$-dependence of $m_k$, will merely
reshuffle the higher order terms. It clearly makes sense to use for $m_k$,
and indeed for $m_j$, the most accurately calculated mass. For $SU(N)$
this is usually the string tension, i.e. we use eqn(\ref{eqn_mimjcont})
with $am_j=am_k=a\surd\sigma$ and determine the ratio $m_i/\surd\sigma$
in the continuum limit for each $m_i$. For $SO(N)$ the lightest scalar glueball
is equally accurate and one can use that to set the scale just as well.

Since the coupling $g^2$ has dimensions of mass, we can perform an
alternative continuum extrapolation for individual masses as follows:
\begin{equation}
\frac{\beta}{2N}am_i(a) = \frac{m_i}{g^2} + \frac{c_i}{\beta} + ...
\label{eqn_bmicont}
\end{equation}
where we have used $\beta=2N/ag^2$. Note that here the leading correction
is $O(a)$ rather than $O(a^2)$. This has to be so because different
couplings are related by $ag^2_p =  ag^2_q + c(ag^2_q)^2 + ...$. 
In fact we shall often make use of this freedom to replace 
$\beta$ in eqn(\ref{eqn_bmicont}) by the `mean-field improved' coupling
$\beta_I=\beta \langle Tr U_p \rangle /N$
\cite{GP_MFI}. 
While the approach to the continuum in eqn(\ref{eqn_bmicont})
is slower than in eqn(\ref{eqn_mimjcont}),
the error on $\beta am_i$ is smaller than that on $m_i/\surd\sigma$ because 
there is no error on the value of $\beta$. We shall use both forms of
extrapolation in this paper.

\subsection{large N limit}
\label{subsection_largeNlimit}

Once we have continuum limits of mass ratios for various $SO(N)$ groups 
we can extrapolate these to $N=\infty$ just as for $SU(N)$ except that
the leading correction is expected to be $O(1/N)$
\cite{CL_N}:
\begin{equation}
\left.\frac{m_i}{\mu}\right|_N
=
\left.\frac{m_i}{\mu}\right|_\infty
 + \frac{c}{N} + \frac{c^\prime}{N^2} + \cdots
\label{eqn_mimjN}
\end{equation}
Here $\mu$ may be a glueball mass, e.g. the mass gap, or the
string tension $\surd\sigma$, or the coupling $g^2$. In this last
case, large $N$ counting tells us that we should use the 
't Hooft coupling $g^2N$:
\begin{equation}
\left.\frac{m_i}{g^2N}\right|_N
=
\left.\frac{m_i}{g^2N}\right|_\infty
 + \frac{c}{N} + \frac{c^\prime}{N^2} + \cdots
\label{eqn_migN}
\end{equation}

In performing the extrapolations it makes sense to choose an energy scale
$\mu$ which does not possess unusually large $O(1/N)$ corrections.
In particular we will need to be cautious about using the string tension
$\sigma$, since the Lie algebra equivalences tell us that the $SO(N)$
fundamental string tensions at low $N$ map onto higher representation 
string tensions in $SU(N)$. We also need to be cautious about using $g^2N$
since we have seen in Section~\ref{subsection_liealgebra} that the matching
of $g^2$ between $SO(N)$ and $SU(N)$ at small $N$ deviates  
strongly from the leading large-$N$ expectations.  

\subsection{bulk transition}
\label{subsection_bulk}

Lattice gauge theories generally possess a (`bulk') transition between the
strong and weak  coupling regions, where the natural expansion parameters
are $\beta \propto 1/ag^2$ and  $1/\beta \propto ag^2$ respectively.
Since the extrapolation to the continuum limit should, \'a priori, be made from 
values of masses calculated within the weak coupling region, it is 
important to locate any such bulk transition, whether it be a cross-over 
or a genuine phase transition. It is clearly desirable that this
transition should occur at a value of $\beta$ where the value of $a$ on the weak 
coupling side is not very small, otherwise prohibitively large lattices 
may be needed to ensure that the volume is adequately large, in physical units,
when in the weak  coupling confining phase.
The location of the transition will depend on the lattice action used
and all our remarks here are for the simple plaquette action. 

For D=3+1 SU($N$) gauge theories it is known that the transition
is first order for $N\geq 5$ and is a cross-over for smaller $N$
\cite{BLMTUW_SUN}. 
In D=2+1 $SU(N)$ gauge theories it appears 
\cite{FBMT_bulkd3}
to be quite similar to the Gross-Witten transition in D=1+1 
\cite{GW_Nd2}
i.e. a gentle cross-over for all $N<\infty$ developing into a
third-order transition at  $N=\infty$. It's location in $D=3+1$
is such that on the weak coupling side we can readily go to  
$a \sim 1/5T_c$ where $T_c$ is the deconfining temperature 
(taking advantage of the metastable region when the
transition is first order). In $D=2+1$, we can go to even larger $a$,
$a \sim 1/1.6T_c$.  These lattice spacings are large enough that,
for $SU(N)$, the bulk transition presents no 
significant obstacle  to continuum extrapolations. 

For $SO(N)$ lattice gauge theories the situation is different. 
In  $D=3+1$ one finds that the transition is again first order, but now 
for all $N\geq 3$
\cite{FBRLMT_SON}.
Moreover at low $N$ the value of $a$ on the weak coupling side is
very small and can pose an obstacle to accessing the continuum limit
\cite{FBRLMT_SON}. 
(Indeed, in the case of $SO(3)$ this has been known for a long time;  
for a recent discussion see
\cite{deF_SO3}.)
Fortunately in $D=2+1$, the case of relevance here, the corresponding
problems are much less severe. There is a transition 
\cite{FBRLMT_SON,RL_thesis}
whose signature is a finite region of $\beta$ in which the mass gap decreases 
towards zero as we increase $\beta$ towards some  $\beta_b$. The approximate 
values of $\beta_b$ are listed in Table~\ref{table_bulk}, where we also
give the value of the lightest scalar glueball obtained on the weak coupling
side of the transition. One can use this, as discussed in 
Section~\ref{subsubsection_finiteV}, to provide a rough lower bound on the
spatial volume needed to avoid large finite volume effects on the
lightest scalar and tensor glueball states. This bound is also given
in Table~\ref{table_bulk}. We see that quite a large lattice is needed,
especially for $SO(3)$, but since we are in $D=2+1$ this problem is readily
surmountable.

The fact that at low $N$, e.g. for $SO(3)$, the bulk transition occurs at 
a small value of $a$, prompts the question of how precisely `strong-coupling'
manifests itself below the bulk transition. We shall address this in
Section~\ref{subsection_sc}.

The bulk transition in $D=2+1$ $SO(N)$ gauge theories does not
possess the usual features of a second-order phase transition despite
the (nearly) vanishing mass gap. In particular it has a peculiar
volume dependence. Its unusual characteristics presumably flow from
the fact that the $SO(N)$ group is not simply connected.
This makes it an interesting transition which merits a more
detailed investigation.

\subsection{systematic errors}
\label{subsection_errors}

Our aim in this paper is to calculate the string tension and the lighter part of
the glueball spectrum. Such a calculation is affected by a number of `systematic'
errors that cannot be easily quantified. In this Section we briefly 
discuss what we believe to be the most important of these, how they may
affect our results, and how we try to minimise them.

\subsubsection{wrong quantum numbers}
\label{subsubsection_quantumnumber}

Our square spatial lattice is invariant under a subgroup of the full rotation group.
Thus  our rotationally invariant operator will project onto states not only
with continuum spin $J=0$ but also with $|J|=4,8,\ldots$. Similarly the
`$|J|=2$' operator will project onto $|J|=2,6,\ldots$ and `$|J|=1$' projects
onto  $|J|=1,3,\ldots$. To the extent that we only calculate the lowest one or
two masses in each representation it is plausible that labelling the states
by the lowest continuum $J$ that can contribute will often turn out to be correct. 
However there are known counter examples in the closely related case of $SU(N)$
gauge theories. In particular there is very good evidence
that the ground state `$0^-$' state is in fact $4^-$
\cite{HM_thesis,HMMT_J}
and also that the lightest `$J=1$' state is in fact $J=3$
\cite{HM_thesis,HMMT_J}.
In principle one could imagine using parity doubling to distinguish $J=4$ states from
$J=0$ states (more discriminatory in $D=3+1$ because of the varying multiplet
structure) but the anticipated $4^+$ mass is in a region of the `$J^P=0^+$'
spectrum which is already quite dense and identifying near-degeneracies, given
the usual statistical errors, would pose a formidable numerical challenge.
We shall therefore make no attempt here to distinguish spins that differ by a
multiple of 4, and will simply label states by the lowest contributing $J$.
Anyone who uses our spectrum to confront a model calculation needs to be
aware of this potential mislabelling.

\subsubsection{missing states}
\label{subsubsection_states}

Our variational calculation purports to give the spectrum of a number
of states starting with the ground state. How well it does this will of
course depend on how well the basis of operators encodes the desired 
states. The blocking/smearing was designed to produce operators with 
a good projection onto the lightest states, which we expect to have
minimal structure, but we do not know enough about excited glueball
wavefunctionals to be confident that we do well for such states. Since
in practice we have trouble identifying a state (within errors) unless its
overlap (normalised and squared) onto our basis is well above 0.5, it is
quite possible that our calculated spectrum has some missing
intermediate states.

While one way to check for this possibility is to redo the calculations
using ever larger operator bases, another way is to compare the spectrum
with what one obtains using  a completely different basis of operators. 
Since we have two independently produced bases, one used for calculating
the spectra for $N=3,5,6$ (labelled A) and the other for $N=4,7,8,12,16$
(labelled B)
we have been able to perform such a comparison in a few cases.
A typical example is provided in Table~\ref{table_spectrum_opAopB} where
we compare the spectra obtained in $SO(4)$ at $\beta=51$ using the two
different bases A and B. (The lattice volumes are not identical, but are similar
enough that this mismatch can be ignored.) We include here all the states of
the spectrum that we will be considering later on in this paper.
We see that in most cases the energies of corresponding states agree
within $2\sigma$. (We use the same symbol for the standard deviation and
the string tension; which is meant should be clear from the context.)
Only for the $2^\pm$ and the $0^{+\star\star}$
are the differences more than $2\sigma$, although still within $3\sigma$.
We conclude that within the (sometimes quite large) errors this provides
no evidence that there are missing states in the part of the spectrum shown.

This result is not in fact an accident. If we go to higher excited
states then we do find that some states with one operator basis do 
not appear when using the other basis, i.e we have
evidence of missing states. Our choice of where to cut off the
spectrum in our study  was, in fact, partly determined by this comparison.

Finally we should mention that the statistics of the `Ops A' calculation
in Table~\ref{table_spectrum_opAopB} 
is much higher than that of our other calculations, including
the `Ops B' calculation listed in  Table~\ref{table_spectrum_opAopB},
This means that the statistical errors on the correlators $C(t)$ are smaller
and hence one can extract more reliably the range $t\geq t_0$ from
which one determines the mass. This typically leads to a larger
estimate of $t_0$ and, given the positivity of the correlator, to a
lower value for the mass. This is presumably the reason that the
`Ops A' masses in Table~\ref{table_spectrum_opAopB} are all slightly
below those of the `Ops B'calculation.

\subsubsection{identifying energy plateaux}
\label{subsubsection_Eplateaux}

For any given set of quantum numbers our variational calculation produces a set
of $p=0$ operators $\Psi_i \ ; \ i=0,1,..$ that are approximate energy eigenoperators
(ordered in energy). While the eigenvalues give us a rough first estimate of
the energies, to obtain our final best estimate we take the correlators 
$<\Psi^\dagger_i(t)\Psi_i(0)>$ and look for the single exponential 
decays at larger $t$, as in eqn(\ref{eqn_M}), so obtaining the corresponding
energy $E_i$. This is the crucial step in obtaining glueball masses
and it is important to be aware that some of the resulting mass estimates 
will be less reliable than others. 

Consider first extracting the ground state energy $E_0$ in some $J^P$ channel. 
We can define an effective energy $E_{eff}(t)$ by
\begin{equation}
\frac{\langle\Psi^\dagger_0(t) \Psi_0(0)\rangle}{\langle\Psi^\dagger_0(t-a) \Psi_0(0)\rangle}
=
e^{-aE_{eff}(t)}.
\label{eqn_E0eff}
\end{equation}
(We shall sometimes use $M_{eff}(t)$ in place of $E_{eff}(t)$.)
From  eqn(\ref{eqn_M}) we know that $E_{eff}(t\to\infty) = E_0$. However we have
finite statistical errors on $E$ so we cannot access the $t\to\infty$ limit
and instead we search for a $t_0$ such that  
$E_{eff}(t) = {constant}$ for  $t\geq t_0 $, within the errors, and 
this constant then provides our estimate of $E_0$. The range $t\geq t_0$ is where
we have an `effective energy (or mass) plateau' and identifying such a plateau
is a crucial step in estimating $E_0$. The major obstacle here is that
for typical glueball calculations the statistical error on the correlator is 
roughly constant in $t$ while its average value is decreasing exponentially in
$t$, at least as fast as $\propto \exp\{-E_0 t\}$. Thus when the ground state energy 
$aE_0$ is large, there will be very few values of $E_{eff}(t)$ with small
errors and it may be that the apparent plateau is not statistically significant.
So heavier ground states are generally less reliable than lighter ones and 
with our statistics one has to be cautious once $aE > 1$ (roughly speaking).
The obvious danger is that one estimates $E_0$ using a value of $E_{eff}(t)$
at a value of $t$ that is too small. Since $E_{eff}(t)$ decreases montonically
with $t$ (because of the reflection positivity of our plaquette action) this
means that we overestimate the true energy, $E_0$. For light states this error
will usually be insignificant but, as we shall shortly see, for our heaviest
states this may well not be the case. (As we have already noted above,
when discussing the comparison between the two $SO(4)$ calculations in
Table~\ref{table_spectrum_opAopB}.)

For an excited state $\Psi_i$ we can also define an effective energy $E_{eff}(t)$
using eqn(\ref{eqn_E0eff}) with $\Psi_0 \to \Psi_i$, and there is a similar caveat
for heavy states.
However now, even if we can identify the start of a plateau in $E_{eff}(t)$, it is 
possible that $\Psi_i$ may have a small overlap onto one of the lighter states, so
that at sufficiently large $t$  $E_{eff}(t)$ will drop away from the plateau to
a smaller value. Thus our criterion is that there should be an `effective'
energy plateau $E_{eff}(t) = \mathrm{constant} \ : t_0 \leq t \leq t_0+\Delta t$
over some finite range $\Delta t$ that is large enough for the plateau to be 
convincing. An additional physical reason for such a behaviour could be that
the excited glueball is heavy enough to be unstable. If its decay width is small,
so that it is a narrow resonance, it should show a temporary effective energy 
plateau within errors. As $N$ increases it will become more stable and will 
extrapolate to a completely stable glueball at $N=\infty$. So it is certainly 
part of the spectrum that we wish to identify. However a temporary plateau 
clearly makes that identification more ambiguous.

It is clearly important to assess the importance of the above comments for the
mass calculations in this paper. To do this we begin by showing  in
Figs~\ref{fig_Eeff_M0p_so12_b250},\ref{fig_Eeff_M2mp0m1mp_so12_b250}
the effective masses
we obtain in $SO(12)$ at $\beta=250$. (Note that in all such plots in this paper we
stop plotting the effective energies once the error exceeds about $10\%$ of the value,
so as not to confuse the plot with large error bars that usually carry little
information.) This $\beta$ corresponds to the smallest value of the lattice spacing,
and therefore provides an important contribution to our continuum extrapolation.
Moreover $N=12$ is one of our largest values of $N$, so it also plays an important
role in the extrapolation to $N=\infty$. It is clear from the plot in
Fig.~\ref{fig_Eeff_M0p_so12_b250} that we can identify the plateaux for the lightest
two $0^+$ states very accurately, and for at least 2 of the other 3 states reasonably well.
The lightest $2^\pm$ plateaux in Fig.~\ref{fig_Eeff_M2mp0m1mp_so12_b250} are
well defined, and the excited $2^\pm$ identification is plausible although not entirely
compelling.
However the $0^-$ and $1^\pm$ plateaux in Fig.~\ref{fig_Eeff_M2mp0m1mp_so12_b250}
are clearly somewhat optimistic. The lesson is that while our $0^+$ and $2^\pm$
estimates should be reliable, at least at these larger values of $N$, the
$0^-$ and $1^\pm$ estimates should be treated with caution. All this is characteristic
of our mass calculations at larger $N$ when we are on lattices that are close to the
continuum limit. Of course a significant role is also played by the masses at smaller
$\beta$ since they help to determine the coefficient of the $O(a^2)$ lattice spacing
corrections in our continuum extrapolations. Here the uncertainties are naturally larger.
As an example we show in Figs~\ref{fig_Eeff_M0p_so12_b155}, \ref{fig_Eeff_M2mp0m1mp_so12_b155}
effective mass plots at $\beta=155$, which corresponds to our second coarsest lattice spacing.
Clearly the systematic errors aasociated with identifying the effective mass
`plateaux' are much larger on this lattice. While the ground state $0^+$ is unambiguous,
and the first excited $0^+$ is plausible, the higher excitation plateaux are much less
well defined. Identifying the $2^\pm$ ground states requires some optimism, although this
can be argued for on the plausible basis that if the gaps $aM_{eff}(a) - aM_{eff}(2a)$ and
$aM_{eff}(2a) - aM_{eff}(3a)$ are small and rapidly decreasing, and the statistical errors
are small enough for this statement to be meaningful, then one would expect the
effective mass plateau to have a value that is close to $aM_{eff}(3a)$. For the
excited  $2^\pm$ states and the ground state $0^-$ and $1^\pm$  identifying a
plateau from our plots is clearly guesswork. One can make some progress,
by noting that with our basis of iteratively blocked operators, the
overlap of our best variational operator onto the corresponding excited state should
be (very roughly) independent of $\beta$. So where we can estimate such an overlap
reliably at large $\beta$, we can assume roughly the same value at coarser $a(\beta)$
and so estimate at what value of $t=an_t$ we should see an effective mass plateau.
In this way we can sometimes estimate effective masses at smaller $\beta$ even if
the statistical errors do not allow a direct identification of the plateau.
Such an indirect estimate possesses systematic errors that are presumably not
large but are hard to quantify.

We now turn to smaller values of $N$ and, in particular, to $SO(3)$. The reason for doing
so is that our comparisons between $SO(N)$ and $SU(N^\prime)$ involve small
values of $N$ and, as we shall now see, the overlaps turn out to be poorer at
small $N$. We begin by showing in Fig.~\ref{fig_Meff_0p_so3su2} the effective
mass plots for, once again, the lightest five $0^+$ states, this time
in $SO(3)$. The calculation
is at a very small value of $a$, so we are effectively in the continuum limit here.
Comparing to Fig.~\ref{fig_Eeff_M0p_so12_b250} it is clear that the overlaps
of our operators onto the corresponding states are considerably worse in $SO(3)$
than in $SO(12)$, so that the plateaux are pushed out to larger $t$. While the mass
identification for the lightest 3 states looks reliable, it clearly becomes
shaky for the higher 2 states. For comparison we also show the corresponding
$SU(2)$ effective masses, taken from
\cite{AAMT_SUN}
at a similar small value of $a$ (in units of the mass gap). It is clear that
the $SU(2)$ calculation is far more accurate, basically because of the far better
overlaps. In Fig.~\ref{fig_Meff_Jm_so3su2} we provide a similar plot for the
lightest two $2^-$ states, and the lightest $0^-$ and $1^-$ states. Here only the
lightest $2^-$ mass can be extracted reliably, with some hint of a temporary
plateau for the excited $2^-$. For the $0^-$ and $1^-$ we do the best that we can,
which is to use the values at larger $t$ where the energy values overlap within their
very large errors. We remark, that the $SO(3)$ overlaps appear to be the worst,
with a rapid improvement as $N$ increases.

Given that $SO(3)$ and $SU(2)$ share the same Lie algebra, one might wonder
why the overlaps in the former should be so much worse than in the latter.
However we should recall that the fundamental of $SO(3)$ corresponds
to the adjoint of $SU(2)$, so one should really compare with the overlaps
in $SU(2)$ when one uses glueball operators that are in the adjoint representation
rather than in the fundamental. This we do in Fig.~\ref{fig_Eeff_M0p2pFA_su2_b23.5}
where we show the effective masses of the lightest three $0^+$ glueball states,
and the lightest $2^+$, on a $96^2 64$ lattice at $\beta=23.5$ in $SU(2)$.
We also show the effective masses using the same basis of operator loops,
but taken (as one would usually do) in the fundamental representation.
(For this comparison the operator basis is smaller than that used in
Figs.~\ref{fig_Meff_0p_so3su2},~\ref{fig_Meff_Jm_so3su2}.) We see that
the adjoint operators have an overlap that gets worse for the $0^+$ excited states,
and are already much worse for the $2^+$ ground state. Comparing the $SU(2)$ adjoint
correlators in Fig.~\ref{fig_Eeff_M0p2pFA_su2_b23.5} against the fundamental $SO(3)$
correlators in Figs.~\ref{fig_Meff_0p_so3su2},~\ref{fig_Meff_Jm_so3su2}, we see
that in fact $SO(3)$ and $SU(2)$ do have similar overlaps when we compare
corresponding operators.
(Aside: in $SU(2)$ we block our links in the fundamental and only take the
adjoint at the very end: ideally we should transform our elementary link variables
to the adjoint first and then block.)

One plausible argument why adjoint loops in $SU(2)$ might produce poor eigenoperators 
is as follows.  Consider the example of a $2^+$ operator. We suppose
that a single loop $\phi_0$ minus its $\pi/2$ rotation, $\phi_{\pi/2}$ 
provides our best $|J|=2$ adjoint eigenoperator,
$\phi_a^{J=2}={\mathrm{Tr_a}}\phi_0-{\mathrm{Tr_a}}\phi_{\pi/2}$, 
where we indicate that the trace is taken in the adjoint. Using the relation between
adjoint and fundamental traces, $Tr_a l = |Tr_f l|^2 - 1 $, we see that
\begin{equation}
  \phi_a^{J=2}
  =
  \left({\mathrm{Tr_f}}\phi_0-{\mathrm{Tr_f}}\phi_{\pi/2}\right)
  \times
  \left({\mathrm{Tr_f}}\phi_0+{\mathrm{Tr_f}}\phi_{\pi/2}\right)
\label{eqn_phiadj}
\end{equation}
i.e.  $\phi_a^{J=2}$ is actually a composite operator that is a product of scalar and
tensor operators taken in the fundamental. In a correlator this will project onto
states that simultaneously contain scalars and tensors. The lightest scalar state
is the vacuum so at large $t$ we will still see the lightest tensor mass, but
it is plausible (given that the best fundamental operator has a very good projection)
that the overlap of the scalar operator onto scalar glueballs will worsen the overlap
onto the lightest tensor. While this argument becomes more elaborate for linear combinations
of loops, the basic idea is that $SU(2)$ adjoint operators are composite in terms of the
underlying fundamental operators, and that this naturally suggests a poorer overlap if we
use the kind of operator basis that works well for the fundamental. A slight
variation of this argument can be made to apply to $SO(4)$ if one treats it as
$SU(2)\times SU(2)$, but clearly loses applicability as one increases $N$.

We finish with an important practical observation. As remarked already, the reflection
positivity of the action ensures that in an expansion such as eqn(\ref{eqn_M}),
the coefficients of the exponentials are indeed non-negative. This  means that
$E_{eff}(t)$ decreases monotonically with increasing $t$. That is to say, as we
approach the effective energy plateau by increasing $t$ we necessarily do so from
above. Thus the characteristic feature of the error that occurs in misidentifying the
plateau (something that readily occurs when the overlaps are mediocre) is that
the estimate of the energy is larger than the true value. This
should be borne in mind when, later on in this paper, we compare masses in various
$SO(N)$ and $SU(N^\prime)$ theories.

\subsubsection{finite volume corrections}
\label{subsubsection_finiteV}

As we reduce the spatial volume, $l^2$, we eventually encounter a transition 
at $l\sim l_c=1/T_c$ where $T_c$ is the deconfining temperature. This is a
cross-over at finite $N$, becoming a phase transition at $N=\infty$.
It is related to the phase transition that occurs on a space-time volume
$l_xl_yl_t$ when $l_y,l_t \to \infty$ and $l_x=1/T_c$, which is, of course, just
the deconfining transition with an interchange of spatial and temporal labels.
To avoid the dramatic finite volume effects associated with this transition
we shall always choose volumes $l\gg l_c$ for our glueball calculations. Note that
this finite volume effect is one that survives the $N\to\infty$ limit.

Once the spatial volume is sufficiently large, the leading finite volume
correction to the mass of a glueball $G$ arises from the 
virtual emission by $G$ of the lightest glueball, $g$, which then propagates 
around a spatial torus before being reabsorbed by $G$. This contribution is 
$\delta m_G \propto \alpha_ {gGG}\exp\{-cml\}$ where $m$ is the mass gap, 
$l$ is the size of the torus, $c=O(1)$, and $\alpha_{gGG}$ is the triple-glueball 
coupling squared
\cite{ML_V}. 
Since $ml_c\sim 4$ for both $SO(N)$
\cite{RLMT_Tc}
and $SU(N)$ 
\cite{Tc_d3}
theories, this correction is negligibly small if we choose, as we do, to use
$l\gg l_c$. In addition, large $N$ counting tells us that $\alpha_{gGG} \to 0$ as 
$N\to \infty$. So this correction vanishes as $N\to\infty$ for any $l>l_c$.

In practice, in a theory that is linearly confining the important finite
volume corrections have  to do with the finite volume
eigenstates of the Hamiltonian that correspond to flux tubes winding around a 
spatial torus. In $SO(2N)$ gauge theories the $Z_2$ center symmetry ensures
that a single winding flux loop operator has zero overlap onto a contractible glueball 
operator, just as in $SU(N)$. However a pair of winding flux loops can have
a non-zero overlap and will contribute to glueball correlators. These `torelon' 
states will have (after subtracting any vacuum expectation value) an energy 
that is, roughly, twice that of a single flux loop
\begin{equation}
E_T(l) \simeq 2 E(l) \simeq 2\sigma l.
\label{eqn_ET}
\end{equation}
Here we have neglected the interaction energy between the two flux loops 
(the first equality) and also finite $l$ string 
corrections (the second equality). Their interaction means that there may be 
a whole tower of such resonant torelon states. By adding or subtracting
the torelons around the orthogonal spatial tori we obtain torelons with
$J^p=0^+$ and $J^P=2^+$ respectively. The mixing between these orthogonal torelons
means that the $0^+,2^+$ torelon energies will be (slightly) different 
from the value of $E_T(l)$ in eqn(\ref{eqn_ET}). Now, as $l$ is reduced,  
$E_T(l)$ decreases and the torelon may become the  $0^+$ or $2^+$  ground 
state, and even before that may appear as one of the low-lying excited 
glueball states. As an example consider the $2^+$ glueball ground state.
For larger $N$ this satisfies, as we shall see,  $m_{2^+} \sim 7\surd\sigma$.
The torelon
will have a similar energy if we reduce $l$ to $2\sigma l  \simeq 7\surd\sigma$.
This corresponds to $l \sim 3.5 l_c$ using the fact that one finds
$\surd\sigma/T_c = l_c\surd\sigma \sim 1$ at large $N$
\cite{RLMT_Tc,Tc_d3}.
This type of estimate allows us to control this type of finite volume 
correction. For $SO(2N+1)$ there is no centre symmetry and there may be a torelon 
composed of a single winding flux loop. However as we shall see below
such theories are `almost' confining and the overlap of such torelons
onto glueball operators appears to be negligible, except possibly for SO(3).
So the situation is in practice just as for $SO(2N)$. Finally we note that the mixing
between the two flux loops and the local glueball will vanish as $N\to\infty$ since
it involves the matrix element of a single trace operator with a double trace operator.
and so these finite volume corrections should become unimportant at larger $N$.

In the above paragraph we discussed torelon contributions to the $0^+$ and $2^+$. 
There will also be contributions with other quantum numbers, but these will require 
at least one of the flux tubes to be in an excited state. These are much heavier
\cite{AAMT_string,MTstring_old}
and so we assume they will not affect our calculations in this paper.
An important remark is that since the $2^+$ and
$2^-$ glueball spectra are identical in the continuum limit in an
infinite volume, and since the low-lying  $2^-$ spectrum is unaffected by torelons,
a comparison between the two spectra at a finite but small lattice spacing
provides a good tool for identifying those $2^+$ states which are
affected by torelons. For the $0^+$ spectrum we unfortunately have no
analogous procedure. 

There is a subtlety to this that is important for low values of $N$. In addition
to torelons composed of pairs of flux tubes in the fundamental representation, 
there may be torelons composed of pairs of flux tubes in the spinorial
representation. As $N$ increases the dimension of the spinorial representation 
becomes large and so, presumably, will the associated string tension, so
any such torelons will be very massive and irrelevant. At low $N$ 
however they may be light enough to dominate. So, in $SO(3)$ the
spinorial is just the fundamental of $SU(2)$ and the corresponding torelon
should therefore be about $2.5$ times lighter than the fundamental
torelon of $SO(3)$ which corresponds to the adjoint of $SU(2)$. In $SO(6)$
the spinorial representation is the fundamental of $SU(4)$ and so the 
corresponding torelon energy will be about $3/4$ that of the
fundamental. So for $N\leq 6$ this needs to be taken into account
in coming to a conservative estimate of what is a `large enough' spatial
volume for a glueball calculation.

While one can make theoretical estimates of finite volume effects, as we
have done above, additional  numerical finite volume studies are essential
to make sure we have not missed something. So in Table~\ref{table_Vso8} we show
the low-lying glueball spectrum as a function of spatial volume in $SO(8)$.
The value $\beta=84.0$ chosen for this study is representative of the values 
used in our later calculations. (See Table~\ref{table_glueball:so8}.)
We note that strong finite volume corrections appear only for $l<24$
and that they appear most strongly in the $0^+$ and $2^+$ channels as
one would expect from our above discussion. In units of the string
tension this suggests that volumes with $l\surd\sigma \geq 3.2$ are
`large enough' for the purposes of this paper, at least for $N\geq 8$,
 much as predicted by our preceeding theoretical discussion.
So the volumes we use for $N\geq 8$, as listed in
Tables~\ref{table_glueball:so8},\ref{table_glueball:so12},\ref{table_glueball:so16},
are chosen to satisfy this constraint.

For smaller $N$ we incorporate into our choice of volumes, listed in
Tables~\ref{table_glueball:so3}--\ref{table_glueball:so7},
not only the constraint $l\surd\sigma \geq 3.2$ 
but also the more demanding one based on rough estimates of the spinorial torelon
energies. As an explicit check we show in Table~\ref{glueball:tab:finitevolumesoN}
a comparison between the spectra obtained on two different volumes for
each of $SO(3)$, $SO(4)$ and $SO(6)$. In each case one of the two volumes
is the standard volume we use in our later calculations. In the cases
of $SO(3)$ and $SO(6)$ the corresponding masses on the two volumes
are well within $2\sigma$, as they are for nearly all the  $SO(4)$ masses.
In the latter case, only the $0^{+\star}$ values are more than $3\sigma$ apart.
(Note that the respective labelling of the $0^{+\star\star\star}$  and
$0^{+\star\star\star\star}$ states for $l=34$ in $SO(4)$ is determined by the
effective masses at $t=a$, but the actual mass estimates indicate a
reversed ordering, which agree much better with the $l=44$ values.)

We take the above arguments and explicit checks as reasonably good evidence
that the volumes we use, as shown in Table~\ref{table_VsizeN}, are large
enough to avoid significant finite volume corrections.

\subsubsection{multi-glueball states}
\label{subsubsection_multiglueball}

We are primarily interested in the mass spectrum of glueballs, i.e. stable
particles, or resonances that are narrow enough to be unambigous. At large
$N$ resonance widths are expected to vanish so that our $N=\infty$ spectrum,
as calculated by correlators of single trace operators, will indeed consist of
some tower of stable particles. This is what we hope to obtain via our
large-$N$ extrapolations. However at finite $N$ our single trace `glueball'
operators will have a non-zero overlap onto multi-glueball states that are  
naturally represented by multi-trace operators. So for example we may
have in the $p=0$ $0^+$ channel states composed of two scalar glueballs with
equal and opposite momenta. Similarly for the $2^+$
channel (except that now the momenta must be non-zero). In the $p=0$ $J=1$ sector
the two glueballs cannot be the same (otherwise the state will be null)
so these states will be heavier. Since these multiglueball states are heavy and
since they should in any case disappear from the single-trace spectrum as
$N$ increases, we shall ignore here the possibility that some of our
heavier states might not be single glueballs but might be, say, pairs of glueballs.
At smaller $N$ this assumption might not be correct, but an explicit
calculation to check this, by putting a large basis of multi-glueball operators 
in our variational basis would take us beyond the scope of this paper.

A particular caveat attaches to states in $SO(4)$. If the continuum spectrum is
indeed the same as that of $SU(2)\times SU(2)$  then there will be states with
one glueball from each of the $SU(2)$ groups with a possible shift in mass due to
lattice spacing corrections. There is no obvious reason for the projection of
such states onto our operator basis to be any less than that of single glueball
states, and such extra states can both confuse a simple minded extrapolation of
states to $N=\infty$, and complicate the comparison  between the glueball
spectra of $SO(4)$ and $SU(2)$.

\subsubsection{reliability grades}
\label{subsubsection_grades}

The errors that we will provide on the mass estimates in our Tables will be purely statistical
since only these can be easily quantified. However, as will be apparent from the above
discussion, we expect the systematic errors to be substantial in many cases. Clearly it
would be useful to provide the reader with some guidance as to how reliable we believe
our individual mass estimates are. To do so we have assigned a grade ranging
from $\alpha$ to $\phi$ to each of our mass estimates, which we shall now define.

Our finite volume checks confirm our expectation that at larger $N$ we do not have
significant finite volume corrections. At smaller $N$ we avoid torelon corrections
by making the volume large enough that two (spinorial) flux tubes are heavier
than our heaviest $0^+$ excitation. Multi-glueball states should only be visible,
if at all. at small $N$, and it is only the 4'th $0^+$ excitation that is heavy
enough to be possibly affected. Similarly for two glueball states in $SO(4)$. There
may be mis-identifications of the spin, but these do not impinge on the quoted masses.

So the main systematic error arises in trying to identify the `effective mass plateau'
as described in Section~\ref{subsubsection_Eplateaux}, and this is what our grades are
designed to reflect. We denote by $\alpha$ mass estimates for which we believe any 
such error is insignificant and much smaller than the quoted statistical error. Thus
we grade the lightest 3 $0^+$ states in Fig.~\ref{fig_Eeff_M0p_so12_b250} as $\alpha$,
as well as the lightest $2^\pm$ states in Fig.~\ref{fig_Eeff_M2mp0m1mp_so12_b250}.
Also the ground state $0^+$ on the coarser lattice in Fig.~\ref{fig_Eeff_M0p_so12_b155}.
The grade $\beta$ indicates a systematic error that is possibly significant, but
probably smaller than the statistical error. Thus the 4'th excited $0^+$ state
in  Fig.~\ref{fig_Eeff_M0p_so12_b250}, the excited  $2^\pm$ states in
Fig.~\ref{fig_Eeff_M2mp0m1mp_so12_b250}, as well as the first excited $0^+$
state in Fig.~\ref{fig_Eeff_M0p_so12_b155}. With the grad $\gamma$ we indicate
a systematic error that may be significant although the pattern of effective
masses suggests it will not be much larger than the statistical error. Thus
the 3rd excited $0^+$ in Fig.~\ref{fig_Eeff_M0p_so12_b250}, the $0^-$ in
Fig.~\ref{fig_Eeff_M2mp0m1mp_so12_b250}, the second excited  $0^+$ in
Fig.~\ref{fig_Eeff_M0p_so12_b155} and the lightest $2^\pm$ in
Fig.~\ref{fig_Eeff_M2mp0m1mp_so12_b155}. With grades $\delta$ and $\phi$
we indicate mass estimates where we largely lose control of the systematic
error, but assume we can extract a mass from the furthest point
at which the statistical errors allow a mass to be extracted. This is likely to
provide some kind of estimate as long as the overlpa is not anomalously small.
(Where what is anomalous is determined by the oerlaps of the lighter states whose
overlaps can be estimated.) States that we label $\delta$ are the $1^\pm$
in  Fig.~\ref{fig_Eeff_M2mp0m1mp_so12_b250},and states we label $\phi$ are the
excited $2^\pm$, the $0^-$ and the $1^\pm$ in Fig.~\ref{fig_Eeff_M2mp0m1mp_so12_b155}.

To avoid cluttering up the Tables of masses we do not show the individual grades
but instead provide an overall grade for the continuum limit of each state.
This grade takes all the grades at the various lattice spacings into account, but
with a weighting  to the grades at the smaller three lattice spacings. Note however
that this overall grade does not attempt to reflect any systematic error in making
the continuum (or large $N$) extrapolations. The quality of these fits is indicated
separately. 

Finally we remark that our calculations of flux loop masses, from which we extract
the string tensions, are all $\alpha$ except for $SO(3)$, which is discussed
in much more detail in Section~\ref{subsection_oddN_confinement}, and
sometimes for the very coarsest value of $a$, particularly in the case
of $SO(4)$ and $SO(5)$.

\section{Confinement}
\label{section_confinement}
We begin with our calculation of the string tension in $SO(N)$ gauge theories.
This is the energy per unit length of the confining flux tube carrying flux
in the fundamental representation. Of course this assumes that the theories are
linearly confining, just like $SU(N)$ gauge theories, and this needs to
be established. The question is particularly delicate for
$SO(2N+1)$ since these groups have a trivial centre and in
$SU(N)$ theories the deconfined and confined phases are the ones
in which the $Z_N$ centre symmetry is or is not spontaneously broken.
We sketch the usual simple argument. Consider (the trace of) an operator
$\phi_t(x,y)$ which consists of the trace of a product of link matrices
on a path that winds on a timelike loop once around the timelike $t$-torus
of length $l_t$. This is just a Polyakov
loop and is the operator associated with the world line of a static fundamental
source located at the point $(x,y)$. Now suppose we change the field on the lattice
by multiplying all the link matrices emanating in the $t$-direction from lattice
sites at a fixed value of $t$, say $t=t_0$, by a non-trivial element of the
centre, $z\in Z_N$. Clearly $\phi_t\to z\phi_t$.
But this new field has exactly the same weight in the path integral as the
original field since the Haar measure is invariant, ${\cal{D}}(U_l)={\cal{D}}(zU_l)$,
and so is the action because it is composed of contractible closed loops that
acquire a factor of $z^\dagger$ for each factor of $z$. That
is to say, this transformation is a symmetry of the theory. So if we assume
the vacuum is unchanged under the symmetry transformation (i.e. that the
symmetry is not spontaneously broken) we immediately deduce that
$  \langle \phi_t \rangle = z \langle \phi_t \rangle = 0 $
and hence that
\begin{equation}
  \langle \phi^\dagger_t(x,0) \phi_t(0,0) \rangle
  \stackrel{x\to\infty}{\to} |\langle \phi_t \rangle|^2 = 0.
\label{eqn_expE}
\end{equation}
Since this correlator is just $\propto \exp\{-E(x)l_t\}$ where $E(x)$ is
the (free) energy of a pair of conjugate fundamental sources a distance
$x$ apart, this tells us that $E(x) \to \infty$ as $x\to\infty$, i.e.
the theory is confining. (Of course, whether it is {\it linearly} confining is
another matter.) In this way the non-trivial centre symmetry leads to
confinement when it is not spontaneously broken.
One can carry this kind of argument further. When we calculate the string
tension we use lattices of size $l_xl_yl_t$, we make $l_y,l_t$ large and
we then calculate the energy of a flux tube winding around the $x$-torus from
correlators of operators $\phi_x(y,t)$ defined on paths that wind around the
$x$-torus. If we apply the same centre symmetry transformation as
above, but with $t\to x$, then $\phi_x\to z\phi_x$. On the other
hand  an operator $\phi_o(x,y,t)$ defined on a closed contractible spatial loop
is unchanged. Hence
\begin{equation}
  \langle \phi_x(y,t) \phi_o(x,y,0) \rangle
  =
  z \langle \phi_x(y,t) \phi_o(x,y,0) \rangle =0.
\label{eqn_ZNtransform}
\end{equation}
Since operators of the form $\phi_o$ provide a basis for glueballs and for the
vacuum, this tells us that when the centre symmetry (around the $x$ torus) is
not broken, the winding flux tubes cannot break into glueballs and/or the vacuum.
Thus we see that in the familiar context of pure $SU(N)$ gauge theories
the fate of the centre symmetry is tied to that of confinement.
Of course things are less simple in the even more familiar case of QCD, where
the quarks break the centre symmetry and break flux tubes, but nonetheless
the theory possesses quasi-stable flux tubes and is confining.
The fact that $SO(2N+1)$ gauge theories have a trivial centre makes us ask
in what sense they might be confining and this is something
we will address at some length in this section, after first remarking on the more
straightforward case of $SO(2N)$ gauge theories.

\subsection{SO(2N) and confinement}
\label{subsection_evenN_confinement}
$SO(2N)$ gauge theories have a $Z_2$ centre symmetry that ensures confinement,
as long as the symmetry is not spontaneously broken: as described above, the expectation
value of a Polyakov loop is zero so (at least naively) the free energy of an
isolated fundamental charge is infinite. Of course what we are really interested
in is something more: linear confinement. To demonstrate this we need to show
that the energy of the ground state that couples to the Polyakov loop of length $l$
grows linearly with $l$, up to $O(1/l)$ corrections, i.e. that it is a flux `tube'.
To do this we take the correlator of a Polyakov loop type of operator that 
winds once around the $x$-torus of an $l_x\times l_y \times l_t$ lattice, where  
$l_y$ and $l_t$ are sufficiently large that finite volume corrections are
negligible, and we calculate the energy of the ground state energy $E(l)$ as a function of
$l=l_x$ to see if it grows (roughly) linearly with $l$.

We provide two examples of such calculations, one in $SO(6)$  at $\beta=46.0$
and  one in $SO(8)$  at $\beta=86.0$. These $\beta$ values correspond 
to small enough values of $a$ that the $O(a^2)$  lattice corrections should be
very small. The lattices that we use and the values of $E(l)$ that we extract
are listed in Table~\ref{table:Estrings_so6so8}. We plot these values in
Fig.~\ref{fig_Ek1_so6_so8}. It is clear that at larger $l$ the energy, $E(l)$,
grows roughly linearly with $l$ in both cases. This is an approximate statement
because there are $O(1/l)$ corrections to the linear dependence. However for a
string-like flux tube it is known that up to $O(1/l^5)$ the corrections to $E(l)$
are universal
\cite{OA_string}
and coincide with what one gets by expanding the Nambu-Goto formula
\begin{equation}
E(l) = \sigma l \left(1- \frac{\pi}{3\sigma l^2}\right)^{1/2} 
\label{eqn_NGgs}
\end{equation}
in powers of $1/\sigma l^2$. We therefore fit our calculated values of $E(l)$
to the expression in eqn(\ref{eqn_NGgs}). We find that for $SO(8)$ we obtain a
good fit to all our values of $l$, with a resulting string tension
\begin{equation}
a^2\sigma = 0.017065(57)     \qquad l\in [12,40], \,\, \chi^2/n_{dof}= 0.82 \quad : SO(8),
\label{eqn_NGgsSO8}
\end{equation}
where $n_{dof}$ is the number of degrees of freedom in the fit.
It is interesting to note that in order to get a good fit down to a flux tube
as short as $l=12a$ one really needs the higher order universal string corrections
encoded in eqn(\ref{eqn_NGgs}): one cannot get a good fit with just the leading
$O(1/l)$ Luscher correction
\cite{ML_string}.
In the more accurate case of $SO(6)$ we again have a very good fit
\begin{equation}
a^2\sigma = 0.016014(27)     \qquad l\in [18,42], \,\, \chi^2/n_{dof}= 0.89 \quad : SO(6).
\label{eqn_NGgsSO6}
\end{equation}
We note that in this case a fit with just an $O(1/l)$ Luscher correction is perfectly
adequate, essentially because the range of $l\surd\sigma$ does not include values
as small as in the case of $SO(8)$.
All this provides convincing evidence that $SO(2N)$ gauge theories are indeed
linearly confining, as one might expect.

Of course these results are only as robust as our identification of the effective
energy plateaux, from which we obtain our values of $E(l)$. We display the effective
energies in Fig.~\ref{fig_Eeffk1_so8_b86} and Fig.~\ref{fig_Eeffk1_so6_b46}.
It should be clear we do indeed have unambiguous plateaux for all our values of $l$.

The fact that the centre of $SO(2N)$ is $Z_2$ in contrast to the $Z_N$ centre
of $SU(N)$, does create some differences in the confining properties of the two
sets of theories. In $SU(N)$ we have stable flux tubes in higher representations.
Consider a source that transforms as $z^k$ under a global gauge transformation 
$z\in Z_N$. If $k\leq N/2$ the flux tube carrying the flux from that source
will be stable since the gluons in the vacuum that might screen the source
in fact transform trivially under $z$. So the $Z_N$ symmetry imples that
we have $N/2$ $k$-strings, with $k=1$ corresponding to the fundamental. In
$SO(N)$ our $Z_2$ centre only leads to a stable $k=1$ string. This raises an
interesting question of how the $SO(2N)$ gauge theories recover the $Z_N$ physics
as $N\to\infty$, given the common planar limit. This is a subtle question
because even in $SU(N)$ the $k$ strings, which one might think of as bound states
of $k$ fundamental strings, become unbound at $N=\infty$. We do not pursue this
issue further in this paper, but hope to address it in detail elsewhere.

\subsection{SO(2N+1) and confinement}
\label{subsection_oddN_confinement}

The question of confinement in $SO(2N+1)$ gauge theories is  more delicate,
since here the centre is trivial and we cannot argue on symmetry grounds that
there is a phase in which the expectation value of the Polyakov loop will vanish.
Moreover we expect that at small values of $N$ the flux tubes may be unstable.

In particular consider $SO(3)$. Here the fundamental flux tube corresponds to the
$SU(2)$ adjoint flux tube which can be broken by pairs of gluons from the vacuum,
and hence can decay into glueballs,
just as the $QCD$ flux tube is broken by quark-antiquark
pairs from the vacuum, and can decay into mesons. In addition estimates of the $SU(2)$
adjoint string tension suggest $\sigma^{SU2}_{adj} > 2 \sigma^{SU2}_f$
\cite{AAMT_string}.
So in $SU(2)$ an adjoint flux tube is energetically capable of decay into a fundamental
flux tube and its conjugate and will acquire an additional decay width due to this
process. One naively expects that $\sigma^{SO3}_f = \sigma^{SU2}_{adj}$ and that the
$SO(3)$ fundamental flux tube should have a decay width that is identical
to the decay width of the adjoint flux tube in $SU(2)$. (Recall that the fundamental of
$SU(2)$ corresponds to the spinorial of $SO(3)$.) Of course if this decay width is
very small, as is the case in $QCD$, then it should still be possible to identify numerically
an approximate linear growth of the flux tube energy with its length. And in any case
since $SU(2)$ is exactly confining in the sense that colour is confined, irrespective of
the adjoint flux tube instability, one can presumably think of the $SO(3)$ theory as also
being exactly confining even if, just like QCD, it does not provide us with an asymptotic
area law for Wilson loops in the fundamental representation.

More generally, we recall that the above relationship between $SO(3)$ and $SU(2)$, with
$SU(2)$ being the largest simply connected group that has the same Lie algebra as
$SO(3)$, is a special case of the same relationship between $SO(N)$ and $Spin(N)$
\footnote{We thank Michele Pepe for emphasising to us the relevance of this in the
  present context}. And $Spin(N)$, the double cover of $SO(N)$, has a nontrivial centre
that is $Z_2$, $Z_4$ or  $Z_2\times Z_2$ depending on the value of $N$. Thus for
$Spin(N)$ we have the usual link between the centre symmetry and confinement and
if that confinement is `linear' there will be absolutely stable flux tubes in that
theory, corresponding to spinorial flux tubes in $SO(N)$. If this is so then this
suggests that $SO(2N+1)$ theories should be thought of as confining, just like $SO(3)$,
even if the fundamental flux tube is not perfectly stable. The deconfinement
transition for $Spin(5)$ and $Spin(6)$ has been investigated in
\cite{Pepe}.

In $SO(3)$ the fact that the flux tube can break is obvious since the fundamental of
$SO(3)$ is also the representation of the gluons in the theory so that it can
be broken by gluon pairs from the vacuum. For larger odd values of $N$
this is no longer the case and it is less clear to us what might be the mechanism
for any instability of the fundamental flux tube. For example, as $N$ increases the
spinorial flux tubes should have a rapidly increasing string tension and hence energy 
(using the quadratic Casimir as a guide) and will become irrelevant.
It is therefore interesting to investigate what happens to linear confinement in
$SO(2N+1)$ gauge theories. Our investigation here does not attempt to be definitive,
something that in any case is beyond the capability of a numerical calculation,
but we will at least try to establish whether it makes sense to calculate a confining
string tension for such theories.

We begin with $SO(5)$. We perform a calculation of $E(l)$ on lattices at $\beta=27.5$.
As we shall shortly see, this corresponds to a string tension $a^2\sigma \simeq 0.02$,
so the $O(a^2$) lattice corrections to continuum confining physics should be negligible.
We list the ground state flux tube energies in Table~\ref{table:Estrings_so3so5}.
In Fig.~\ref{fig_Ek1_so5_b27.5} we display these energies as a function of $l$.
It is clear that $E(l)$ does indeed rise (roughly) linearly at larger $l$. In fact
we find that we get an acceptable fit to all $l$ using the Nambu-Goto formula in
eqn(\ref{eqn_NGgs}) 
\begin{equation}
a^2\sigma = 0.019185(35)    \qquad l\in [14,42], \,\, \chi^2/n_{dof}=1.70  \quad : SO(5).
\label{eqn_NGgsSO5}
\end{equation}
(and a very good fit, with $a^2\sigma = 0.019208(36)$, if we exclude the smallest value of 
$l$). Moreover there is no difference to these results whether we use correlators of vacuum
subtracted operators or not. Thus it certainly looks as if $SO(5)$ is linearly confining.
Of course this conclusion is only reliable if our extraction of $E(l)$ is reliable.
So in Fig.~\ref{fig_Eeffk1_so5_b27.5} we display the effective energies, as defined
in eqn(\ref{eqn_E0eff}), from whose `plateaux' we extract the values of $E(l)$.
We show separately the values obtained from the correlators of operators with and
without the vacuum expectation value subtracted. As one can see, any difference is much
smaller than the statistical errors. That is to say, any overlap of the ground state
flux tube onto the vacuum is invisible within our errors. Indeed the `plateaux'  appear
to be just as well-defined as they are for $SO(6)$ or $SO(8)$, and  we see no indication
that any of these flux tubes is unstable.

The calculations above demonstrate that the ground state of the flux tube in $SO(5)$ has at
most a very small overlap onto the vacuum. So it is interesting to ask if we have any
good evidence that it is in fact non-zero. Another interesting question is whether the
vacuum has a non-zero overlap onto our whole basis of winding operators. 
To address the first of these questions we take at each $l$ the ground state winding
operator $\phi_{gs}(l)$ that arises from our variational procedure and calculate
its vacuum expectation value, $\langle \phi_{gs} \rangle$. To answer the second question
we take the orthonormal set of operators $\{\phi_i\}$ that are produced by our variational
calculation, and which therefore span our space of winding operators, and contruct the total
projection onto that space by  $\sum_i|\langle \phi_{i} \rangle|^2$. The values of both these
measures are listed in Table~\ref{table:overlaps_so5}. For each $l$ we show the values as
obtained with two bases of operators, characterised by the largest blocking level $bl_{max}$.

Of course our best variational operator $\phi_{gs}(l)$ does not have a $100\%$ overlap
onto the real ground state, $|gs\rangle$, because our basis is finite. We show estimates
of the overlap $O^2_{gs}=|\langle vac|\phi_{gs}|gs \rangle|^2$ in the Table,
and $1-O^2_{gs}$ is then the size of the overlap onto the non-ground-state contributions
to the correlator of  $\phi_{gs}$. So it is only if we find that
\begin{equation}
|\langle \phi_{gs} \rangle|^2 > 1-O^2_{gs}
\label{eqn_bound}
\end{equation}
that we can claim that the vacuum overlap cannot be just onto the small excited state component 
of  $\phi_{gs}$, but must be, at least in part, onto the ground state as well. Turning back 
to  Table~\ref{table:overlaps_so5} we see that
most of the overlaps onto $\phi_{gs}$ are consistent with zero and those that appear not to be 
are very small and  much too small to satisfy the bound in eqn(\ref{eqn_bound}), despite the
fact that $O^2_{gs}$ is very close to unity. That is to say,
we have no evidence in these calculations that the true ground state wave-functional has a non-zero
overlap onto the vacuum.

On the other hand we see from  Table~\ref{table:overlaps_so5} that
$\sum_i|\langle \phi_{i} \rangle|^2$, the vacuum overlap onto our whole
basis of winding operators, is non-zero. In fact this non-zero overlap arises
entirely (within errors) from the most highly blocked operators, which extend
in the transverse directions right around the lattice torus, and further. It might appear
that the values of the overlaps for the lower value of $bl_{max}$ are also non-zero compared
to the errors, even if they are very small. This is true, but it is consistent
with being the result of the fact that we are summing here positive definite quantities,
so the fluctuations will appear to give non-zero results. We find that we get
comparable results in our $SO(6)$ calculation, which supports this interpretation.
We also show the effective energy at $t=a$ for each basis. The fact that
the pairs of values are equal within errors indicates that the smaller basis
already contains all of the ground state that one finds with the larger basis. 
So we conclude that while there certainly is a non-zero vacuum overlap onto our
winding operators, this overlap arises entirely, within our errors, from
the operators that are so highly smeared that they wrap right around the transverse
directions and, moreover, these operators do not contribute to the wave-functional of the
ground state of the $SO(5)$ flux tube.

We now turn to the more delicate case of $SO(3)$ where we expect that the ground state flux 
tube will be unstable because it can be broken by the gluons in the vacuum. We perform
at $\beta=9.0$ a similar calculation to the one described above for $SO(5)$. Here the
lattice spacing is much smaller (in units of the mass gap) than in our $SO(5)$ example,
so we expect lattice corrections to be even smaller. In $SO(3)$ the vacuum
overlap onto our operator basis turns out, as expected, to be much larger than
in the case of $SO(5)$, so we choose to calculate $E(l)$ separately for correlators
with and without an explicit vacuum subtraction. In the latter case the ground state
of our variational procedure may be the vacuum and in that case we take the first
excited state to be the flux tube ground state. This we need to do
for $l=34,38,62$. (Recall that our `ground state' operator is defined to be the one
that maximises $E_{eff}(t=a)$ and if the overlap of the vacuum onto the basis is small,
then its $E_{eff}(t=a)$ will be large, even though $E_{eff}(t)$ will drop to zero at
larger $t$.) In addition, for our smallest value of $l$ the vacuum mixes into both the
ground state and the first excited state so in that case we drop the operators
at the highest blocking level (which have most of the vacuum projection) and use the
remaining basis for the calculation. Obviously the vacuum subtracted correlators do
not need to be tweaked in this way. The resulting ground state flux tube energies
are listed in Table~\ref{table:Estrings_so3so5} and displayed in
Fig.~\ref{fig_Ek1_so3_b9.0}. We show what one obtains with and without vacuum
subtraction and one sees that there is no appreciable difference. In both cases
the approximate linear growth with $l$ at larger $l$ is evident. The
behaviour at smaller $l$ appears to be more complicated. Our expectation that
the flux tube should be unstable, makes it useful to display the effective energies,
which we do in Fig.~\ref{fig_Eeffk1_so3_b9.0} for the vacuum unsubtracted case. 
We see that the determination of an effective energy plateau is much harder than in 
the case of $SO(5)$ or $SO(6)$. The main culprit is not the instability, if any, of the 
flux tube, but rather a mediocre overlap of our operator basis onto the
ground state. This means that at the larger values of $l$ our correlator
disappears into the noise before we can be confident that we have identified a plateau.
Nonetheless the plateau identification appears plausible for $l \leq 46$ and perhaps
also for $l=52$, but one cannot put it more strongly than that.
For $l=62,\, 82$ the choices are clearly speculative and are essentially motivated by 
assuming that the plateau will start at roughly the same value of $t=an_t$ as at the 
lower values of $l$ where it is more clearly identifiable.

In fact the mediocre overlaps in $SO(3)$ should not come as as a surprise since we
have already seen in Figs~\ref{fig_Meff_0p_so3su2},\ref{fig_Meff_Jm_so3su2} that
the same is the true for glueballs. Moreover we can  see the same behaviour
in $SU(2)$ when we perform calculations with flux tubes in the adjoint representation.
We show in Fig.~\ref{fig_Eeff_k1FA_su2_b16.0} a comparison
of the effective energies of ground state flux tubes in the fundamental  and
adjoint representations in $SU(2)$ at $\beta=16$, which corresponds to a value
of the lattice spacing similar to the one we have in $SO(3)$ at $\beta=9$.
For the purposes of comparison we have renormalised the adjoint
$E_{eff}(t)$ values at each value of $l$ to asymptote to the same value
as the fundamental effective energies at that $l$. We see from
Fig.~\ref{fig_Eeff_k1FA_su2_b16.0} that the adjoint overlaps are much worse
than the fundamental ones, which means that the plateau in the effective energies
is reached much more slowly and is much harder to identify, just as in $SO(3)$.

A further issue arises when one looks at the excited states of the flux tube. One finds
that several of these have effective energies that decrease at larger $t$, possibly
to values comparable to or below that of our `ground state'. To understand this
behaviour (which we do not see in $SO(5)$ or with any other value of $N$) it is
again useful to consider the adjoint flux tube in $SU(2)$ where one sees a very
similar behaviour
\cite{AAMT_string}.
In the latter case there is a ready interpretation: we are seeing states composed of
a pair of winding fundamental flux tubes with various equal and opposite transverse
momenta. (And possibly glueballs as well.) We can carry this interpretation over
to $SO(3)$, with the fundamental flux tubes of $SU(2)$ becoming spinorial flux tubes. 
Although this is reasonable and provides a resolution of the puzzle of the
numerous low-lying states, we also need to assume
that it is our (variational) ground state, rather than one of the excited
states, that is the ground state of the flux tube. The argument is that
the states composed of two spinorial flux tubes will have a suppressed coupling
to our fundamental operators, which is why they have larger values of
$E_{eff}(t=a)$ and appear as excited states in the variational calculation.
This is essentially a large-$N$ argument and is only partially 
convincing here, since the value $N=3$ is not large.

All this suggests that in $SO(3)$ we do indeed have some kind of partially
stable ground state flux tube, although the mediocre quality of the effective
energy `plateaux', does leave room for substantial instability. To proceed
further we need to be more quantitative, so we calculate the same overlaps that
we calculated for $SO(5)$. We list the results in Table~\ref{table:overlaps_so3}.
The vacuum overlap onto our whole winding basis is clearly larger than in $SO(5)$
but, just as in the latter case, most of the overlap is onto operators at the very
largest blocking level. In most cases if we discard these highly smeared operators,
the remaining basis has an overlap onto the ground state that is very nearly the same,
as we infer from the values of $E_{eff}(t=a)$ listed in Table~\ref{table:overlaps_so3},
and this slightly reduced basis has essentially no overlap on to the vacuum.
If we just look at $|\langle \phi_{gs}\rangle|^2$ we see that it is remarkably small, even
when we keep the largest blocking level. Although it is usually non-zero, the overlap
of  $\phi_{gs}$ on to excited states, as measured by $1-O^2_{gs}$ is always very much
larger so we cannot infer that the true ground state flux tube wave-functional has a
non-zero overlap onto the vacuum. This conclusion is of course almost inevitable
given the mediocre overlap of the ground state onto our operator basis in $SO(3)$, and so
one should not read too much significance into it.

We have remarked that the vacuum overlap typically finds its way into states
that our variational criterion labels as highly excited. 
As an example of this we show in Fig.~\ref{fig_Cork1_so3_b11} some
correlation functions obtained on a $100^280$ lattice at $\beta=11.0$ in $SO(3)$,
with the quantum numbers of the ground state of the flux tube. We show those of our
4 lightest states (including the ground state) none of which exhibit any significant
instability, together with corresponding exponential fits. We then show the 8'th excited
state, which is the first one to exhibit a significant vacuum expectation value.
We also show the 10'th which shows the maximum vacuum expectation value. The 
accompanying exponential lines are the fits to the 8'th and 10'th excited states 
that one obtains if one explicitly subtracts the vacuum expectation value from the
operators.

We have seen above that the signals for confinement in $SO(5)$ are much more
convincing than for $SO(3)$. This prompts the question; 
what happens as we increase $N$ further, with $N$ odd?
To address this question we have calculated the properties of flux loops of
length $l=36$ for $N=3,5,7,9,11$. We choose values of $\beta$ such that
the lattice spacing is comparable (in physical units) between these different
gauge groups. We perform the same analysis of vacuum overlaps as we did above for 
$SO(3)$ and $SO(5)$, and we display our results in Table~\ref{table:overlaps_soNodd}.
It is striking that even though the overlap squared of our variational ground state
operator onto the true ground state is at least $99.5\%$ at larger $N$, any
vacuum expectation of that operator is so tiny that it provides no evidence for the
instability of the true ground state. We also see that even when our basis includes 
operators that wrap around the boundaries, with $bl_{max}=6$, the total projection
of the vacuum onto the basis decreases very rapidly with increasing $N$, apparently 
much faster than a low inverse power of $N$. We also see that for $N\geq 5$ the
$bl_{max}=6$ basis does not have (within small errors) any more projection onto the
ground state than
the smaller  $bl_{max}=5$ basis. And for $N\geq 5$ the vacuum expectation value of
our variationally selected ground state operator in this latter basis is consistent 
with being exactly zero within the very small errors.

In summary, the message from these calculations is that for odd $N\geq 5$  we have a
(ground state) flux tube that is stable within our very small errors so that, for all
practical purposes, we have linear confinement with a well-defined string tension.
For $SO(3)$ our errors are much larger, although here
too we can identify a flux tube, albeit one that may not be very stable. We note
from Tables~\ref{table:overlaps_so5},~\ref{table:overlaps_so3} that if we
restrict ourselves to operators up to a given maximum blocking level, the vacuum
overlaps decrease quite rapidly as $l$ increases, leaving open the interesting
if speculative scenario that the flux tubes might become exactly confining in the
large $l$ limit.

It would of course be useful to go beyond this focus on vacuum overlaps,
and calculate overlaps onto glueballs as well. (This would require enlarging our
basis so as to include cross-correlations with glueball operators and although 
straightforward would represent a quite different calculation and 
would take us outside the scope of the present paper.) It is, after all, well-known
that in practice it is very difficult to obtain statistically significant
evidence for string breaking in $QCD$ without explicitly including in the basis of
flux tube operators the typical meson pairs into which the flux tube breaks.
While we cannot compare our results to most of those calculations, which involve
flux tubes ending at sources, there do exist calculations using blocked
Polyakov loop operators similar to ours. An example is
\cite{AHMT}
where the calculations are performed with moderately heavy quarks on $L_s=16$ lattices.
As we see from Fig.1 of
\cite{AHMT}
the vacuum expectation values of the blocked Polyakov loops are, while small,
very  visible even for the lower blocking levels where the smearing does not
extend around the whole lattice. In comparison our vacuum expectation values are
very much smaller, except possibly for $SO(3)$. That is to say, the indications are
that any string breaking in $SO(N\geq 5)$ theories is much weaker than in $QCD$.

\section{String tensions}
\label{section_stringtension}

\subsection{lattice results}
\label{subsection_strings_results}

We list in Tables~\ref{table_string:so3}-\ref{table_string:so16} the result of our calculations
of the ground state flux tube energy for various values of $\beta$ and for our various $SO(N)$
groups. The length $l$ of the flux tube is equal to the size, $L_s$,  of the spatial
direction which our winding operator encircles.

For most values of $N$ we use the
same lattices for our glueball and for our string tension calculations. However for 
$SO(3)$, $SO(4)$ and $SO(6)$ we use smaller spatial volumes for the string tension calculations
than for the glueball ones. One reason is that we need to use larger physical volumes for
glueballs at the smallest values of $N$ in order to minimise the systematic errors 
discussed in Section~\ref{subsubsection_finiteV}. In addition, the string tensions $\sigma$
turn out to be larger at small $N$ when expressed in units of the mass gap, so for the ground
state flux tube energy not to become so large that its determination becomes
susceptible to the systematic errors discussed in Section~\ref{subsubsection_Eplateaux},
we need to choose values of $l$ that are
small enough for the correlator not to have disappeared into the statistical noise
before the desired effective energy plateau can be identified. This is particularly
important for $SO(3)$ and $SO(4)$ because the overlap of the ground state
onto our operator basis is quite mediocre for these two small values of $N$, which means
that the energy plateaux will only appear at larger $t$. For these reasons we perform
our $SO(3)$ and $SO(4)$ string tension calculations on separate series of smaller
lattices. For $SO(3)$ we choose a lattice size that corresponds to $l=46$ at
$\beta=9.0$. As we see from our finite volume study in Fig.~\ref{fig_Ek1_so3_b9.0}
this value of $l$ is large enough to be well fit by the Nambu-Goto expression in
eqn(\ref{eqn_NGgs}). That is to say, we can extract a string tension
using that formula for this value of $l$. Our glueball calculations, on the other
hand, are performed on spatial lattice sizes that corresponds to $l=82$ at $\beta=9.0$,
and, as should be apparent from Fig.~\ref{fig_Eeffk1_so3_b9.0}, extracting a flux
tube energy on such a large lattice would be beset with potential systematic errors.
Similar comments apply to our $SO(4)$ calculations.

Given our caveats about the low $N$ calculations, it is useful to display the effective
energies from which we estimate the `plateau' energies listed in Table~\ref{table_string:so3}
and Table~\ref{table_string:so4}. This we do in Fig.~\ref{fig_Eeffk1_vsm_so3} and
Fig.~\ref{fig_Eeffk1_sm_so4}. It appears plausible that these plateaux have been correctly
identified for at least the four smallest lattice spacings, and these are the ones that
will dominate our continuum extrapolations. When we move to $SO(5)$ we find that the overlaps
have become much better, as we can see from the effective energies plotted in
Fig.~\ref{fig_Eeffk1_so5}. Here the spatial volumes used are still quite large, and
therefore so are the flux tube energies, but since the overlaps are now very good, the
identification of an energy plateau is convincing. (Except for the coarsest
$a(\beta)$, which is largely irrelevant for the continuum limit.) For $SO(6)$ the
calculations on our smaller volumes are at least as convincing. For $N\geq 7$ we use
smaller volumes for the glueballs and since the overlaps are now very good (as we can
infer from Fig.~\ref{fig_Eeffk1_so6_b46} and Fig.~\ref{fig_Eeffk1_so8_b86})
the energy estimates become quite unambiguous. 

Our finite $l$ studies for various $N$ make us confident that the Nambu-Goto expression
in eqn(\ref{eqn_NGgs}) encodes all the finite-$l$ corrections that are significant,
at our value of $l$, given our errors. As remarked earlier, this is to be expected
since we know that all correction terms to $\sigma l$ up to $O(1/l^5)$ are universal
\cite{OA_string,SD_string}
for any effective string action describing flux tubes and these terms
are precisely what one obtains when expanding the  Nambu-Goto expression
in powers of $1/\sigma l^2$. Of course when the flux tube is not stable it is not
clear that it should be described by an effective string action, and here
our numerical tests in $SO(3)$ and $SO(5)$ are useful.
We therefore use the Nambu-Goto formula to extract the
string tension values, $a^2\sigma$, from the flux tube masses, $am_p$, and we list
these in Tables~\ref{table_string:so3}-\ref{table_string:so16}.

\subsection{continuum limit and $N$-dependence}
\label{subsection_strings_contN}

What we ultimately want of course is not so much lattice values of $\sigma$, but the
continuum limit in some physical units. Since $g^2$ has dimensions of mass
in $D=2+1$, we can use that to set our units, i.e. we calculate
the continuum limit of the dimensionless ratio  $\surd\sigma/g^2N$.
To do so we could use the standard lattice coupling defined through
$\beta=2N/ag^2$, but we instead choose to use the mean-field improved
coupling
\cite{GP_MFI},
$g^2_I$, defined by $\beta_I= \beta \langle u_p \rangle = 2N/ag_I^2$,
with  $\langle u_p \rangle = \langle \mathrm{Tr} U_p \rangle /N$
the average plaquette. (In the case of
$SU(N)$ gauge theories this is found to provide a slightly more rapid approach
to the continuum limit.) Using this we extrapolate
to the continuum limit using
\begin{equation}
  \frac{\beta_I}{2N^2} {a\surd\sigma}
  \equiv
  \left.\frac{\surd\sigma}{g_I^2N}\right|_{\beta_I}
  \stackrel{\beta_I\to\infty}{=}
  \left.\frac{\surd\sigma}{g^2N}\right|_{\infty}
    + \frac{c}{\beta_I} + O(\frac{1}{\beta_I^2})
\label{eqn_kg}
\end{equation}
We show our resulting continuum extrapolations in Fig.~\ref{fig_kg_soNd3}
where we plot the values of ${\surd\sigma}/{g_I^2N}$ against $ag_I^2N$.
In general we obtain good
fits with just the linear $O(1/\beta_I)$ correction in eqn(\ref{eqn_kg}).
We list the resulting continuuum values
of  ${\surd\sigma}/{g^2N}$ in Table~\ref{table_string:continuum}.

We now plot these values against $1/N$ in Fig.~\ref{fig_kgNcont_soNd3}.
We expect the $N\to\infty$ limit to be some finite non-zero number
and we expect the finite-$N$ corrections to begin at $O(1/N)$.
We find that a single $O(1/N)$ correction will not do for all $N\geq 3$, and that one
needs at least a further  $O(1/N^2)$ term to obtain a barely acceptable fit.
%
%
%
Without $SO(3)$ we obtain a better fit, and can even get by with a single correction term.
In summary, one obtains
\begin{equation}
  \frac{\surd\sigma}{g^2N} 
  =\begin{cases}
  0.09821(57) - \frac{0.142(6)}{N} - \frac{0.048(31)}{N^2} &
  \qquad N\geq 4 \quad \frac{\chi^2}{n_{dof}} \simeq 1.67  \\
  0.09897(25) - \frac{0.1542(19)}{N}  &
  \qquad N\geq 4 \quad \frac{\chi^2}{n_{dof}} \simeq 1.76 \\
  0.09749(43) - \frac{0.129(6)}{N} - \frac{0.100(17)}{N^2} &
  \qquad N\geq 3 \quad \frac{\chi^2}{n_{dof}} \simeq 2.32  \\
  \end{cases}.
\label{eqn_kgN_N}
\end{equation}
Given the fact that the $SO(3)$ string may have a significant decay width,
it seems safer to discount the last of these fits. The two fits for $N\geq 4$
almost overlap within errors and one can regard the difference as providing
a measure of the systematic error in the choice of large-$N$ extrapolation.
Given that the much more accurate $SU(N)$ extrapolation (see below) requires
the inclusion of an $O(1/N^4)$ correction, we are inclined to take the
higher order fit to the $SO(N\geq 4)$ string tension as being the most realistic.

Given that we expect $SU(N)$ and $SO(N)$ to have a common planar limit,
it is interesting to compare our $SO(N)$ results to those obtained in $SU(N)$.
We therefore list in Table~\ref{table_string:continuum} some $SU(N)$ values of
${\surd\sigma}/{2g^2N}$ taken from 
\cite{AAMT_SUN}.
The extra factor of 2 encodes the fact that in identifying the planar limits of 
$SO(N)$ and $SU(N)$ one equates $g^2N$ in the former with  $2g^2N$ in the latter
\cite{CL_N}.
We plot these $SU(N)$ values in Fig.~\ref{fig_kgNcont_soNd3}. Since the leading
large-$N$ correction in $SU(N)$ is $O(1/N^2)$, the points near $N=\infty$ will
fall on a quadratic curve. To get a good fit one needs to include a subleading
correction, giving for our best fit
\begin{equation}
  \frac{\surd\sigma}{2g^2N}
  =
  0.09818(6) - \frac{0.0543(16)}{N^2}- \frac{0.0143(58)}{N^4}
  \qquad N\geq 2 \quad \frac{\chi^2}{n_{dof}} \simeq 1.25  \, :\quad SU(N)
\label{eqn_kgN_suN}
\end{equation}
which we plot in  Fig.~\ref{fig_kgNcont_soNd3}. We observe that the large $N$
limits of $SO(N)$ and $SU(N)$ do indeed coincide (within errors) despite the
very different values at finite $N$.

Another interesting question is whether the values of the string tension in
$SO(2N)$ and $SO(2N+1)$ gauge theories form a single `continuous' sequence
or two separate sequences that only coincide at $N=\infty$. The fact that
we can perform a single smooth extrapolation to $N=\infty$ using all our
values of ${\surd\sigma}/{2g^2N}$ confirms, within small errors, that there
is in fact a single sequence
as we see in Fig.~\ref{fig_kgNdiff_soNd3} where we plot the difference
between our calculated values of $\surd\sigma/g^2N$ and the fit in
eqn(\ref{eqn_kgN_N}), normalised by the former.
It appears that the fact that $SO(2N+1)$ gauge theories have only a trivial
centre does not affect the value of the confining string tension.

\subsection{Lie algebra equivalences}
\label{subsection_Lie_strings}

As we remarked earlier, it is interesting to see whether pairs of
$SO(N),SU(N^\prime)$ gauge theories that share the same Lie algebra
lead to the same physics. Here we look at $\surd\sigma/g^2N$, taking
advantage of the fact that in $D=2+1$, $g^2$ has dimensions of mass.

There are two subtleties: firstly the fundamental string tension in the
$SO(N)$ gauge theory will correspond to a higher representation
string tension in the  $SU(N^\prime)$ theory. Secondly the
relation between the $g^2$ will also be non-trivial. 
There is also a caveat. As discussed in Section~\ref{subsection_continuum},
the continuum extrapolation of $\surd\sigma/g^2$ begins with a correction
at $O(1/\beta) \sim O(a)$ rather than at $O(a^2)$ which means that
the systematic error is larger than where we consider ratios of physical
energies, and may well be significantly larger than our quoted statistical error.

We begin by comparing the adjoint string tension in $SU(2)$ with the fundamental
string tension in $SO(3)$. This is the case where we encounter
the largest uncertainties since in both cases the overlaps of our operators are
mediocre and this affects how reliable is the identification of the plateaux
in the effective energies. In  $SU(2)$ the 
relevant coupling is the adjoint coupling, $g_a^2$, which is related to
the fundamental couping $g_f^2$ in eqn(\ref{eqn_gfgaSU2}). So taking
the $SO(3)$ and $SU(2)$ values of $\surd\sigma_f/g_f^2N$ from
Table~\ref{table_string:continuum}, using
$\surd\sigma_f/g_f^2|_{so3}=\surd\sigma_{adj}/g_a^2|_{su2}$,
and imposing $g_a^2|_{su2}= 4 g_f^2|_{su2}$ from eqn(\ref{eqn_gfgaSU2}),
we obtain the prediction for $SU(2)$ 
\begin{equation}
  \left.\frac{\surd\sigma_f}{g^2}\right|_{so3} = 0.1281(20)
  \quad \Longrightarrow \quad
    \left.\frac{\sigma_{adj}}{\sigma_f}\right|_{su2} = 2.34(8).
\label{eqn_k_su2so3}
\end{equation}
This can be compared to the value one obtains directly in $SU(2)$,
$\sigma_{adj}/\sigma_f|_{su2} = 2.24(3)$.
(See Table B2 of
\cite{AAMT_SUN}.)
We see that the two values are in reasonable agreement given the statistical
errors -- perhaps surprisingly so given the systematic errors discussed above.

We turn now to  $SO(4)$ and $SU(2)\times SU(2)$. Taking the $SO(4)$ value
of  $\surd\sigma_f/g_f^2$ from Table~\ref{table_string:continuum}, and
imposing the relations eqns(\ref{eqn_Kso4},\ref{eqn_gso4}) we obtain
a predicted value for $SU(2)$
\begin{equation}
  \left.\frac{\surd\sigma_{f}}{g^2}\right|_{so4} = 0.2408(22)
  \Longrightarrow
  \left.\frac{\surd\sigma_{f}}{g^2}\right|_{su2} = 0.3405(31)
\label{eqn_k_su2so4}
\end{equation}
which is to be compared to the directly calculated $SU(2)$ value of $\simeq 0.3349(3)$
shown in  Table~\ref{table_string:continuum}. The two values are numerically
very close, and while there is a  $\sim 1.8$ standard deviation discrepancy, this
includes only the statistical errors and given our earlier comments about
the possible systematic errors, one need not regard this tension as being
significant.

Finally we compare $SO(6)$ with $SU(4)$. As discussed in
Section~\ref{subsubsection_su4andso6} the fundamental $SO(6)$ string tension
corresponds to the $k=2a$ $SU(4)$ string tension and the $SO(6)$ coupling
is related to the  $SU(4)$ coupling as in eqn(\ref{eqn_gso6}). Using the
latter, together with the $SO(6)$ value in Table~\ref{table_string:continuum},
we obtain a prediction 
\begin{equation}
  \left.\frac{\surd\sigma_{f}}{g^2}\right|_{so6} = 0.4403(9)
  \Longrightarrow
  \left.\frac{\surd\sigma_{k=2a}}{g^2}\right|_{su4} = 0.8806(18)
\label{eqn_k_su4so6}
\end{equation}
which is entirely consistent with the directly calculated $SU(4)$ value
$\surd\sigma_{k=2a}/{g^2}=0.8833(11)$. (See tables B1,B2 in
\cite{AAMT_SUN}.)

We conclude that within our quite small uncertainties, there does indeed appear
to be consistency between the values of  $\surd\sigma/g^2$  within pairs of
$SO(N)$ and $SU(N^\prime)$ theories that share a common Lie algebra.

\section{Glueball spectrum}
\label{section_glueballs}

Most of our calculations in this paper concern the glueball spectrum.
The glueballs we focus upon are the ground state glueballs with
$J^P=0^\pm, 2^\pm, 1^\pm$, the first four $0^+$ excited states and the
first $2^\pm$ excited states.
We have chosen the lattice sizes with a view to avoiding significant finite volume 
corrections to these states, as discussed in Section~\ref{subsubsection_finiteV}.

\subsection{SO(N) glueball masses}
\label{subsection_glueballs}

We list in Tables~\ref{table_glueball:so3}-\ref{table_glueball:so16} our
results for a number of the lightest glueballs in $SO(N)$ gauge theories
for $N=3,4,5,6,7,8,12,16$. This range of $N$ is designed to allow us to make
plausible large-$N$ extrapolations and also to compare the $N$ dependence of
$SO(2N)$ and $SO(2N+1)$ gauge theories.

We shall mostly extrapolate ratios of physical masses to the continuum limit since
the corrections will be $O(a^2)$ and will therefore converge much faster
to $a=0$ than if we were to express our masses in units of $g^2N$, where the
correction is $O(a)$. Since for most values of $N$ the string tension is
the most accurately calculated physical quantity (with a caveat for $SO(3)$)
we shall usually extrapolate the values of $aM_G/a\surd\sigma$, for our various
glueballs $G$.

We use a variant of eqn(\ref{eqn_mimjcont}) to extrapolate to the continuum
limit
\begin{equation}
\frac{aM_G(a)}{a\surd\sigma(a)}=\frac{M_G(a)}{\surd\sigma(a)} 
=\frac{M_G(0)}{\surd\sigma(0)} + c a^2 \sigma(a) 
\label{eqn_MKcont}
\end{equation}
suppressing obvious indices. We truncate to just the leading correction
since that usually suffices in practice. The fits are to all the values
listed in Tables~\ref{table_glueball:so3}-\ref{table_glueball:so16} except
in a few cases where the value at the largest $a$ is excluded from the fit.
The resulting continuum limits, and the corresponding values of  $\chi^2/n_{dof}$
for the extrapolations, are listed in
Tables~\ref{table_glueball:N3-6continuum}, \ref{table_glueball:N7-16continuum}.
Most values of  $\chi^2/n_{dof}$ are modest, but a few are large. However 
given that typically $n_{dof}=4$ this is not fatal. We illustrate all this
in Fig.~\ref{fig__M0p2pK_contd3}  for the lightest $0^+$ and $2^+$ glueballs
in $SO(3)$, $SO(6)$ and $SO(12)$.

We observe that our results are reassuringly consistent with the parity
doubling expected in $D=2+1$, i.e. $2^+$ is degenerate with $2^-$ (also
$2^{+\star}$ with $2^{-\star}$) and $1^+$ with $1^-$. A striking contrast
is provided by the  $J=0$ sector where  no parity doubling is expected
or observed.

It will also be useful for our analysis below to calculate the mass gap
in units of the 't Hooft coupling $g^2N$. 
To calculate the continuum value of $M_{0^+}/g^2N$ we perform an extrapolation
\begin{equation}
  \frac{\beta_I}{2N^2} {aM_{0^+}}
  \equiv
  \left.\frac{M_{0^+}}{g_I^2N}\right|_{\beta_I}
  \stackrel{\beta_I\to\infty}{=}
  \left.\frac{M_{0^+}}{g^2N}\right|_{\infty} + \frac{c}{\beta_I} 
\label{eqn_MgN}
\end{equation}
using the mean-field improved lattice coupling $\beta_I=\beta\langle u_p \rangle$
that we introduced and used in Section~\ref{subsection_strings_contN}. (We use
the values of $aM_{0^+}$ in Tables~\ref{table_glueball:so3}-\ref{table_glueball:so16}
and the plaquette values in Tables~\ref{table_string:so3}-\ref{table_string:so16}.)
It turns out that we can obtain reasonable fits to all our lattice data using just
the leading $O(1/\beta_I)$ correction, and the results of these fits are listed in
Table~\ref{table_m0pggn_N}.

\subsection{large N: odd and even N}
\label{subsection_largeN_glueballs}

We now perform an extrapolation of our continuum results to $N=\infty$ where
we can compare to the known $SU(N\to\infty)$ values
\cite{AAMT_SUN}.
We note that the large-$N$ expansion can be performed at a non-zero lattice
spacing
\cite{tHooft_Nlat}
so we could in principle perform the comparison between  $SO(N\to\infty)$ and
$SU(N\to\infty)$ at a fixed $a$. There are however some extra complications
(and hence errors) in fixing $a$ across these theories, and we do not choose
to make use of this possibility here.

We have already discussed the $N$-dependence of the string tension (expressed
in units of the 't Hooft coupling) in Section~\ref{subsection_strings_contN}. We saw
that the large-$N$ limit of $\surd\sigma/g^2N$ is, within errors, the same as that of
$SU(N)$, up to a predicted factor of two. We also saw that to have a reasonable
fit we had to add an $O(1/N^2)$ correction to the leading $O(1/N)$ correction.
One can treat the mass gap $M_{0^+}/g^2N$ in the same way. Using the continuum
values listed in Table~\ref{table_m0pggn_N} we find that
\begin{equation}
  \frac{M_{0^+}}{g^2N}
  =
  0.4017(14) - \frac{0.802(7)}{N}
  \qquad  N\geq 3 ,\quad \chi^2/n_{dof}=0.81
\label{eqn_MgN_N}
\end{equation}
provides a good fit to all our values of $N$. It is surprising that in
this case the leading correction suffices, given the relatively large
coefficient of the correction term. No doubt our substantial statistical
errors obscure the need for the inclusion of a further $\propto 1/N^2$
correction term. If we nonetheless perform a fit with such an extra term we
obtain
\begin{equation}
  \frac{M_{0^+}}{g^2N}
  =
  0.4079(35) - \frac{0.871(36)}{N} + \frac{0.165(82)}{N^2}
  \qquad  N\geq 3 ,\quad \chi^2/n_{dof}=0.36 ,
\label{eqn_MgN_NN}
\end{equation}
which provides us with a measure of the systematic error associated with
truncating the large-$N$ expansion in eqn(\ref{eqn_MgN_N}). The corresponding
$SU(N\to\infty)$ value, as given in Table B1 of
\cite{AAMT_SUN},
is ${M_{0^+}}/{2g^2N}=0.4051(6)$, which is in reasonable agreement with the 
above $SO(N)$ values. Note that we include the factor of 2 in the matching of the
couplings as prescribed by the large-$N$ diagrammatic analysis
\cite{CL_N}.

Repeating the exercise in units of the string tension, we obtain 
\begin{equation}
  \frac{M_{0^+}}{\surd\sigma}
  =
  4.179(16) - \frac{3.17(11)}{N}
  \qquad N\geq 3 ,\quad \chi^2/n_{dof} \simeq 1.30.
\label{eqn_M0pK_N}
\end{equation}
From Table B10 of
\cite{AAMT_SUN}
we obtain the value $M_{0^+}/\surd\sigma\simeq 4.116(6)$ for $SU(\infty)$.
This is very close to our above $SO(\infty)$ value. Although the difference
is at the slightly uncomfortable level of $\sim 3.7\sigma$, the error is
statistical, so we can regard the values as reasonably compatible. For example,
if we perform an extrapolation to $N=\infty$ with an extra $\propto 1/N^2$
correction term, and if we exclude the $SO(3)$ value (which relies on a
significantly unstable string tension) we obtain $M_{0^+}/\surd\sigma=4.106(39)$
with a much better fit, $\chi^2/n_{dof} \simeq 0.58$.
This is entirely consistent with the $SU(\infty)$ value. Given the
substantial coefficient of the $\propto 1/N$ correction term in
eqn(\ref{eqn_M0pK_N}) it is very likely that a significant $\propto 1/N^2$
is present even if our relatively large statistical errors do not force us
to include it in the extrapolation.

We now repeat such extrapolations for various other states and present the results 
in Table~\ref{table_MK_largeN}. In many cases a reasonable leading order fit in $1/N$ is
acceptable but, once again, given that the coefficient of the $O(1/N)$ term
is usually substantial, one should worry that the only reason we can get by with
a leading order fit for some of the more massive states is that the errors are
so large. So we present in all cases the results of fits with just a leading correction
and, separately, with an added non-leading correction. In each case we indicate the
range of $N$ fitted as well as the $\chi^2$ per degree of freedom. (This information
allows the reader to estimate the $p$-value in each case.) We see that in many cases
the $N=\infty$ values differ between the two fits by more than the statistical
errors. We may, again, regard the difference between the two fits as providing
an estimate of the systematic error
associated with the large-$N$ extrapolation. One also sees that in all cases the
lower-order fit has a larger extrapolated value than the higher order fit and has
smaller statistical errors.

We also display, in the same Table, the results for $SU(N\to\infty)$ taken from
Tables B10-B12 in
\cite{AAMT_SUN}.
We see reasonable consistency between the $SO(\infty)$ and $SU(\infty)$ mass ratios
in almost all cases, especially if one uses the higher order fits, although in most
cases the errors are substantial and this obviously
reduces the significance one can read into the agreement.
The only comparison that works badly is the one for the first excited $0^+$.
Here the discrepancy is a worrying $\sim 7.4$ standard deviations if one uses the leading
order fit, and is only reduced to $\sim 4.7$ standard deviations if one takes the error
from the higher order extrapolation. The first excited $2^-$ comparison is also
quite poor, but here we grade the reliability of the mass estimate as quite poor,
and this is supported by the apparent gap between the mass of the $2^{-\star}$ and
that of the $2^{+\star}$ with which it should be degenerate.

As we remarked above, the variation with $N$ of the typical glueball mass
when expressed in units of the
string tension is quite large compared to what one sees in $SU(N)$ gauge theories. As an
example we show in Fig.~\ref{fig__M0pgsex1K_soNsuN} the values of $M/\surd\sigma$ for
the lightest two $0^+$ glueballs, in both $SO(N)$ and $SU(N)$. One can ask whether
one can get a weaker $N$-dependence by using different units in which to express
the mass. As an example we show in Fig.~\ref{fig_m2pmu_soNd3} the dependence
one obtains using the 't Hooft coupling, $g^2N$, the string tension, $\surd\sigma$,
and the mass gap, $M_{0^+}$ for the lightest $2^+$ glueball. We see that the
weakest $N$-dependence occurs when using the mass gap as the scale, and this
is in fact typical. This motivates extrapolating the ratio $M_G/M_{0^+}$ to
$N=\infty$, and we show in Table~\ref{table_glueball_largeN} what one obtains
in that case. Here we only show what one obtains with just a leading $O(1/N)$ correction,
$M_G/M_{0^+} = c_0 + c_1/N$, and we see that $c_1/c_0$, the normalised coefficient of
the correction, is invariably very small. We find reasonable agreement between the
$SO(N\to\infty)$ and $SU(N\to\infty)$ values, except for the  $0^{+\star}$ and
the  $2^{-\star}$, although now the $0^{+\star}$ difference is $\sim 4\sigma$.

We have obtained acceptable fits without distinguishing between odd and even values
of $N$. This already indicates that these values are compatible with lying  on
a single smooth interpolating curve. Our most accurate mass is the mass gap,
and we show in Fig.~\ref{fig_M0pgsKdiff_soN} how the odd and even values differ
from our interpolation in eqn(\ref{eqn_M0pK_N}). As we can see, just as for
the string tension, there is no indication that the odd and even values of
$N$ differ in some systematic fashion.

\subsection{Lie algebra equivalences: mass gap and couplings}
\label{subsection_Lie_glue_g}

In Section~\ref{subsection_Lie_strings} we compared the ratio $\surd\sigma/g^2$
within the three pairs of $SO(N)$ and $SU(N^\prime)$ gauge theories that share
a common Lie algebra. We assumed the relationship between the couplings
and the matching between flux tube representations given in
Section~\ref{section_sunsonconstraints} and found that the resulting values
of the string tension were consistent between all three pairs of gauge theories.
We turn now to the mass gap, the lightest scalar glueball. We will first calculate
the ratios $M_{0^+}/\surd\sigma$ in $SO(3)$, $SO(4)$ and $SO(6)$ and compare them
to the values that one obtains in $SU(2)$, $SU(2)$ and $SU(4)$ respectively.
Since we have seen that the string tensions are in agreement, this will
provide us with a test of whether the mass gap agrees or not. We shall see that
they do in fact agree, and we then move on to a comparison of the values of
$M_{0^+}/g^2$ which will provide us with a test of the relationship between
the couplings.

We begin with $SO(6)$ and $SU(4)$. From the $SO(6)$ value of $M_{0^+}/\surd\sigma$
listed in Table~\ref{table_glueball:N3-6continuum} we obtain the predicted value in
$SU(4)$ as follows:
\begin{equation}
  \left.\frac{M_{0^+}}{\surd\sigma_f} \right|_{so6} = 3.656(13)
  =  \left.\frac{M_{0^+}}{\surd\sigma_{2A}} \right|_{su4}
  \Rightarrow
   \left.\frac{M_{0^+}}{\surd\sigma_{f}} \right|_{su4} = 4.259(16),
\label{eqn_MK_su4so6}
\end{equation}
where in the last step we use $\left.\surd\sigma_{2A}/\surd\sigma_{f}\right|_{su4} = 1.1649(11)$
\cite{AAMT_SUN}.
This is entirely consistent with the value ${M_{0^+}}{\surd\sigma_{f}} = 4.242(9)$ that
one obtains by direct calculation in $SU(4)$
\cite{AAMT_SUN}.

Similarly we obtain a prediction for $SU(2)$ from our calculation in $SO(4)$:
\begin{equation}
  \left.\frac{M_{0^+}}{\surd\sigma_f} \right|_{so4} = 3.343(23)
    \Rightarrow
   \left.\frac{M_{0^+}}{\surd\sigma_{f}} \right|_{su2} = 4.728(33),
\label{eqn_MK_su2so4}
\end{equation}
where we use $\left.\sigma_f\right|_{so4} = 2\left.\sigma_{f}\right|_{su2}$.
This is to be compared to the value  ${M_{0^+}}{\surd\sigma_{f}} = 4.737(6)$
that one obtains by direct calculation in $SU(2)$
\cite{AAMT_SUN}.

Finally we obtain another prediction for $SU(2)$, this time from our calculation
in $SO(3)$:
\begin{equation}
  \left.\frac{M_{0^+}}{\surd\sigma_f} \right|_{so3} = 3.132(34)
  =  \left.\frac{M_{0^+}}{\surd\sigma_{adj}} \right|_{su2}
  \Rightarrow
   \left.\frac{M_{0^+}}{\surd\sigma_{f}} \right|_{su2} = 4.689(60),
\label{eqn_MK_su2so3}
\end{equation}
where in the last step we use $\sigma_{adj}/\sigma_f|_{su2} = 2.24(3)$.
(See Table B2 of
\cite{AAMT_SUN}.)
Again this is consistent with the value  ${M_{0^+}}{\surd\sigma_{f}} = 4.737(6)$
that one obtains by direct calculation in $SU(2)$
\cite{AAMT_SUN}.

Having established that the mass gap is the same (within errors) in the above
pairs of theories, we can now ask whether the values we obtain for
${M_{0^+}}/{g^2}$ predict the expected relations between the couplings.

We begin with $SO(3)$ and $SU(2)$. From Table~\ref{table_m0pggn_N} we
obtain ${M_{0^+}}/{g^2}|_{so3} = 0.4071(42) 0.4080(41)$ whereas
for $SU(2)$ one finds $M_{0^+}/g^2|_{su2} = 1.5860(22)$. (See Table B1 of
\cite{AAMT_SUN}.)
Assuming the mass gap is identical in $SU(2)$ and $SO(3)$, we obtain a
relationship between the couplings
\begin{equation}
  \frac{g^2_{f,so3}}{g^2_{f,su2}} = \frac{g^2_{a,su2}}{g^2_{f,su2}} = 3.896(41)
\label{eqn_g_su2so3}
\end{equation}
which should be compared to the expected value of $4$ in eqn(\ref{eqn_gfgaSU2}).
The agreement is reasonably good.

We repeat the comparison for $SO(4)$ and $SU(2)$. Using the value in 
Table~\ref{table_m0pggn_N} we obtain ${M_{0^+}}/{g^2}|_{so4} = 0.8032(72)$.
Then, assuming the mass gap is identical in $SU(2)\times SU(2)$ and $SO(4)$,
we obtain a relationship between the couplings
\begin{equation}
  \frac{g^2_{f,so4}}{g^2_{f,su2}} = 1.975(18)
\label{eqn_g_su2so4}
\end{equation}
which agrees reasonably well with the expected value of 2 given in eqn(\ref{eqn_gso4}).

Finally we repeat the exercise for $SO(6)$ and $SU(4)$. Again using the values in
Table~\ref{table_m0pggn_N} we obtain ${M_{0^+}}/{g^2}|_{so6} = 1.6074(84)$.
Averaging the large and medium lattice values for $SU(4)$ in Table B1 of
\cite{AAMT_SUN},
we find $M_{0^+}/g^2|_{su4} = 3.2212(68)$. We thus obtain a relationship
between the couplings
\begin{equation}
  \frac{g^2_{f,so6}}{g^2_{f,su4}} = 2.004(11)
\label{eqn_g_su4so6}
\end{equation}
which agrees well with the expected value of 2 derived in eqn(\ref{eqn_gso6}).

The above results appear to show, within small errors, that the ratios
of the mass gap, string tension and couplings are not influenced
by the differing global properties of these pairs of groups.

As an aside, we remark that it can be interesting to perform such
comparisons not only in the continuum limit but also at finite lattice spacing,
because we expect the global properties of the group to make some
difference at larger $a$ (smaller $\beta$), since the field fluctuations
will be larger there. We limit our comparison to comparing the values
of $M_{0^+}/\surd\sigma_f$ in $SO(6)$ with those of  $M_{0^+}/\surd\sigma_{k=2A}$
in $SU(4)$. We find these ratios to be independent of $\beta$ in both cases,
and equal to each other, within errors. Since our coarsest $a(\beta)$ in
$SO(6)$ is close to the strong-to-weak coupling bulk transition, this
suggests that the lattice physics is insensitive to the global properties
of the group as soon as we are on the weak coupling side of the bulk transition.

\subsection{Lie algebra equivalences: mass ratios}
\label{subsection_Lie_glue_M}

As discussed above, it is interesting to ask whether the single particle mass spectra
of the pairs of $SO(N)$ and $SU(N^\prime)$ theories that share the same Lie algebra
are in fact the same. We have just seen that this appears to be the case for the
mass gap. So if the differing global properties of the groups are unimportant
we  would expect the value of $M_G/M_{0^+}$ to be the same for the
$SO(3)$, $SO(4)$ and $SU(2)$ gauge theories and, separately, for the $SO(6)$ and
$SU(4)$ gauge theories. 
To address this question we list in Table~\ref{table_glueball_SONvsSUM}
the continuum ratio of the mass of each of our glueballs $M_G$ to that of the
lightest scalar glueball $M_{0^+}$, for the groups $SO(3)$,  $SO(4)$ and  $SO(6)$,
together with the corresponding mass ratios for $SU(2)$ and $SU(4)$ as obtained from 
\cite{AAMT_SUN}.

\subsubsection{$SO(3)$ and $SU(2)$}
\label{subsubsection_SO3SU2_M}

We begin by comparing $SO(3)$ with $SU(2)$. The lightest and hence best determined
mass ratios are those of the $0^{+\star}$ and of the $2^\pm$ glueballs and these we
see from Table~\ref{table_glueball_SONvsSUM} are 
consistent between $SO(3)$ and $SU(2)$. The only other state graded as reliable,
i.e. $\alpha$ or $\beta$, is the  $0^{+\star\star}$ which also quite consistent,
albeit within its large errors. The remaining masses do appear to differ,
but mostly at no more than the $\sim 2\sigma$ level. 
Since we know that the overlaps of the glueballs onto our operators are much poorer
in $SO(3)$ than in $SU(2)$, one possibility is that the effective energy plateaux
are not being accurately identified in $SO(3)$ for these heavier states.
This is indeed indicated by our reliability grading in Table~\ref{table_glueball_SONvsSUM}.
And the fact that where there are significant differences, it is the the $SO(3)$
masses in  Table~\ref{table_glueball_SONvsSUM} that are in every case heavier
than the corresponding $SU(2)$ ones is consistent with this interpretation.

Since the resolution of our correlators and the associated effective masses is greatest
when the lattice spacing is smallest, and since in any case it is this calculation
that is closest to the continuum limit, it is useful to look at the effective masses
of our $SO(3)$ calculation at $\beta=11$ and compare it to an $SU(2)$ calculation
at a similar lattice spacing (in units of the mass gap). This we have already done
in Figs~\ref{fig_Meff_0p_so3su2} and \ref{fig_Meff_Jm_so3su2}. From 
Fig.~\ref{fig_Meff_0p_so3su2} we see that while the overlap of the lightest scalar
on our basis is almost as good for $SO(3)$ as for $SU(2)$, the $SO(3)$ overlaps
rapidly worsen for the heavier scalars. Nonetheless it seems clear from
the figure that the ratios $M_{0^{+\star}}/M_{0^+}$,  $M_{0^{+\star\star}}/M_{0^+}$ are
quite consistent between $SO(3)$ and $SU(2)$. It is also
clear from Fig.~\ref{fig_Meff_0p_so3su2} that we cannot be confident that
there is any real mismatch between  $SO(3)$ and $SU(2)$ for the ratios
$M_{0^{+\star\star\star}}/M_{0^+}$ and $M_{0^{+\star\star\star\star}}/M_{0^+}$.
Similar comments apply to the ratios $M_{0^{-}}/M_{0^+}$
and $M_{1^{-}}/M_{0^+}$ shown in Fig.~\ref{fig_Meff_Jm_so3su2}. Only the $SO(3)$
$M_{2^{-\star}}/M_{0^+}$ appears to show a sign of a plateau around $t=4a,5a$
that is quite different from the $SU(2)$ one, but even in that case the effective
masses at large $t$ show some indication of convergence. We also show the
effective masses for the $M_{2^{+\star}}/M_{0^+}$, which should be degenerate
with $M_{2^{-\star}}/M_{0^+}$ (in the large volume continuum limit),
but here we do not see much sign of this plateau, suggesting that it may well be a
fluctuation in the case of the $2^{-\star}$. So, in summary, what this
brief analysis of our `best' $SO(3)$ calculation shows is that  for all but
the lightest states the systematic errors that arise in our identification
of the effective mass plateau are quite possibly large compared to the quoted statistical
error, something which our grading of the states is intended to capture.
The culprit is the poor overlap onto our basis in $SO(N)$ gauge theories
when $N$ is small, the effect of which is greater at larger value of $a(\beta)$,
and this can of course affect the reliability of the continuum extrapolation.

\subsubsection{$SO(4)$ and $SU(2)$}
\label{subsubsection_SO4SU2_M}

We now turn to a comparison of the $SO(4)$ and $SU(2)$ spectra 
in Table~\ref{table_glueball_SONvsSUM}. Here the mass ratios that match
within, say, a relatively innocuous $\sim 2\sigma$ are the $M_{0^{+\star}}/M_{0^+}$ 
and the $M_{2^{\pm}}/M_{0^+}$. Since these are precisely the states that our grading
judges to be reasonably reliable, this is reassuring. However the mismatch in
the other cases is typically much larger than for $SO(3)$ which is, at first sight,
surprising since the $SO(4)$ overlaps, while not very good, are significantly better
than those in $SO(3)$ -- which should imply a more reliable identification of the
effective energy plateaux. However we also note, comparing Table~\ref{table_glueball:so4}
to Table~\ref{table_glueball:so3}, that the $SO(3)$ calculation is at much
smaller lattice spacings than the $SO(4)$ calculation, when expressed in units
of the mass gap. (We recall that the $SO(3)$ choice of lattice spacings was forced
upon us by the the awkward location of the `bulk' transition.) Since locating the
energy plateau is less reliable at larger $a(\beta)$, this will serve
to counteract the effect of better overlaps. This is well exemplified
in Fig.~\ref{fig_Eeffk1_vsm_so3} and Fig.~\ref{fig_Eeffk1_sm_so4} where
we plotted the effective energies for the flux tube energy in our
$SO(3)$ and $SO(4)$ calculations respectively. For a given $a(\beta)$,
i.e. a given $aE_{eff}(t_0)$,  the $SO(4)$ plateau is indeed better identified,
but the calculations at the smallest values of $a$ are indeed better
for $SO(3)$ than $SO(4)$. This effect is even more marked for the glueball
masses, since the $SO(3)$ string tension is larger than the $SO(4)$ one, and in
comparing  Fig.~\ref{fig_Eeffk1_vsm_so3} and Fig.~\ref{fig_Eeffk1_sm_so4}
we are effectively comparing values of $a(\beta)$ expressed in units of the
string tension.

Given these uncertainties, it is interesting to compare the $SO(4)$ and $SU(2)$ 
mass calculations for a specific and similar lattice spacing. For this we return
to Table~\ref{table_spectrum_opAopB} where we list $SU(2)$ masses taken from
\cite{AAMT_SUN}
in a calculation with a mass gap similar to the $SO(4)$ one in the table.
The listed $SU(2)$ masses have been rescaled by the ratio of $SU(2)$ and $SO(4)$
mass gaps, so the apparent exact agreement for the mass gap is not significant.
In addition to the `Ops B' calculation, which is used in our continuum
extrapolations, we list masses obtained with much higher statistics and with
an operator basis that is the same as used in the $SU(2)$ calculation. The
higher statistics should allow a somewhat better identification of the
energy plateaux and so the `Ops A' calculation should better represent the
true $SO(4)$ spectrum at this value of $a(\beta)$. Comparing the `Ops A'
masses to the $SU(2)$ ones we see much less disagreement than for
the continuum extrapolations listed in Table~\ref{table_glueball_SONvsSUM}.
The only discrepancy (slightly) greater than $\sim 3\sigma$ is for the
$2^{-\star}$. This comparison suggests that the difference between the
continuum extrapolations in Table~\ref{table_glueball_SONvsSUM}
of $SO(4)$ and $SU(2)$ may indeed be driven, at least
in large part, by a poor identification of the effective energy plateaux,
particularly at the larger lattice spacings.

Finally, one needs to recall that the Lie algebra of $SO(4)$ is identical
to that of $SU(2)\times SU(2)$ and not to that of $SU(2)$. Thus we would
expect the spectrum of $SO(4)$ to contain states composed of two non-interacting
glueballs, one from from each of the $SU(2)$ colour groups. Of course $SU(2)$
also contains two glueball states, but their contribution is likely
to be somewhat suppressed by the usual `large-$N$' suppression of
single-trace/double-trace matrix elements. Thus we might expect the spectrum
of $SO(4)$ to possess extra states, compared to $SU(2)$, once we are
looking at masses $\geq 2m_{0^+}$. (Up to some shift due to interactions induced
by lattice spacing corrections.) We note that almost all the
states in Table~\ref{table_glueball_SONvsSUM} where we have disagreement
do indeed lie above this bound.

\subsubsection{$SO(6)$ and $SU(4)$}
\label{subsubsection_SO6SU4_M}

For $SO(6)$ the overlaps are reasonably good and the calculation extends
to a smaller value of $a(\beta)$ than for $SO(4)$ (in units of the mass gap).
Hence the ambiguities in
identifying an effective energy plateau should, in principle, be much less. And
indeed the level of agreement between the  $SO(6)$ and  $SU(4)$ mass ratios
listed in Table~\ref{table_glueball_SONvsSUM} is much better than that between
$SO(4)$ and $SU(2)$. In fact the only states for which the discrepancy
is greater than $\sim 2\sigma$ are the $0^{+\star\star}$ and the $0^{+\star\star\star}$.
These are states which, by our grading, should not be regarded as being
very reliable.

The liklihood is that there is in fact a problem with our identification
of the effective energy plateaux for these heavier states, and that this
will be worse at the larger lattice spacings, where we have a coarser
resolution of the correlators. The resulting systematic errors could then be
magnified by the extrapolations involved in taking the continuum limit. This
suggests that it might be useful to perform a comparison of the mass spectra
at our smallest values of $a(\beta)$, rather than taking a continuum
limit. We now do this, plotting the effective energies obtained from various
glueball correlators in $SO(6)$ on a $62^270$ lattice at $\beta=60.0$ and in
$SU(4)$ on a $70^280$ lattice at $\beta=86.0$
\cite{AAMT_SUN}
which correspond to our smallest values of $a(\beta)$ in each case.
We normalise the effective masses to our estimate of the mass gap in each case.
We can now directly compare the effective mass plots. This we do for the
lightest five $0^+$ states in Fig.~\ref{fig_Meff_0p_so6su4} and for the
lightest two $2^+$ states, and the lightest $0^-$ and $1^+$ in
Fig.~\ref{fig_Meff_J_so6su4}. (The lightest $J=1,2$ negative parity
states are essentially the same as their positive parity partners.)
What we observe in Fig.~\ref{fig_Meff_J_so6su4} is that in all the
cases shown, it is entirely plausible that the $SO(6)$ effective energies
are approaching the corresponding $SU(4)$ energies at large $t$. It is
of course not possible to be confident in all cases that they actually do so,
given what are clearly poorer overlaps in $SO(6)$ than in $SU(4)$.
In  Fig.~\ref{fig_Meff_0p_so6su4} we see that the $0^{+\star}$ masses are
certainly consistent between $SO(6)$ and $SU(4)$, as are the
$0^{+\star\star\star\star}$ masses. (The agreement of the $0^+$ ground state
masses is of course enforced by our choice of normalisation.) 
However there is a substantial mismatch in the case of the  $0^{+\star\star}$
and of the  $0^{+\star\star\star}$ which shows no sign of disappearing as
$t$ increases.

So while we can conclude from our comparison at the smallest value of
$a(\beta)$ that most of the masses we calculate in $SO(6)$ are consistent
with their  $SU(4)$ counterparts, there is an issue with
the second and third excitations in the  $0^+$ sector. Here the effective
masses do appear to display reasonably convincing plateaux, but these are different
from the corresponding $SU(4)$ plateaux. One possibility is that due to quite
different overlaps what we identify as the $0^{+\star\star}$ is in fact the
$0^{+\star\star\star}$ and vice-versa, so that the $SO(6)$ state we label
as the  $0^{+\star\star}$  should in fact be compared to the $SU(4)$
$0^{+\star\star\star}$. As we can see from Fig.~\ref{fig_Meff_0p_so6su4}, these
states do in fact agree. One then has to speculate that what we label as
the  $SO(6)$  $0^{+\star\star\star}$ will at large enough $t$ become consistent
with the $SU(4)$ $0^{+\star\star}$. The effective masses in Fig.~\ref{fig_Meff_0p_so6su4}
of the former state do show some hint that they decrease at larger $t$, but
this is no more than a speculation with the calculations as they stand.
So whether these mismatches are
due to a sensitivity to the difference in the global properties of
the $SO(6)$ and $SU(4)$ groups, or there are some significant systematic errors
that we have not taken into account, is a question that needs to be addressed.

\section{Other results}
\label{section_other}

\subsection{large-$N$ scaling}
\label{subsection_Nscaling}

We have assumed throughout that the large-$N$ limit requires keeping $g^2N$ fixed
and that the leading correction is $O(1/N)$, following the all-orders analysis
of diagrams
\cite{CL_N}
The fact that, as we have seen, such fits work well provides good evidence for their
non-perturbative validity. Here we will try to be a little more quantitative.

We begin with the obvious comment that keeping $g^2N$ fixed
is necessary if one wants to obtain an $SO(\infty)$ 
theory that is perturbative (and asymptotically free) at short distances.
If the correct limit was to keep $g^2N^\gamma$ constant with $\gamma < 1$
then at any distance, however short, the theory would not be perturbative; and
if  $\gamma > 1$ then the theory would be free on all scales. In neither case
could one regard that as a `smooth' large-$N$ limit. So we begin by assuming
this scaling of the coupling and determine the power of the leading correction
for the mass gap in units of the 't Hooft coupling:
\begin{align}
  &&\left.\frac{M_{0^+}}{g^2N} \right|_{SO(N \rightarrow \infty)}
  = c_0 + \frac{c_1}{N^\alpha}&& ,SO(N\geq 3) .
\end{align}
We find that $\alpha = 0.94\pm 0.08$. So if the power of the correction is an
integer, then this confirms that it must be $O(1/N)$.

It is worthwhile to see if our results also demand that $g^2N$ should be kept constant.
We can fit 
\begin{align}
  &&\left.\frac{M_{0^+}}{g^2N^\gamma} \right|_{SO(N \rightarrow \infty)}
  = c_0 + \frac{c_1}{N}&& ,SO(N\geq 3) .
\end{align}
and doing so we find a tight constraint $\gamma=1.015\pm 0.017$, just as expected.

\subsection{spinorial states in the spectrum}
\label{subsection_spinorial}

In contrast to $SU(N)$, not all representations of  $SO(N)$ can be obtained
from  products of the fundamental. In particular operators that project onto a
single winding flux tube containing spinorial flux cannot be constructed using
products of fundamental $SO(N)$ fields and so one might be tempted to assume
that they do not appear in the spectrum of the $SO(N)$ Hamiltonian (or transfer
matrix). On the other hand we would expect that a state consisting of a winding
spinorial flux tube together with its conjugate (widely separated) would appear
in the finite volume spectrum of the $SO(N)$ theory, because the net
winding flux is a singlet, and so it will have a non-zero projection
onto the contractible loops used for the  glueball spectrum. (Such a state
is often called a `torelon'.) This may appear mildly surprising and in this
subsection we shall provide evidence that this is indeed the case.

We consider $SO(4)$ where, as we have seen in Sections~\ref{subsubsection_su2andso4}
and \ref{subsection_Lie_strings} the string tension is twice that of $SU(2)$, i.e.
$\sigma^{so4}_f = 2 \sigma^{su2}_f$ when expressed in units of, say, the 
lightest scalar glueball mass. Torelon states composed of a winding flux tube
and its conjugate will have an energy $E_T(l) \sim 2 \sigma l + O(1/l)$,
neglecting their interaction. On a symmetric $l\times l$ spatial
volume we can consider $J^P=0^+, 2^+$ states obtained by adding or
subtracting the torelons winding around the $x$ and $y$ spatial tori.
Once $l$ is small enough that the torelon mass is smaller than the large 
$l$ glueball mass, the torelon will become the lightest glueball with that 
$J$ and from then on we will find that the glueball mass decreases
with decreasing $l$ roughly like $2 \sigma l$. In $SU(2)$ this will occur with
$\sigma=\sigma^{su2}_f$ and so for $2\sigma_f^{su2} l \leq M_G(l=\infty)$
we expect the glueball mass to decrease (roughly) linearly with $l$,
as indeed one observes. In the case of  $SO(4)$ one would naively
expect the decrease to set in when  $2\sigma_f^{so4} l \leq M_G(l=\infty)$,
i.e. at a much smaller value of $l$ than in $SU(2)$. (We need two winding
flux tubes since a single one transforms non-trivially under the
$Z_2$ centre symmetry, and this is why we consider $SO(4)$ rather than $SO(3)$.)
If however the glueball mass in  $SO(4)$ is the same
as in $SU(2)$ then it should show the same $l$-dependence. This decrease
in the lightest glueball mass will thus  set in at a much larger value 
of $l$ than one would naively expect by using $\sigma=\sigma^{so4}_f$ 
and if observed provides a signature of the presence 
of pairs of such spinorial flux tubes in the $SO(4)$ spectrum.

In Fig.~\ref{fig_Vso4su2} we display the dependence on $l$ of the lightest $0^+$
and $2^+$ glueball masses in $SO(4)$ and in $SU(2)$. The $SO(4)$  calculation 
is at a value of the lattice spacing  $a\surd\sigma^{so4}_f \simeq 0.160$ and
the $SU(2)$ one at $a\surd\sigma^{su2}_f \simeq 0.118$. Given that
$\surd\sigma^{SO4}_f = 2 \surd\sigma^{SU2}_f$, this means that the $SO(4)$ 
lattice spacing differs by only about $8\%$ from the $SU(2)$ one, and so the calculations
should be comparable once $l$ and the glueball masses are expressed in physical 
units, as they are in   Fig.~\ref{fig_Vso4su2}. (For this we use the $l=\infty$ 
value of the mass gap.) We also plot the energy of twice the fundamental flux loop
in both cases (as a rough estimate of the corresponding torelon masses).
We observe that the $l$-dependence is very similar in $SU(2)$ and $SO(4)$.  
(The slight difference may be due to slightly different $O(a^2)$ corrections.) 
We also observe that the rapid decrease with decreasing $l$ of the $2^+$ glueball 
mass sets in, in the case of $SO(4)$, when the fundamental torelon is very
much more massive,
and presumably irelevant, while the spinorial (i.e. fundamental $SU(2)$) torelon
is comparable in mass. This provides quite compelling evidence that at
small $l$ the $SO(4)$ mass is dominated by a spinorial torelon.

Some comments. The dimension of the spinorial representation grows rapidly with
$N$ and the associated string tension presumably becomes larger than the fundamental 
string tension. So it can be ignored at larger $N$, as far as finite volume
corrections are concerned. We have chosen not to perform the comparison for
$SO(3)$ both for the reason already given and also because the $SO(3)$ fundamental flux
tube is not stable (it corresponds to the
adjoint of $SU(2)$) and so can, by itself, have a non-zero overlap onto glueball 
operators. Finally, using $SO(6)$ intead of $SO(4)$ would have been elegant, but
since  $\sigma^{SO6}_f = \sigma^{SU4}_{k=2} \simeq 1.35 \sigma^{SU2}_{k=2}$
the effect would have been less striking than with $SO(4)$ unless one
had significantly greater statistical accuracy.

\subsection{strong coupling}
\label{subsection_sc}

We see from Table~\ref{table_bulk} that at low $N$ the `bulk transition' occurs at a 
small value of the lattice spacing when the latter is expressed in physical units
such as the string tension or the mass gap. Given this fact and the fact that 
the physics, or most of it, appears 
to be continuous across the transition region, it is tempting to ask if any aspect
of the continuum behaviour already appears on the strong coupling side of the
transition. This question is peripheral to the main focus of this paper,
so we address it only briefly.

We consider $SO(3)$ where the bulk transition is at the weakest coupling.
We have calculated some glueball masses and the string tension 
for a range of couplings in the `strong coupling' region, including 
the bulk cross-over region. As we go deeper into strong coupling,
the value of a mass in lattice units, $aM$, will become larger,  and so 
the correlators will fall more steeply and our estimate of the glueball mass
becomes less reliable. We therefore focus here on the lightest $J^P=0^+$ 
and $J^P=2^-$ glueball masses and the string tension. We focus on glueballs
with $J^P=2^-$ rather than $J^P=2^+$ because our volumes in the strong 
coupling region are not large enough to guarantee that
the $2^+$ is unaffected by finite volume torelon contributions.

We display in Fig~\ref{fig_kg_mg_so3d3} our calculated values in $SO(3)$ of  
\begin{equation}
  \frac{\beta_I}{2N^2} {a\mu}
  \equiv
  \frac{\beta \langle u_p \rangle}{2N^2} {a\mu}
  =
  \frac{\mu}{g_I^2N} \quad ; \quad \mu=\surd\sigma, M_{0^+}
\label{eqn_mg_kg}
\end{equation}
where $\langle u_p \rangle$ is the average plaquette, and the subscript $I$
denotes the mean-field improvement of the (inverse) couplings.
Note that in the cross-over region we do not use the anomalous, 
nearly massless and weakly coupled scalar glueball mass but rather the
first excitation which is the one
that is a clear continuation of the strong coupling mass gap.
We show the best fits to the weak coupling values that are linear in $1/\beta_I$,
and remark that attempting to include some of the strong coupling values
with quadratic fits simply does not work.
It is very clear from all this that the strong coupling dependence of $\mu/g^2$
is qualitatively different from that on the weak coupling side, and does not
provide useful information about the continuum limit of this ratio.

It is of course not surprising that the dependence on $\beta$ of $\beta\mu$
should not be well captured by a low order expansion in powers of $1/\beta$
when we are on the strong coupling side of the bulk transition where
the natural expansion is presumably in positive powers of $\beta$. 
A more relevant question to ask is whether any ratios of physical quantities
are better behaved when plotted against $a^2$. The point here is that
perhaps $a^2$ is small enough on at least part of the strong coupling side for an
operator product expansion in powers of $a^2$ to make sense even if an
expansion in positive powers of $ag^2$ does not work. To address this question
we plot in  Fig~\ref{fig_Mm_scwc_so3d3} the ratios $m_{0^+}/\surd\sigma$,
$m_{2^-}/\surd\sigma$ and  $m_{2^-}/m_{0^+}$. We see that in this last case,
where we have the ratio of two glueball masses, there is in fact no
marked difference in the $a^2$ dependence between weak and strong coupling,
and the weak-coupling linear extrapolation  to the continuum limit
appears to describe the physical mass ratios at strong coupling just
as well, to a good approximation. In contrast to this, the continuum
extrapolations of $m_{0^+}/\surd\sigma$ and $m_{2^-}/\surd\sigma$
clearly do not pass through the strong coupling values. However
what is interesting is that it is possible to get good linear
fits to (most of) the strong coupling values by themselves. These
are shown as dashed lines in  Fig~\ref{fig_Mm_scwc_so3d3}. We
recall a very similar observation in
\cite{RLMT_Tc}
for the deconfining temperature in units of the string tension. (In that
case for $N\geq 4$ since no calculation for $SO(3)$ was attempted.)
These strong coupling `continuum' extrapolations differ from the
weak-coupling ones, albeit by not very much. All this suggests that
perhaps the expansion in $a^2$ makes sense on the strong coupling side,
and perhaps it even makes sense to talk of a separate continuum limit
to the strongly coupled gauge theory. 

As an aside we remark that in $D=3+1$ the
bulk cross-over becomes a first-order transition, and the mass ratios appear
to take very different values on either side of the transition. So here
the strong-weak coupling transition appears to be much more conventional.

\section{Conclusions}
\label{section_conclusion} 

In this paper we set out to address several interesting questions about
$SO(N)$ gauge theories. We have chosen to work in $D=2+1$ rather than in $D=3+1$,
since the strong-to-weak coupling transition is awkwardly placed in the latter
case, and because these questions apply equally well in $D=2+1$. The quantities
we focus upon are the low-lying mass (`glueball') spectrum, the confining string
tension and the coupling. (Elsewhere
\cite{RLMT_Tc}
we have analysed the finite temperature deconfining transition.)

The main questions we address have to do with various relationships between $SO(N)$
and $SU(N)$ gauge theories. At $N=\infty$ diagrammatic
\cite{CL_N}
and orbifold
\cite{Orb_equiv,Orb_phys}
arguments lead us to expect that the theories have the same local physics 
in the $C=+$ sector, and we would like to confirm by explicit calculation
that this is so for non-perturbative physics. At low $N$, we know that $SO(3)$,
$SO(4)$ and $SO(6)$ have the same Lie algebras as $SU(2)$, $SU(2)\times SU(2)$ and
$SU(4)$ respectively, so we would like to see if this means that the physics within
each pair of theories is the same, or if the differing global properties of the
groups leads, for example, to some glueball masses being different.

Another question we address is whether $SO(N)$ gauge theories are linearly
confining. This question is particularly interesting when $N$ is odd, because in
that case the centre of the group is trivial, and one loses the standard connection
between confinement and the unbroken centre symmetry.

To give convincing answers to these questions requires accuracy
and precision, particularly because the $SU(N)$ spectra and string tensions
with which one compares are now known with quite good accuracy
\cite{AAMT_SUN,AAMT_string}.
Unfortunately, as it turns out we are only able to achieve the intended
accuracy in some of the lightest states. In the case of our $N=\infty$
extrapolations it is apparent from
Tables~\ref{table_MK_largeN},\ref{table_glueball_largeN}
that the statistical errors on our $SO(\infty)$
mass ratios are typically some  3 to 5 times larger than the errors on
the $SU(\infty)$ calculations despite the fact that nominally the `statistics'
of the two sets of calculations is comparable. While at small $N$ one can point
to the mediocre overlaps of our operators onto the glueballs as contributing to
this, at larger $N$ this ceases to be the case and it is our larger
$N$ results that should dominate the extrapolation to $N=\infty$. This suggests that
our Monte Carlo algorithm for $SO(N)$ is less efficient than the
one used for $SU(N)$ despite the superficial similarity of the heat bath.
While this may be in part because we have not included some analogue of the $SU(N)$
over-relaxation in our $SO(N)$ update, this might not be the whole story.
In any case, this is clearly something that needs to be addressed if one wishes
to achieve real precision in $SO(N)$. Even more important for our comparisons
is the mediocre overlap of our operator basis onto the states of interest,
particularly at small $N$. This magnifies certain systematic errors and a significant
part of our work in this paper has been devoting to deciding whether the observed
mismatches between certain mass ratios in various $SO(N)$ and $SU(N^\prime)$
theories reflect a real difference or an underestimate of these systematic errors.
To help guide the reader as to which states are most reliably calculated,
we introduced a grading system. The masses of states labelled by $\alpha$ or $\beta$
should be reliable, within the quoted errors, but this becomes less likely as we
move to $\gamma$ and beyond. The string tension calculations should be reliable,
with some caution for $SO(3)$.

Despite the above caveats, we have been able to come to a number of useful
conclusions. In the context of the $N$-dependence
we found that when extrapolating to $N=\infty$ various ratios of string
tensions, glueball masses and the 't Hooft coupling, a single
fitting function encompasses both the odd $N$ and even $N$ values. That is to say
the physics appears to be a  single function of $N$, rather than there being separate
functions for odd and for even $N$, which only converge at $N=\infty$.
As for the $N=\infty$ limit, we found
that our most accurately extrapolated ratios, those involving the mass gap, $m_{0^+}$,
the string tension, $\sigma$, and the coupling $g^2$, agree well with the
corresponding values for $SU(N\to\infty)$, with errors at the level of $1\%$
or less. (In the case of $g^2$ the comparison includes, and so confirms, the factor
of 2 that is predicted by the diagrammatic analysis.) Of the other states
which we believe to be reliable (graded as $\beta$), the $2^+$ and $2^-$
are consistent with the $SU(\infty)$ values, but the $0^{+\star}$ is not.
The discrepancy of the latter is around 4.7 standard
deviations when using the more plausible higher order extrapolations to
$N=\infty$. Whether this is a real mismatch between $SO(N\to\infty)$ and
$SU(N\to\infty)$, or is due to unexpectedly large systematic errors remains an
open question at this stage. As for the other, less well-determined mass ratios,
these  are in reasonable agreement with those in  $SU(N\to\infty)$, except
for the $2^{-\star}$. But given our grade of $\delta$ for the reliability
of this state, this need not be a cause for concern.

Our comparison of the mass gap, string tension and coupling of $SU(2)$ with
those of $SO(3)$ and $SO(4)$, and of $SU(4)$ with $SO(6)$, shows that the ratios
agree once one translates the couplings and the string tension representations
appropriately. (For example, the fundamental flux in $SO(6)$
corresponds to the $k=2$ antisymmetric flux in $SU(4)$.) One should note,
however, that when comparing the values of $\surd\sigma/g^2$
in $SO(3)$ and $SU(2)$ the errors are large due to the instability of the
flux tube, which reduces the significance of the apparent agreement.
For the continuum glueball mass ratios, $M/m_{0^+}$, listed in
Table~\ref{table_glueball_SONvsSUM}, the masses that are calculated
reliably (i.e. $\alpha,\beta$) do agree within $2\sigma$.
This is also the case for the other states in the comparison
between $SO(3)$ and $SU(2)$ and in that between $SO(6)$ and $SU(4)$
except, in the latter case, for the $0^{+\star\star}$ and $0^{+\star\star\star}$.
There are more pronounced disgreements in the comparison between
$SO(4)$ and $SU(2)$, which may be due to the unsuppressed presence
of states that simultaneously contain glueballs from the two $SU(2)$
colour groups. In summary, it is striking that the physical quantities which we
best control do in fact agree between the $SO(N)$ and $SU(N^\prime)$
theories that share the same Lie algebras. So in these cases at least,
it appears that the difference in the global properties of the groups
plays no role.

Our above discussion assumes a well defined string tension which presupposes that $SO(N)$
theories are linearly confining -- if not exactly then at least for `all practical
purposes'. This we have attempted to verify by calculating the ground state energy, 
$E_f(l)$, of a flux tube that winds once around a spatial torus, as a function of
the length $l$ of the torus and hence of the flux tube. For even
$N$ we found that $E_f(l)$ is well described by a fit that includes a linear
term modified by the known universal corrections to the linear term,
as in Fig.~\ref{fig_Ek1_so6_so8}. This provides numerical evidence for linear
confinement that is more-or-less as convincing as what one has in $SU(N)$ gauge
theories. Odd $N$ is more controversial because here the centre of the group is trivial.
In the $N\to\infty$ planar limit we expect to recover linear confinement, so
the question is what happens at smaller odd values of $N$. For $SO(3)$ the
situation, as shown in Figs~\ref{fig_Ek1_so3_b9.0} and \ref{fig_Eeffk1_so3_b9.0},
is far from clear-cut, no doubt due to the significant instability of
the flux tube, which is expected since it can be broken by gluon pairs
from the vacuum.  Nonetheless we can  extract an effective string
tension, albeit with large errors, similarly to the way one can
extract an effective adjoint string tension in $SU(2)$
\cite{AAMT_SUN,AAMT_string}. For $SO(5)$ the evidence for linear confinement,
as displayed in Figs~\ref{fig_Ek1_so5_b27.5} and \ref{fig_Eeffk1_so5_b27.5},
looks as convincing as what we find for even $N$. Of course, for neither
odd nor even $N$ does
this demonstrate asymptotic linear confinement, something that a numerical
calculation cannot do. However what it tells us is that even if the flux
tube is not stable for odd $N$, its decay width is so small that we can safely
extract a string tension within our errors -- rather as if we were assigning
a mass to a resonance with a decay width that while non-zero was so small
as to be invisible in our calculations. What a small but non-zero decay width
of the flux tube might imply for absolute confinement is not obvious. Indeed
that is what one has in $QCD$ with light quarks (string-breaking by $q\bar{q}$ pairs)
and that theory is believed to be absolutely confining. Moreover it appears
that in $QCD$ the vacuum expectation value of similarly blocked Polyakov loops
is substantially larger than our bounds on that quantity in $SO(5)$.
(See Figs 1,2 of
\cite{AHMT}.)
The fact that we do observe a non-zero overlap of the vacuum onto
our basis of winding operators in $SO(5)$, albeit not onto the ground state
of the flux tube, motivates seeing what happens to this overlap
as $N$ grows.  We therefore performed a more detailed study in
Section~\ref{subsection_oddN_confinement} of the overlap of the vacuum onto the
ground state of the flux tube  for $N=3,5,7,9,11$. This confirmed that apart,
perhaps, from the case of $SO(3)$, any instability of the flux tube ground state
is so small as to be consistent with zero within the small errors of our calculation,
and, moreover, the overlap onto the whole basis of winding operators does decrease rapidly
with increasing $N$. While this provides evidence that $SO(2N+1)$ theories are linearly
confining `for all practical purposes' (at least for $2N+1\geq 5$) it does not
answer the question of whether they are exactly confining. The lack of a
non-trivial centre symmetry makes the question similar to the one for $QCD$
with light quarks. Although $QCD$ is believed to be exactly confining (i.e. the
energy of an isolated coloured charge is infinite) 
our usual non-local order parameters (e.g. asymptotic exponential decay
of Wilson loops) are not useful. Here we expect that the route to follow will
be similar to that for $SO(3)$. The exact confinement of fundamental charges
in $SO(3)$ is believed to follow from the exact confinement of adjoint charges
in $SU(2)$, even though the adjoint Wilson loop in $SU(2)$ does not possess
an asymptotic area decay. The natural generalisation of this argument is to
note that $SO(N)$ and $Spin(N)$ share the same Lie algebra and that the
latter has a non-trivial center and so it would be no surprise if it possesses
exact linear confinement. To develop this argument in detail would be very
interesting but would take us beyond the scope of the calculations in this paper.

We also briefly touched upon the interesting question of whether the spinorial
flux tubes in $SO(4)$, which correspond to the fundamental flux tubes of
$SU(2)$, affect the physics of the  $SO(4)$ theory. By comparing the finite volume
dependence of the tensor glueball mass in this pair of theories we obtained
what we believe to be quite good evidence that the $SO(4)$ theory does
indeed contain states composed of a spinorial flux tube and its conjugate,
even if, somewhat paradoxically, it does not contain states composed of
a single such spinorial flux tube. The issue being highlighted here is
clearly much more general. For example, does $SO(6)$ contain states that correspond
to a pair of $J^{PC}=0^{--}$ $SU(4)$ glueballs (which together will form a
$J^{PC}=0^{++}$ state for even orbital angular momentum) even though it does not
possess a state consisting of a single such glueball? After all, such states
contribute to the decay width of a sufficiently excited $0^{++}$ glueball in
$SU(4)$ and if the decay width is the same in $SO(6)$ then they are surely
present in the latter theory? That is to say, a field theory may
contain more states than is apparent from the `evident' symmetries of
the theory? We intend to address these interesting questions separately elsewhere.

\section*{Acknowledgements}

Our interest in this project was originally motivated by Aleksey
Cherman and Francis Bursa. We acknowledge, in particular, the key contributions 
made by Francis Bursa to this project in its early stages.
The project originated in a number of discussions during the 2011 Workshop on 
`Large-N Gauge Theories'  at the Galileo Galilei Institute in Florence,
and we are grateful to the Institute for providing such an ideal environment 
within which to begin collaborations.
The numerical computations were carried out on the computing cluster 
in Oxford Theoretical Physics.

\clearpage

\begin{table}
\centering
\begin{tabular}{|c|c|c|c|} \hline
group 	&  $\beta_b \in$ &  $a\surd\sigma \sim$ &  $l^2 >$ \\ \hline
SO(3)   & [5.75,6.25]  &  0.15  & $50^2$  \\
SO(4)   & [9.1,10.2]   &  0.22  & $30^2$  \\
SO(5)   & [13.5,15.4]  &  0.27  & $22^2$  \\
SO(6)   & [18.0,21.3]  &  0.31  & $16^2$  \\
SO(7)   & [23.5,28.0]  &  0.35  & $12^2$  \\
SO(8)   & [31,35]      &  0.39  & $10^2$  \\
SO(12)  & [65,73]      &  0.46  & $9^2$   \\
SO(16)  & [111,124]    &  0.54  & $8^2$  \\  \hline
\end{tabular}
\caption{ Location $\beta=\beta_b$ of bulk transition, with string tension
on weak coupling side and smallest useful spatial volume on weak coupling side.}
\label{table_bulk}
\end{table}

\begin{table}
\centering
\begin{tabular}{|c||l|l||l|}
  \hline
  $J^P$  & \multicolumn{2}{|c||}{ $SO(4)$ , $\beta=15.1$} & $SU(2)$ , $\beta=16.0$ \\ \hline
     	 &  Ops A on $50^256$ & Ops B on $48^256$ & Ops A on $68^248$ \\ \hline
$0^{+}$	        & 0.4807(13)    & 0.480(4)   &  0.4807(19) \\
$0^{+*}$	& 0.706(4)     & 0.712(6)   &  0.692(4) \\
$0^{+**}$	& 0.886(9)     & 0.925(11)  &  0.857(6) \\
$0^{+***}$	& 0.982(14)    & 1.020(19)  &  0.939(7) \\
$0^{+****}$	& 1.026(17)    & 1.028(19)  &  0.987(6) \\
$2^{+}$		& 0.8003(25)   & 0.820(6)   &  0.789(5) \\
$2^{-}$         & 0.8005(51)   & 0.816(4)   &  0.799(4) \\
$2^{+*}$	& 0.963(10)   & 1.029(42)  &  0.919(10) \\
$2^{-*}$	& 0.985(5)    & 0.992(19)  &  0.939(7) \\
$0^{-}$		& 1.005(17)   & 1.043(8)   &  1.005(8) \\
$1^{+}$	        & 1.045(76)   & 1.180(26)  &  1.084(9) \\
$1^{-}$	        & 1.137(28)   & 1.176(27)  &  1.068(9) \\ \hline
$am_p$	        & 1.037(6)    & 0.993(4)   &   \\ \hline
\end{tabular}
\caption{Spectrum at $\beta=15.1$ in $SO(4)$ using two different operator bases. Also an $SU(2)$
  calculation using basis A, with masses rescaled up by a factor 1.169 to a common mass gap.
Also the flux loop mass, $am_p$.}
	\label{table_spectrum_opAopB}
\end{table}

\begin{table}
  \centering
  \begin{tabular}{|llllllll|}
    \hline
    $L_s^2L_t$ &$12^236$&$16^236$ &$20^236$ &$24^236$ &$28^236$ &$32^236$ &$36^236$\\
    $l\surd\sigma$ & 1.61 & 2.15 & 2.68  & 3.22 & 3.76  & 4.30 & 4.83 \\ 
    \hline
    $am_P$&0.1349(15)&0.2366(19)&0.3306(19)&0.4078(41)&0.4834(22)&0.5606(19)&0.6354(22)\\
    $a\surd\sigma$&0.1243(5)&0.1303(5)&0.1338(4)&0.1339(6)&0.1340(4)&0.1343(3)&0.1344(3)\\
    \hline
    $0^+$&0.247(10)&0.447(5)&0.494(9)&0.492(9)&0.506(2)&0.514(4)&0.511(3)\\
    $0^{+*}$&0.252(11)&0.508(26)&0.675(17)&0.737(16)&0.769(4)&0.789(8)&0.773(7)\\
    $0^{+**}$&0.489(29)&0.558(23)&0.836(38)&0.935(21)&0.968(30)&1.021(14)&0.994(11)\\
    $0^{+***}$&0.482(29)&0.854(29)&0.883(19)&0.996(33)&0.992(10)&1.021(6)&1.028(12)\\
    $0^{+****}$&0.523(25)&0.947(7)&1.005(20)&1.091(35)&1.098(11)&1.124(18)&1.088(17)\\

    $2^{+}$&0.228(6)&0.385(31)&0.874(11)&0.875(8)&0.859(7)&0.857(8)&0.850(8)\\
    $2^{+*}$&0.668(34)&0.681(46)&0.899(18)&1.031(26)&1.058(10)&1.040(15)&1.054(13)\\
    $2^{-}$&0.471(10)&0.769(11)&0.854(12)&0.892(21)&0.852(5)&0.837(17)&0.869(4)\\
    $2^{-*}$&0.783(19)&0.824(11)&1.035(24)&1.001(25)&1.052(8)&1.073(15)&1.071(15)\\

    $0^{-}$&1.124(7)&1.126(24)&1.117(22)&1.134(38)&1.095(11)&1.110(7)&1.118(19)\\

    $1^{+}$&1.308(33)&1.322(36)&1.277(36)&1.260(43)&1.252(14)&1.280(20)&1.265(23)\\
    $1^{-}$&1.286(26)&1.225(24)&1.274(33)&1.295(48)&1.270(12)&1.237(25)&1.246(20)\\
    \hline
  \end{tabular}
  \caption{Glueball masses, $am_G$, and the flux loop energy, $am_P$, with corresponding
string tensions $a\surd\sigma$ extracted using the Nambu-Goto formula in eqn(\ref{eqn_NG}). 
On various volumes at $\beta=84$ in $SO(8)$.}
  \label{table_Vso8}
\end{table}

\vspace{0.8cm}

\begin{table}[h]
\centering
\begin{tabular}{|l|ll|ll|ll|}
  \hline
  & \multicolumn{2}{|c}{$SO(3)$} & \multicolumn{2}{|c}{$SO(4)$} & \multicolumn{2}{|c|}{$SO(6)$}  \\ 
\hline
$L_s^2L_t$  & $46^246$  & $62^248$  & $34^244$  & $44^252$  & $36^244$ & $54^244$ 	\\
  \hline
$a\surd\sigma$	& 0.1333(8) & 0.1353(11) & 0.1617(4) & 0.1621(3) & 0.12658(22) & 0.12680(28)\\
$l\surd\sigma$	& 6.13 & 8.39 & 5.50 & 7.13 & 4.56 & 6.85 \\
	\hline
$0^+$		& 0.3963(25)  & 0.3964(34) & 0.5353(35) & 0.5395(16)	 &0.457(3)&0.467(2)\\
$0^{+*}$	& 0.571(11)   & 0.5958(42) & 0.770(14)  & 0.801(3)	 &0.695(7)&0.708(5)\\
$0^{+**}$	& 0.716(11)   & 0.716(12)  & 0.910(26)  & 1.030(5)	 &0.891(19)&0.910(17)\\
$0^{+***}$	& 0.784(12)   & 0.801(16)  & 1.138(9)   & 1.099(26)      &0.920(19)&0.929(15)\\
$0^{+****}$	&  --         & 0.869(20)  & 1.046(14)  & 1.143(23)	 &1.014(7)&1.012(22)\\

$2^{+}$		& 0.6783(21) & 0.6660(51) & 0.898(8)   & 0.908(9)   &0.788(6)&0.768(9)\\
$2^{+*}$	& 0.796(15)  & 0.815(8)   & 1.088(6)   & 1.091(19)  &0.966(7)&0.961(20)\\
        
$2^{-}$		& 0.6750(37) & 0.6756(45) & 0.9145(36) & 0.900(9)  &0.777(4)&0.772(9)\\
$2^{-*}$	& 0.779(16)  & 0.785(12)  & 1.097(16)  & 1.090(22) &0.970(22)&0.974(8)\\

$0^{-}$		& 0.882(10)  & 0.845(20)  & 1.183(24) & 1.195(9)   &1.003(9)&0.972(23)\\

$1^{+}$		& 0.967(9)   & 0.977(12)  & 1.311(34) &	1.404(15)  &1.167(8)&1.165(10)\\
$1^{-}$		& 0.938(28)  & 0.966(13)  & 1.322(31) &	1.371(14)  &1.158(10)&1.156(9)\\
  \hline
\end{tabular}
\caption{Finite volume test for glueball masses and the string tension in
  $SO(3)$, $SO(4)$ and $SO(6)$ at $\beta=7.0, 13.7, 46.0$ respectively.}
	\label{glueball:tab:finitevolumesoN}
\end{table}

\vspace{0.8cm}

\begin{table}[h]
\centering
\begin{tabular}{ |c|cccccccc|}
  \hline
 $N$ & 3 & 4 & 5 & 6 & 7 & 8 & 12 & 16 \\
 $l\surd\sigma\sim$ & 8.0(6.5) & 7.0(4.5) & 6.0 & 4.5 & 4.2 & 3.7 & 3.5 & 3.5 \\
  \hline
\end{tabular}
	\caption{Approximate values of $l\surd\sigma$ for $SO(N)$ mass spectrum calculations.
In brackets the smaller volumes used at low $N$ for the string tension.}
	\label{table_VsizeN}
\end{table}

\begin{table}[h]
\centering
\begin{tabular}{ |ccc|ccc|}
  \hline
  \multicolumn{3}{|c}{$SO(6)$} & \multicolumn{3}{|c|}{$SO(8)$}  \\ \hline
  $l_x$ & $l_y\times l_t$ & $aE(l_x)$ & $l_x$ & $l_y\times l_t$ & $aE(l_x)$ \\ \hline
  $18$ & $48\times 64$ & 0.2582(11)  & $12$  & $48\times 80$  & 0.1504(35)  \\
  $22$ & $42\times 52$ & 0.3280(13)  & $16$  & $36\times 64$  & 0.2360(39)  \\
  $26$ & $40\times 52$ & 0.3972(16)  & $20$  & $32\times 48$  & 0.3149(47)  \\
  $30$ & $40\times 50$ & 0.4596(28)  & $24$  & $24\times 36$  & 0.3844(29)  \\
  $36$ & $36\times 44$ & 0.5586(23)  & $32$  & $32\times 32$  & 0.5317(35)  \\
  $42$ & $42\times 44$ & 0.6606(30)  & $40$  & $40\times 32$  & 0.6727(40)  \\ \hline
\end{tabular}
\caption{Energy of the ground state flux tube versus its length, in $SO(6)$
  at $\beta=46$, and in $SO(8)$ at $\beta=86$.}
	\label{table:Estrings_so6so8}
\end{table}

\begin{table}[h]
\centering
\begin{tabular}{|cccc|ccc|}
  \hline
  \multicolumn{4}{|c}{$SO(3)$} & \multicolumn{3}{|c|}{$SO(5)$}  \\ \hline
  $l_x$ & $l_y\times l_t$ & $aE(l_x)_{sub}$ & $aE(l_x)_{nosub}$ & $l_x$ & $l_y\times l_t$ & $aE(l_x)$ \\ \hline
  $26$ & $80\times 140$ & 0.2133(36)  & 0.2292(84) & $14$  & $54\times 72$  & 0.2245(17)  \\
  $30$ & $80\times 100$ & 0.2501(55)  & 0.2508(71) & $22$  & $40\times 50$  & 0.3980(17)  \\
  $34$ & $70\times 86$  & 0.2724(82)  & 0.2724(75) & $30$  & $36\times 44$  & 0.5572(22)  \\
  $38$ & $70\times 86$  & 0.3462(49)  & 0.3462(45) & $36$  & $36\times 44$  & 0.6780(18)  \\
  $46$ & $70\times 86$  & 0.4354(72)  & 0.4335(71) & $42$  & $42\times 44$  & 0.7859(84)  \\
  $52$ & $70\times 86$  & 0.4964(62)  & 0.4963(62) &   &   &   \\ 
  $62$ & $62\times 60$  & 0.5849(81)  & 0.5854(81) &   &   &   \\ 
  $82$ & $82\times 64$  & 0.839(34)   & 0.824(33)  &   &   &   \\ \hline
\end{tabular}
\caption{Energy of the ground state flux tube versus its length, in $SO(3)$
  at $\beta=9.0$, both vacuum subtracted and unsubtracted, and in $SO(5)$ at $\beta=27.5$.}
	\label{table:Estrings_so3so5}
\end{table}

\begin{table}[h]
\centering
\begin{tabular}{ |ccc|ccc|}
  \hline
  \multicolumn{6}{|c|}{$SO(5)$}  \\ \hline
  $l$ & $O_{gs}^2\sim$ & $bl_{max}$ & $\langle \phi_{gs} \rangle$ & $\sum_i|\langle \phi_i \rangle|^2$ & $aE_{eff}(t=a)$ \\ \hline
14  & 0.994  & 4  & 0.0083(31)  & 0.00026(7)  &  0.2298(7)  \\
    &        & 5  & 0.0122(31)  & 0.1438(5)   &  0.2290(7)  \\ \hline
22  & 0.998  & 4  & 0.0026(22)  & 0.000038(15)&  0.4049(8)  \\
    &        & 5  & 0.0033(23)  & 0.0452(3)   &  0.4008(9)  \\ \hline
30  & 0.993  & 4  & -0.0017(15) & 0.000024(8) &  0.5711(9)  \\
    &        & 5  & -0.0018(15) & 0.00395(8)  &  0.5642(8)  \\ \hline
36  & 0.991  & 5  & 0.0006(9)   & 0.00016(2)  &  0.6873(7)  \\
    &        & 6  & 0.0037(9)   & 0.2244(4)   &  0.6869(7)  \\ \hline
42  & 0.970  & 5  & -0.0016(10) & 0.000023(6) &  0.8063(9)  \\
    &        & 6  & -0.0015(10) & 0.0431(4)   &  0.8061(9)  \\ \hline
\end{tabular}
\caption{Various overlaps versus length, as described in Section~\ref{subsection_oddN_confinement},
  in $SO(5)$ at $\beta=27.5$.}
	\label{table:overlaps_so5}
\end{table}

\begin{table}[h]
\centering
\begin{tabular}{ |ccc|ccc|}
  \hline
  \multicolumn{6}{|c|}{$SO(3)$}  \\ \hline
  $l$ & $O_{gs}^2\sim$ & $bl_{max}$ & $\langle \phi_{gs} \rangle$ & $\sum_i|\langle \phi_i \rangle|^2$ & $aE_{eff}(t=a)$ \\ \hline
   26 & 0.91 & 5 & 0.059(3) & 0.3758(9) & 0.2509(8) \\
      & 0.81 & 4 & 0.033(2) & 0.0135(3) & 0.2791(6) \\  \hline
   30 & 0.83 & 5 & 0.023(3) & 0.1306(5) & 0.3027(9) \\
      & 0.71 & 4 & 0.011(3) & 0.0014(1) & 0.3360(8) \\  \hline
   34 & 0.78 & 6 & -0.019(3) & 0.7579(4) & 0.3427(12) \\
      & 0.67 & 5 &  0.010(3) & 0.0421(3) & 0.3499(11) \\  \hline
   38 & 0.90 & 6 & -0.019(3) & 0.5733(3) & 0.3952(10) \\
      & 0.91 & 5 &  0.000(3) & 0.0118(2) & 0.4015(9) \\  \hline
   46 & 0.89 & 6 & 0.012(2) & 0.2472(7)  & 0.4896(8) \\
      & 0.89 & 5 & 0.003(2) & 0.00014(2) & 0.4973(7) \\  \hline
   52 & 0.86 & 6 & -0.0011(13) & 0.0645(4)  & 0.5615(8) \\
      & 0.87 & 5 & -0.0006(13) & 0.00004(1) & 0.5674(7) \\  \hline
   62 & 0.83 & 7 & -0.006(1) & 0.4975(5)   & 0.6689(11) \\
      & 0.83 & 6 & -0.001(1) & 0.0089(2)   & 0.6744(10) \\
      &      & 5 & -0.000(8) & 0.000007(2) & 0.6850(8) \\  \hline
   82 & 0.85 & 7 & 0.014(1)  & 0.1821(5)  & 0.9121(10) \\
      & 0.82 & 6 & 0.0014(7) & 0.00009(1) & 0.9130(11) \\
      &      & 5 & 0.0012(7) & 0.00002(1) & 0.9201(11) \\  \hline
\end{tabular}
\caption{Various overlaps versus length, as described in Section~\ref{subsection_oddN_confinement},
  in $SO(3)$ at $\beta=9.0$.}
	\label{table:overlaps_so3}
\end{table}

\clearpage

\begin{table}[h]
\centering
\begin{tabular}{ |cccc|ccc|}
  \hline
  \multicolumn{7}{|c|}{$SO(2N+1)$; $l=36$}  \\ \hline
  group  & $\beta$ & $bl_{max}$ & $O_{gs}^2\sim$ & $\langle \phi_{gs} \rangle$ & $\sum_i|\langle \phi_i \rangle|^2$ & $aE_{eff}(t=a)$ \\ \hline
SO(3)  & 7.0    & 5  & 0.869 & -0.0014(8)  & 0.0030(2)   & 0.6958(14) \\
       &        & 6  & 0.886 & -0.0028(8)  & 0.5338(15)  & 0.6890(18)  \\ \hline
SO(5)  & 27.5   & 5  & 0.991 & 0.0006(9)   & 0.00016(2)  &  0.6873(7) \\
       &        & 6  & 0.991 & 0.0037(9)   & 0.2244(4)   & 0.6869(7)  \\ \hline
SO(7)  & 64.0   & 5  & 0.995 & 0.0006(11)  & 0.000011(3) & 0.5979(6) \\
       &        & 6  & 0.995 &  0.0004(11) &  0.0606(2)   & 0.5979(6) \\ \hline
SO(9)  & 106.0  & 5  & 0.995 & -0.0016(9)  &  0.000017(4) & 0.6764(6) \\
       &        & 6  & 0.995 & -0.0015(9)  &  0.00913(8)  & 0.6764(6) \\ \hline
SO(11) & 164.0  & 5  & 0.995 & 0.0011(9)   & 0.000012(4)  & 0.6753(6) \\
       &        & 6  & 0.995 & 0.0011(9)   & 0.00121(3)   & 0.6753(6) \\ \hline
\end{tabular}
\caption{Various overlaps versus $N$, as described in Section~\ref{subsection_oddN_confinement},
  at the couplings shown, on $36^244$ lattices for $N\geq 5$ and on $36\times 52\times 64$ for $N=3$.}
	\label{table:overlaps_soNodd}
\end{table}


\begin{table}[h]
\centering
\begin{tabular}{ |ccccc|}
  \hline
 $L_xL_yL_t$ &  $\beta$ & $\frac{1}{N}\text{tr}(U_p)$ & $am_{P}$ & $a\surd\sigma$\\
  \hline
$30.46.60$  &6.5	&  0.8335266	&  0.5846(87)  & 0.1417(10) \\
$34.52.64$  &7.0 	&  0.8465520	&  0.5499(103) & 0.1290(12) \\
$42.62.80$  &8.5	&  0.8755635	&  0.4370(82)  & 0.1035(10) \\
$46.70.86$  &9.0	&  0.8829150	&  0.4331(100) & 0.0983(11) \\
$50.76.90$  &10.0	&  0.8952681	&  0.3532(68)  & 0.0853(8) \\
$54.80.100$ &11.0	&  0.9052486	&  0.3120(71)  & 0.0772(9) \\
$62.90.110$ &12.0	&  0.9134837	&  0.2913(87)  & 0.0696(10) \\
  \hline
\end{tabular}
\caption{$SO(3)$ average plaquette values, masses of flux loops winding around the $x$-torus, and resulting string tensions
  on the lattices shown.}
	\label{table_string:so3}
\end{table}

\vspace{0.8cm}

\begin{table}[h]
\centering
\begin{tabular}{ |ccccc|}
  \hline
  	$L_s^2L_t$ 	&$\beta$ &$\tfrac{1}{N}\text{tr}(U_p)$ &$am_{P}$&$a\surd\sigma$\\
  \hline
  	$20^228$ 	&11.0	&0.80135	&0.8229(63)	&0.2061(8)\\
	$24^232$	&12.2	&0.82295	&0.7810(46)	&0.1829(5)\\
	$28^236$	&13.7	&0.84402	&0.6960(41)	&0.1598(5)\\
	$32^240$	&15.1	&0.85955	&0.6336(51)	&0.1425(6)\\
	$36^244$	&16.5	&0.87223	&0.5858(39)	&0.1292(5)\\
	$40^248$	&18.7	&0.88808	&0.4946(27)	&0.1127(4)\\
  \hline
\end{tabular}
	\caption{$SO(4)$ average plaquette values, flux loop masses, and string tensions.}
	\label{table_string:so4}
\end{table}

\vspace{0.8cm}

\begin{table}[h]
\centering
\begin{tabular}{ |ccccc|}
  \hline
  	$L_s^2L_t$ 	&$\beta$ &$\tfrac{1}{N}\text{tr}(U_p)$ &$am_{P}$&$a\surd\sigma$ \\
  \hline
  	$26^230$ 	&17.5	&0.79176	&1.4282(88)	&0.23603(72) \\
	$30^236$	&20.0	&0.82045	&1.1955(55)	&0.20109(46) \\
	$36^240$	&23.5	&0.84929	&0.9793(44)	&0.16616(37) \\
	$42^244$	&27.5	&0.87257	&0.7857(85)	&0.13786(73) \\
	$52^256$	&32.0	&0.89138	&0.6962(49)	&0.11655(40) \\
	$58^264$	&36.0	&0.90396	&0.6017(70)	&0.10262(59) \\
  \hline
\end{tabular}
	\caption{$SO(5)$ average plaquette values, flux loop masses, and string tensions.}
	\label{table_string:so5}
\end{table}

\vspace{0.8cm}

\begin{table}[h!]
\centering
\begin{tabular}{ |ccccc|}
  \hline
  	$L_s^2L_t$ 	&$\beta$ &$\tfrac{1}{N}\text{tr}(U_p)$ &$am_{P}$&$a\surd\sigma$\\
  \hline
$20^228$	&28.0	&0.80677	&0.9798(18)		&0.22431(20)\\
$24^232$	&33.0	&0.83851	&0.7938(27)		&0.18438(30)\\
$32^240$	&41.0	&0.87194	&0.6452(25)		&0.14381(27)\\
$36^244$	&46.0	&0.88656	&0.5620(19)		&0.12657(21)\\
$42^248$	&54.0	&0.90406	&0.4623(15)		&0.10634(17)\\
$46^254$	&60.0	&0.91400	&0.4004(17)		&0.09463(19)\\
  \hline
\end{tabular}
	\caption{$SO(6)$ average plaquette values, flux loop masses, and string tensions.}
	\label{table_string:so6}
\end{table}
\vspace{0.8cm}
\begin{table}[h!]
\centering
\begin{tabular}{ |ccccc|}
  \hline
  	$L_s^2L_t$ 	&$\beta$ &$\tfrac{1}{N}\text{tr}(U_p)$ &$am_{P}$&$a\surd\sigma$\\
  \hline
$16^224$	&35.0	&0.78116	&1.0714(9)  &0.26275(11)\\
$20^228$	&42.0	&0.82117	&0.8584(10) &0.21035(12)\\
$24^232$	&49.0	&0.84862	&0.7143(23) &0.17517(27)\\
$28^236$	&57.0	&0.87112	&0.5925(14) &0.14778(17)\\
$32^240$	&64.0	&0.88592	&0.5245(11) &0.13004(13)\\
$36^244$	&70.0	&0.89613	&0.4832(17) &0.11761(20)\\
  \hline
\end{tabular}
	\caption{$SO(7)$ average plaquette values, flux loop masses, and string tensions.}
	\label{table_string:so7}
\end{table}

\begin{table}
\centering
\begin{tabular}{ |ccccc|}
  \hline
  	$L_s^2L_t$ 	&$\beta$ &$\tfrac{1}{N}\text{tr}(U_p)$ &$am_{P}$&$a\surd\sigma$\\
  \hline
$16^224$	&51.0	&0.80206	&0.8721(24) &0.23789(31)\\
$20^228$	&62.0	&0.84000	&0.6888(11) &0.18914(15)\\
$24^232$	&73.0	&0.86561	&0.5672(9)  &0.15672(12)\\
$28^236$	&84.0	&0.88409	&0.4841(15) &0.13405(20)\\
$32^240$	&94.0	&0.89696	&0.4323(11) &0.11845(14)\\
$36^244$	&105.0	&0.90815	&0.3831(10) &0.10513(13)\\
  \hline
\end{tabular}
	\caption{$SO(8)$ average plaquette values, flux loop masses, and string tensions.}
	\label{table_string:so8}
\end{table}

\begin{table}
\centering
\begin{tabular}{ |ccccc|}
  \hline
  	$L_s^2L_t$ 	&$\beta$ &$\tfrac{1}{N}\text{tr}(U_p)$ &$am_{P}$&$a\surd\sigma$\\
  \hline
$16^224$	&132.0	&0.82178	&0.7263(13)	&0.21791(19)\\
$20^228$	&155.0	&0.85007	&0.6290(13)	&0.18107(18)\\
$24^232$	&175.0	&0.86820	&0.5760(14)	&0.15788(18)\\
$28^236$	&200.0	&0.88546	&0.4999(16)	&0.13614(21)\\
$32^240$	&225.0	&0.89871	&0.4409(19)	&0.11958(25)\\
$36^248$	&250.0	&0.90920	&0.3954(15)	&0.10675(19)\\
  \hline
\end{tabular}
	\caption{$SO(12)$ average plaquette values, flux loop masses, and string tensions.}
	\label{table_string:so12}
\end{table}

\begin{table}
\centering
\begin{tabular}{ |ccccc|}
  \hline
  	$L_s^2L_t$ 	&$\beta$ &$\tfrac{1}{N}\text{tr}(U_p)$ &$am_{P}$&$a\surd\sigma$\\
  \hline
$16^224$	&247.0	&0.82758	&0.6960(18)	&0.21353(26)\\
$20^228$	&302.0	&0.86092	&0.5450(15)	&0.16909(22)\\
$24^232$	&353.0	&0.88192	&0.4627(20)	&0.14216(29)\\
$28^236$	&408.0	&0.89848	&0.3940(18)	&0.12147(26)\\
$32^240$	&456.0	&0.90954	&0.3524(13)	&0.10740(19)\\
$36^248$	&512.0	&0.91974	&0.3113(14)	&0.09519(20)\\
  \hline
\end{tabular}
	\caption{$SO(16)$ average plaquette values, flux loop masses, and string tensions.}
	\label{table_string:so16}
\end{table}

\begin{table}
\centering
\begin{tabular}{ |ccc|cc|}
  \hline
  \multicolumn{3}{|c}{$SO(N)$} & \multicolumn{2}{|c|}{$SU(N)$} \\ \hline
$N$ & $\surd\sigma/(g^2N)$ & $\bar{\chi}_{\text{dof}}^2$ &
$N$ & $\surd\sigma/(2g^2N)$ \\
  \hline
3       &0.04273(68)    &0.66 & 2 & 0.08373(6)  \\    
4	&0.06019(54)	&0.38 & 3 & 0.09195(8) \\
5       &0.06763(36)    &0.72 & 4 & 0.09479(8) \\    
6	&0.07339(15)	&0.45 & 6 & 0.09665(6) \\
7	&0.07686(10)	&1.09 & 8 & 0.09743(8) \\
8	&0.07991(14)	&0.70 & 12 & 0.09779(12) \\
12	&0.08625(19)	&0.06 & 16 & 0.09775(14) \\
16	&0.08898(22)	&0.98 & & \\ \hline
$\infty$ &0.09897(25)$^a$   &1.76 & $\infty$ & 0.09818(6) \\
         &0.09821(57)$^b$   &1.67 & &  \\
         &0.09749(43)$^c$   &2.32 & &  \\
  \hline
\end{tabular}
\caption{$SO(N)$ continuum string tensions, $\surd\sigma$, in units of the 't Hooft coupling,
  $g^2N$, and the $\chi^2$ per degree of freedom of each continuum extrapolation. (Labels
  $a$,$b$,$c$ denote different extrapolations: $O(1/N)$ extrapolation for $N\geq 4$, 
  $O(1/N^2)$ extrapolation for $N\geq 4$,$O(1/N)$ extrapolation for $N\geq 3$ respectively.)
  Also shown are $SU(N)$ values from \cite{AAMT_SUN}, in units of $2g^2N$.}
	\label{table_string:continuum}
\end{table}

\clearpage

\begin{table}
\centering
\begin{tabular}{|lc|lllllll|}
  \hline
$L_s^2L_t$ &	&$54^240$  &$62^248$  &$74^260$  &$82^264$  &$90^270$  &$100^280$  &$88^290$  \\
$\beta$	&	&6.5	&7.0	&8.5	&9.0	&10.0	&11.0  &12.0 \\
  \hline
$l\surd\sigma$ & &  8.0  &  8.4 &  7.9 & 8.3 & 8.0 & 8.1  & 6.5 \\
	\hline
$0^+$	        & $\alpha$	& 0.434(3)  & 0.397(4) & 0.3185(25) & 0.2996(22) & 0.2650(21) & 0.2421(15) & 0.218(2) \\
$0^{+*}$	& $\alpha$	& 0.652(11)   & 0.595(8) & 0.4652(65) & 0.4364(42) & 0.4003(45) & 0.3540(49) & 0.318(6) \\
$0^{+**}$	& $\beta$	& 0.768(33)   & 0.706(23)  & 0.587(12)  & 0.5595(57) & 0.484(10)  & 0.4366(73) & -- \\
$0^{+***}$	& $\gamma$	& 0.882(20)   & 0.782(32)  & 0.663(15)  & 0.631(11)  & 0.553(16)  & 0.492(12) & -- \\
$0^{+****}$	& $\delta$	& 0.962(12)   & 0.869(20)  & 0.7081(84) & 0.636(14)  & 0.591(11)  & 0.5143(96) & -- \\

$2^{+}$		& $\alpha$	&  0.734(11)  & 0.666(5) & 0.5362(45) & 0.472(8)   & 0.4504(47) & 0.3918(91) & -- \\
$2^{+*}$	& $\gamma$	&  0.839(22)  & 0.809(14)  & 0.606(28)  & 0.584(16)  & 0.542(10)  & 0.478(11)  & -- \\
$2^{-}$		& $\alpha$	&  0.728(11)  & 0.673(9) & 0.5282(66) & 0.4898(53) & 0.4438(44) & 0.3986(30) & 0.364(4) \\
$2^{-*}$	& $\gamma$	&  0.857(22)  & 0.785(12)  & 0.6284(59) & 0.581(20)  & 0.5464(81) & 0.485(11)  & 0.440(8) \\

$0^{-}$		& $\delta$	&  0.966(32) & 0.845(19) & 0.715(23) & 0.661(27) & 0.581(17) & 0.5172(77) & 0.472(7) \\

$1^{+}$		& $\phi$	&  1.012(58) & 0.979(31) & 0.793(15) & 0.741(24) & 0.650(14) & 0.517(27) & 0.514(20) \\
$1^{-}$		& $\phi$	&  1.020(52) & 0.838(57) & 0.818(13) & 0.670(46) & 0.669(29) & 0.527(25)  & 0.508(18) \\
  \hline
\end{tabular}
	\caption{$SO(3)$ glueball masses $am_G$.Grades, e.g.$\alpha$, explained in Section~\ref{subsubsection_grades}.}
	\label{table_glueball:so3}
\end{table}

\begin{table}
\centering
\begin{tabular}{|lc|llllll|}
  \hline
$L_s^2L_t$ &		&$34^242$ 	&$38^246$ 	&$44^252$ 	&$48^256$	&$54^262$	&$62^270$\\
$\beta$	 &	&11.0	&12.2		&13.7		&15.1		&16.5		&18.7\\
  \hline
$l\surd\sigma$ & &7.2 &7.0 &7.1 &6.9 &7.0 &7.0 \\
	\hline
$0^+$	&  $\alpha$	&0.6963(18) &0.6170(25) &0.5394(23) &0.4801(35) &0.4308(26) &0.3787(23)\\
$0^{+*}$&  $\alpha$	&1.045(10) &0.9137(76) &0.7970(68) &0.7116(55) &0.6354(92) &0.5733(75)\\
$0^{+**}$&  $\gamma$	&1.286(26) &1.123(16) &1.028(16) &0.952(28) &0.806(20) &0.716(11)\\
$0^{+***}$&  $\delta$	&1.462(46) &1.258(39) &1.143(23) &1.020(20) &0.943(17) &0.8249(88)\\
$0^{+****}$&  $\phi$	&1.547(72) &1.288(42) &1.105(25) &1.028(20) &0.955(14) &0.841(12)\\

$2^{+}$&  $\beta$		&1.188(15) &1.018(13) &0.9099(87) &0.8200(63) &0.734(13) &0.6390(86)\\
$2^{+*}$&  $\delta$	&1.426(50) &1.268(32) &1.091(20) &0.987(17) &0.897(31) &0.794(18)\\
$2^{-}$	&  $\beta$	&1.186(13) &1.035(14) &0.8997(89) &0.8132(88) &0.724(14) &0.6371(71)\\
$2^{-*}$&  $\delta$	&1.355(45) &1.187(31) &1.090(22) &0.988(21) &0.923(36) &0.773(21)\\

$0^{-}$&  $\gamma$		&1.522(70) &1.297(38) &1.188(27) &1.015(21)&0.908(15) &0.8374(95)\\

$1^{+}$&  $\phi$		&1.60(9) &1.41(8) &1.412(47) &1.175(26) &1.067(19) &0.968(11)\\
$1^{-}$&  $\phi$		&1.56(13) &1.43(7) &1.336(54) &1.176(27) &1.106(20) &0.945(10)\\
  \hline
\end{tabular}
	\caption{$SO(4)$ glueball masses $am_G$. Grades, e.g.$\alpha$, explained in Section~\ref{subsubsection_grades}.}
	\label{table_glueball:so4}
\end{table}

\begin{table}
\centering
\begin{tabular}{|lc|llllll|}
  \hline
$L_s^2L_t$ & 	&$26^230$  &$30^236$  &$36^240$  &$42^244$  &$52^256$  &$58^264$  \\
$\beta$	 &	&17.5	&20.0	&23.5	&27.5	&32.0	&36.0 \\
  \hline
$l\surd\sigma$  & & 6.1  & 6.0  & 6.0  & 5.8  & 6.1 & 6.0  \\
	\hline
$0^+$	&  $\alpha$	& 0.8354(29) & 0.7087(20) & 0.5878(22) & 0.4889(22) & 0.4146(22) & 0.3624(13)  \\
$0^{+*}$&  $\beta$	& 1.242(14)  & 1.065(11) & 0.861(22) & 0.7133(83) & 0.6145(51) & 0.5412(46)  \\
$0^{+**}$&  $\gamma$	& 1.590(60)  & 1.340(22)  & 1.037(36)  & 0.927(21) & 0.787(11) & 0.6917(77)  \\
$0^{+***}$&  $\gamma$	& 1.623(70)  & 1.430(37)  & 1.191(16)  & 1.005(12) & 0.850(14) & 0.732(10)  \\
$0^{+****}$&  $\phi$	& 1.78(12)   & 1.489(43) & 1.257(22)  & 1.031(14)  & 0.8998(79) & 0.7970(59)  \\

$2^{+}$	&  $\beta$	& 1.384(24)  & 1.182(16) & 0.9804(94)  & 0.8184(59) & 0.6944(44) & 0.6153(32)  \\
$2^{+*}$&  $\delta$	& 1.644(89) & 1.414(30)   & 1.209(18)  & 0.991(26)  & 0.808(15)  & 0.7275(90)  \\
$2^{-}$	&  $\beta$	& 1.394(28) & 1.200(18)  & 0.933(33)  & 0.802(12) & 0.693(10)  & 0.6110(65)  \\
$2^{-*}$&  $\delta$	& 1.710(14) & 1.439(7)   & 1.174(23) & 1.012(12) & 0.8447(46)  & 0.7518(42)  \\

$0^{-}$&  $\delta$		& 1.80(11) & 1.524(40) & 1.284(31) & 0.998(35) & 0.893(18) & 0.8033(92)  \\

$1^{+}$	&  $\phi$	& 1.95(15)  & 1.653(72) & 1.475(50) & 1.193(18) & 0.988(11) & 0.8979(81)  \\
$1^{-}$	&  $\phi$	& 1.85(15)  & 1.739(71) & 1.22(17) & 1.118(45) & 0.971(30) & 0.877(12)  \\ 
  \hline
\end{tabular}
	\caption{$SO(5)$ glueball masses $am_G$.Grades, e.g.$\alpha$, explained in Section~\ref{subsubsection_grades}.}
	\label{table_glueball:so5}
\end{table}

\begin{table}
\centering
\begin{tabular}{|lc|llllll|}
  \hline
$L_s^2L_t$ 	&	&$26^228$	&$30^232$ 	&$42^244$ 	&$46^248$ 	&$56^260$ 	&$62^270$\\
$\beta$		&	&28.0		&33.0		&41.0		&46.0		&54.0		&60.0\\
  \hline
$l\surd\sigma$ & &5.8 &5.5 &6.0 &5.8 &6.0 &5.9 \\
	\hline
$0^+$	        & $\alpha$	& 0.8212(20) & 0.6756(22)  & 0.5258(19)  & 0.4633(24)  & 0.3893(18)  &  0.3454(17) \\ 
$0^{+*}$	& $\beta$	& 1.232(13) & 1.021(11)  & 0.7938(67)  & 0.7030(53)  & 0.5959(31)  &  0.5255(33) \\
$0^{+**}$	& $\gamma$	& 1.583(37) & 1.291(20)  & 1.026(12)  &  0.8965(91) & 0.7553(45)  & 0.6823(48)  \\
$0^{+***}$	& $\delta$	& 1.631(43) & 1.346(21)  & 1.041(13)  &  0.948(10) & 0.7931(59)  & 0.7023(36)  \\
$0^{+****}$	& $\phi$	& 1.651(54) & 1.415(32)  & 1.092(52)  &  0.933(27) & 0.844(15)  & 0.727(19)  \\

$2^{+}$		& $\beta$	& 1.352(19) & 1.124(14)  & 0.8767(78)  & 0.7718(51)  & 0.6546(44)  & 0.5852(36)  \\
$2^{+*}$	& $\gamma$	& 1.667(51) & 1.372(31)  & 1.082(15)  & 0.907(27)  & 0.800(16)  & 0.708(9)  \\

$2^{-}$		& $\beta$	& 1.370(24) & 1.137(12)  & 0.8805(93)  & 0.7729(63)  & 0.6534(32)  & 0.5813(36)  \\
$2^{-*}$	& $\gamma$	& 1.653(50) & 1.364(26)  & 1.064(10)  & 0.930(25)  & 0.791(14)  & 0.706(8)  \\

$0^{-}$		& $\delta$	& 1.717(67) & 1.506(40)  & 1.134(22)  & 1.025(11)  & 0.827(14)  & 0.770(9)  \\

$1^{+}$		& $\phi$	& 1.835(83) & 1.646(66)  & 1.259(27)  & 1.113(16)  & 0.905(26)  & 0.835(15)  \\
$1^{-}$		& $\phi$	& 1.84(11) & 1.635(60)  & 1.250(29)  & 1.115(14)  & 0.914(19)  & 0.826(14)  \\
  \hline
\end{tabular}
	\caption{$SO(6)$ glueball masses $am_G$.Grades, e.g.$\alpha$, explained in Section~\ref{subsubsection_grades}.}
	\label{table_glueball:so6}
\end{table}

\clearpage

\begin{table}
\centering
\begin{tabular}{|lc|llllll|}
  \hline
$L_s^2L_t$  &		&$16^224$	&$20^228$	&$24^232$ 	&$28^236$ 	&$32^240$ 	&$36^244$\\
$\beta$	   &	&35.0		&42.0		&49.0		&57.0		&64.0		&70.0\\
  \hline
$l\surd\sigma$ & &4.2 &4.2 &4.2 &4.1 &4.2 &4.2 \\
	\hline
$0^+$	  & $\alpha$ 	&0.9823(14)&0.7865(10)&0.6578(16)&0.5518(22)&0.4842(16)&0.4394(20)\\ 
$0^{+*}$   & $\beta$ &1.487(14)&1.198(8)&1.001(9)&0.8329(57)&0.7332(21)&0.6759(40)\\
$0^{+**}$  & $\beta$ &1.907(44)&1.512(27)&1.262(14)&1.080(7)&0.9522(47)&0.8731(60)\\
$0^{+***}$ & $\gamma$	&1.958(66)&1.546(24)&1.301(15)&1.1230(76)&0.9776(48)&0.8876(49)\\
$0^{+****}$ & $\phi$ &2.039(80)&1.646(28)&1.381(24)&1.187(9)&1.052(7)&0.956(9)\\

$2^{+}$	 & $\beta$ 	&1.623(19)&1.310(13)&1.064(24)&0.915(15)&0.8212(70)&0.7352(71)\\
$2^{+*}$  & $\gamma$ 	&1.977(70)&1.594(24)&1.287(54)&1.154(34)&0.991(11)&0.901(12)\\

$2^{-}$	 & $\beta$ 	&1.628(29)&1.240(39)&1.097(22)&0.928(8)&0.824(7)&0.737(9)\\
$2^{-*}$  & $\delta$ 	&2.004(74)&1.573(24)&1.330(16)&1.122(29)&0.996(13)&0.918(14)\\

$0^{-}$	 & $\delta$ 	&1.95(9)&1.655(27)&1.350(53)&1.202(52)&1.024(12)&0.934(21)\\

$1^{+}$	 & $\phi$ 	&2.31(19)&1.951(59)&1.649(31)&1.375(44)&1.167(24)&1.081(20)\\
$1^{-}$	 & $\phi$ 	&2.15(15)&1.904(60)&1.604(32)&1.383(62)&1.135(21)&1.050(28)\\
  \hline
\end{tabular}
	\caption{$SO(7)$ glueball masses $am_G$. Grades, e.g.$\alpha$, explained in Section~\ref{subsubsection_grades}.}
	\label{table_glueball:so7}
\end{table}

\begin{table}
\centering
\begin{tabular}{|lc|llllll|}
  \hline
$L_s^2L_t$ &	&$16^224$	&$20^228$	&$24^232$ 	&$28^236$ 	&$32^240$ 	&$36^244$\\
$\beta$	  &	&51.0		&62.0		&73.0		&84.0		&94.0		&105.0\\
  \hline
$l\surd\sigma$ & &3.8 &3.8 &3.8 &3.7 &3.8 &3.8 \\
	\hline
$0^+$	        & $\alpha$	&0.901(4)&0.7160(19)&0.5959(14)&0.5057(19)&0.4493(21)&0.3973(18)\\ 
$0^{+*}$	& $\alpha$	&1.370(3)&1.0833(74)&0.9140(26)&0.7708(38)&0.6890(30)&0.6108(30)\\
$0^{+**}$	& $\gamma$	&1.681(39)&1.381(17)&1.072(28)&0.9916(24)&0.8704(59)&0.7576(96)\\
$0^{+***}$	& $\gamma$	&1.791(50)&1.433(22)&1.175(12)&1.003(10)&0.898(8)&0.8009(52)\\
$0^{+****}$	& $\phi$	&1.836(66)&1.562(30)&1.281(17)&1.098(11)&0.9750(60)&0.861(13)\\

$2^{+}$		& $\gamma$	&1.502(20)&1.209(11)&1.007(9)&0.8593(72)&0.7555(49)&0.6831(30)\\
$2^{+*}$	& $\gamma$	&1.748(58)&1.434(27)&1.139(33)&1.0573(91)&0.9371(60)&0.8343(55)\\

$2^{-}$		& $\beta$	&1.494(19)&1.202(12)&0.994(7)&0.8520(54)&0.7464(73)&0.6645(59)\\
$2^{-*}$	& $\delta$	&1.877(55)&1.433(27)&1.210(16)&1.053(8)&0.9466(59)&0.8466(55)\\

$0^{-}$		& $\delta$	&1.797(65)&1.537(36)&1.206(53)&1.095(12)&0.978(9)&0.8725(53)\\

$1^{+}$		& $\phi$	&2.17(13)&1.790(50)&1.39(8)&1.204(49)&1.090(27)&0.971(18)\\
$1^{-}$		& $\phi$	&2.21(16)&1.760(61)&1.467(20)&1.270(13)&1.114(10)&0.993(8)\\
 \hline
\end{tabular}
	\caption{$SO(8)$ glueball masses $am_G$. Grades, e.g.$\alpha$, explained in Section~\ref{subsubsection_grades}.}
	\label{table_glueball:so8}
\end{table}

\begin{table}
\centering
\begin{tabular}{|lc|llllll|}
  \hline
$L_s^2L_t$ &	&$16^224$	&$20^228$	&$24^232$ 	&$28^236$ 	&$32^240$ 	&$36^244$\\
$\beta$	  &	&132.0		&155.0		&175.0		&200.0		&225.0		&250.0\\
  \hline
$l\surd\sigma$ & &3.5 &3.6 &3.8 &3.8 &3.8 &3.8 \\  
	\hline
$0^+$	        & $\alpha$	&0.8537(29)&0.7057(23)&0.6170(40)&0.5288(41)&0.4711(42)&0.4128(30)\\ 
$0^{+*}$	& $\beta$	&1.270(18)&1.081(12)&0.951(7)&0.813(8)&0.7209(31)&0.6455(28)\\
$0^{+**}$	& $\gamma$	&1.641(14)&1.362(10)&1.207(5)&1.053(6)&0.9246(37)&0.8127(59)\\
$0^{+***}$	& $\phi$	&1.683(13)&1.424(9)&1.236(6)&1.051(13)&0.9204(93)&0.801(16)\\
$0^{+****}$	& $\phi$	&1.67(10)&1.556(9)&1.317(31)&1.182(6)&0.997(14)&0.916(10)\\

$2^{+}$		& $\beta$	&1.392(28)&1.187(17)&1.036(11)&0.871(10)&0.7833(61)&0.7084(53)\\
$2^{+*}$	& $\delta$	&1.753(16)&1.4602(66)&1.2789(53)&1.110(7)&0.965(12)&0.8655(81)\\

$2^{-}$		& $\beta$	&1.432(8)&1.2013(52)&1.0450(39)&0.886(10)&0.7871(77)&0.7067(65)\\
$2^{-*}$	& $\delta$	&1.772(13)&1.414(35)&1.2686(56)&1.1145(73)&0.974(11)&0.8691(72)\\

$0^{-}$		& $\delta$	&1.73(10)&1.460(40)&1.345(27)&1.157(21)&0.936(26)&0.873(24)\\

$1^{+}$		& $\phi$	&2.112(25)&1.762(14)&1.481(37)&1.300(26)&1.143(13)&0.997(22)\\
$1^{-}$		& $\phi$	&2.105(22)&1.749(13)&1.475(37)&1.295(23)&1.139(17)&1.001(23)\\
  \hline
\end{tabular}
	\caption{$SO(12)$ glueball masses $am_G$. Grades, e.g.$\alpha$, explained in Section~\ref{subsubsection_grades}.}
	\label{table_glueball:so12}
\end{table}

\begin{table}
\centering
\begin{tabular}{|lc|llllll|}
  \hline
$L_s^2L_t$ &	&$16^224$ 	&$20^228$	&$24^232$ 	&$28^236$ 	&$32^240$ 	&$36^244$\\
$\beta$	&	&247.0	&302.0		&353.0		&408.0		&456.0		&512.0\\
  \hline
$l\surd\sigma$ & &3.4 &3.4 &3.4 &3.4 &3.4 &3.4 \\
	\hline
$0^+$	        & $\alpha$	&0.8480(26)&0.6622(44)&0.5671(29)&0.4827(21)&0.4268(21)&0.3779(17)\\
$0^{+*}$	& $\beta$	&1.262(23)&1.006(13)&0.852(11)&0.7398(32)&0.6565(22)&0.5794(35)\\
$0^{+**}$	& $\gamma$	&1.560(61)&1.079(69)&1.068(19)&0.959(11)&0.8130(63)&0.7246(50)\\
$0^{+***}$	& $\delta$	&1.693(74)&1.233(26)&1.087(16)&0.956(11)&0.8549(91)&0.745(13)\\
$0^{+****}$	& $\delta$	&1.59(11)&1.416(33)&1.233(27)&1.066(16)&0.902(25)&0.814(18)\\

$2^{+}$		& $\beta$	&1.421(8)&1.1310(54)&0.935(15)&0.812(6)&0.7176(57)&0.6402(41)\\
$2^{+*}$	& $\gamma$	&1.720(14)&1.390(9)&1.151(27)&0.987(12)&0.869(9)&0.7833(55)\\

$2^{-}$		& $\beta$	&1.412(9)&1.134(6)&0.9529(53)&0.8065(65)&0.7123(47)&0.6411(49)\\
$2^{-*}$	& $\delta$	&1.601(90)&1.387(37)&1.149(20)&1.003(12)&0.872(8)&0.7945(71)\\

$0^{-}$		& $\delta$	&1.569(90)&1.453(11)&1.236(8)&1.044(5)&0.924(13)&0.8187(60)\\

$1^{+}$		& $\phi$	&1.88(18)&1.594(60)&1.356(32)&1.169(17)&0.996(29)&0.911(21)\\
$1^{-}$		& $\phi$	&2.09(20)&1.659(74)&1.208(85)&1.160(44)&1.030(33)&0.905(21)\\
  \hline
\end{tabular}
	\caption{$SO(16)$ glueball masses $am_G$. Grades, e.g.$\alpha$, explained in Section~\ref{subsubsection_grades}.}
	\label{table_glueball:so16}
\end{table}

\begin{table}
\centering
\begin{tabular}{|l|ll|ll|ll|ll|}
  \hline
  \multicolumn{9}{|c|}{$M_G/\surd\sigma$}\\ \hline
  $J^P$ 	&$SO(3)$ &$\bar{\chi}^2_\text{dof}$ &$SO(4)$ &$\bar{\chi}^2_\text{dof}$ &$SO(5)$ & $\bar{\chi}^2_\text{dof}$ & $SO(6)$ & $\bar{\chi}^2_\text{dof}$ \\	\hline

$0^+$		& 3.132(34) &0.43 &  3.343(23) &0.36 & 3.545(17) &0.42 & 3.656(13) & 0.08 \\
$0^{+*}$	& 4.558(70) &1.75 &  4.966(64) &1.06 & 5.249(46) &0.63 & 5.597(31) & 0.51\\
$0^{+**}$	& 5.81(15)  &0.19 &  6.49(13)  &1.75 & 6.760(92) &1.14 & 7.187(54) & 0.71\\
$0^{+***}$	& 6.56(18)  &0.26 &  7.47(14)  &0.36 & 7.30(11)  &0.64 & 7.487(54) & 1.02\\
$0^{+****}$	& 6.73(14)  &1.44 &  7.63(17)  &1.68 & 7.86(10)  &0.71 & 7.864(49) & 1.24\\

$2^{+}$		& 5.13(9)   &4.15  & 5.711(81) &1.36 & 6.008(46) &0.13 & 6.190(38) & 0.36\\
$2^{+*}$         & 6.30(16)  &1.81 & 7.02(20)  &0.23 & 7.07(12)  &1.15 & 7.49(11)  & 0.61\\

$2^{-}$		& 5.16(7)   &1.41  & 5.598(81) &0.44 & 5.919(71) &1.04 & 6.140(38) & 0.19\\
$2^{-*}$	& 6.40(12)  &1.02  & 7.22(22)  &0.60 & 7.301(48) &1.56 & 7.46(10)  & 0.06\\

$0^{-}$		& 6.79(14)  &0.43  &  7.39(15) &2.43 & 7.82(13)  &1.07 & 8.10(11)  & 1.93\\

$1^{+}$		& 7.45(24)  &1.49  &  8.85(21) &1.73 & 8.75(14)  &1.34 & 8.89(17)  & 0.72\\
$1^{-}$		& 7.61(28)  &4.21  &  8.71(21) &0.70 & 8.55(18)  &0.95 & 8.78(17)  & 0.58\\
  \hline
\end{tabular}
\caption{Continuum glueball masses in string tension units, $M_G/\surd\sigma$,
  for $SO(3)$, $SO(4)$, $SO(5)$, $SO(6)$, with the $\chi^2$ per degree of freedom
  of the linear extrapolation. Reliability as graded in
  Tables~\ref{table_glueball:so3}-\ref{table_glueball:so6}.}
	\label{table_glueball:N3-6continuum}
\end{table}

\begin{table}
\centering
\begin{tabular}{|c|ll|ll|ll|ll|}
  \hline
  \multicolumn{9}{|c|}{ $M_G/\surd\sigma$} \\ \hline
  $J^P$ 	&$SO(7)$ &$\bar{\chi}^2_\text{dof}$ &$SO(8)$ &$\bar{\chi}^2_\text{dof}$ &$SO(12)$ & $\bar{\chi}^2_\text{dof}$ & $SO(16)$ & $\bar{\chi}^2_\text{dof}$ \\	\hline

$0^+$		&3.737(10)  &0.94  &3.788(14) &0.92  &3.878(24) &0.73  &3.973(15) &1.22\\
$0^{+*}$	&5.655(25)  &2.42  &5.773(43) &1.23  &6.096(38) &0.45  &6.178(38) &0.45\\
$0^{+**}$	&7.419(57)  &0.93  &7.389(70) &2.60  &7.791(45) &1.89  &7.725(88) &3.14\\
$0^{+***}$	&7.610(60)  &0.96  &7.614(68) &0.61  &7.53(11)  &0.57  &8.30(15)  &0.80\\
$0^{+****}$	&8.237(83)  &0.15  &8.298(94) &0.48  &8.43(12)  &1.01  &8.90(20)  &1.42\\

$2^{+}$		&6.297(54)  &0.72  &6.498(36) &1.21  &6.636(64) &1.60  &6.714(40) &0.41\\
$2^{+*}$         &7.68(11)  &0.35  &8.112(72) &1.43  &8.165(65) &0.27  &8.218(63) &1.10\\
  
$2^{-}$		&6.348(60)  &0.99  &6.346(48) &0.28  &6.622(48) &0.88  &6.706(39) &1.26\\
$2^{-*}$	&7.79(12)   &0.45  &8.144(70) &1.98  &8.33(11)  &0.98  &8.40(11)  &1.28\\

$0^{-}$		&7.98(14)   &0.45  &8.450(81) &0.65  &8.28(24)  &2.49  &8.60(8)  &0.69\\

$1^{+}$		&9.09(21)   &0.77  &9.16(20)  &0.42  &9.43(13)  &0.72  &9.45(30) &0.81\\
$1^{-}$		&8.90(23)   &1.50  &9.49(13)  &0.13  &9.41(14)  &0.44  &9.67(30) &0.47\\
  \hline
\end{tabular}
\caption{Continuum glueball masses in string tension units, $M_G/\surd\sigma$,
  for $SO(7)$, $SO(8)$, $SO(12)$, $SO(16)$, with the $\chi^2$ per degree of freedom
of the linear extrapolation. Reliability as graded in
  Tables~\ref{table_glueball:so7}-\ref{table_glueball:so16}.}
	\label{table_glueball:N7-16continuum}
\end{table}

\begin{table}
\centering
\begin{tabular}{ |ccc|}
  \hline
  \multicolumn{3}{|c|}{$SO(N)$} \\ \hline
  $N$ & $M_{0^+}/(g^2N)$ & $\bar{\chi}_{\text{dof}}^2$ \\  \hline
3       &0.1357(14)     & 0.59   \\    
4	&0.2008(18)     & 0.92  \\
5       &0.2397(13)     & 0.32  \\    
6	&0.2679(14)     & 0.21  \\
7	&0.2869(11)	& 0.34 \\
8	&0.3024(18)	& 0.77 \\
12	&0.3330(32)	& 0.71 \\
16	&0.3542(20)	& 1.61 \\ \hline
$\infty$ &0.4017(14)    & 0.81  \\ \hline
\end{tabular}
\caption{Lightest $SO(N)$ glueball, $M_{0^+}$, in units of the 't Hooft coupling, $g^2N$, and the $\chi^2$ per degree of freedom of each continuum extrapolation. Also shown is the large-$N$ extrapolation.}
	\label{table_m0pggn_N}
\end{table}

\begin{table}
\centering
\begin{tabular}{|cr||lll|lll||l|}
  \hline
  \multicolumn{9}{|c|}{ $M_G/\surd\sigma$ } \\ \hline
  $J^P$ & & \multicolumn{3}{|c|}{ $O(1/N^2)$ } & \multicolumn{3}{|c|}{ $O(1/N)$ } & $SU(\infty)$ \\ \hline
  & & $SO(\infty)$ & $\bar{\chi}^2_\text{dof}$ & $N\geq$ & $SO(\infty)$ & $\bar{\chi}^2_\text{dof}$ & $N\geq$ & \\	\hline
$0^{+}$	   & $\alpha$   &  4.150(33) & 1.34  & 3  & 4.179(16) & 1.30   &  3  &  4.116(6)  \\
$0^{+*}$    & $\beta$	&  6.628(68) & 2.07  & 3  & 6.578(35) & 2.07   &  3  &  6.308(10)  \\
$0^{+**}$   & $\gamma$	&  7.93(16)  & 1.44  & 3  & 8.276(81) & 2.40   &  3  &  7.844(14)  \\
$0^{+***}$  & $\phi$	&  8.12(15)  & 3.05  & 3  & 8.171(70) & 2.56   &  3  &  8.147(19)  \\
$0^{+****}$  & $\phi$	&  9.12(22)  & 1.45  & 3  & 9.11(11)  & 2.01   &  3  &  8.950(25) \\
$2^{+}$	    & $\beta$	&  6.987(88) & 0.87  & 3  & 7.129(43) & 1.47   &  3  &  6.914(13) \\
$2^{-}$     & $\beta$   &  7.044(84) & 0.55  & 3  & 7.090(40) & 0.51   &  3  &  6.930(13)  \\
$2^{+*}$    & $\gamma$	&  8.61(14)  & 3.63  & 3  & 8.77(7)   & 3.25   &  3  &  8.423(15) \\
$2^{-*}$    & $\delta$	&  9.14(19)  & 2.53  & 3  & 8.98(9)   & 2.35   &  3  &  8.488(21) \\
$0^{-}$	   & $\delta$	&  8.84(18)  & 1.46  & 3  & 9.11(10)  & 1.76   &  3  &  8.998(28)  \\
$1^{+}$   & $\phi$	&  9.53(34)  & 1.11  & 3  & 9.98(16)  & 1.35   &  3  &  9.912(26)  \\
$1^{-}$   & $\phi$	&  9.90(36)  & 2.02  & 3  & 10.10(16) & 1.70   &  3   &  9.886(27)  \\
  \hline
\end{tabular}
\caption{Continuum glueball masses in units of the string tension extrapolated
  to $N=\infty$, with the $\chi^2$ per degree of freedom and range in $N$: for the best linear
  and quadratic fits in $1/N$. Also the values for $SU(\infty)$ from \cite{AAMT_SUN}.
  Grades, e.g.$\alpha$, explained in Section~\ref{subsubsection_grades}.}
	\label{table_MK_largeN}
\end{table}

\begin{table}
\centering
\begin{tabular}{|c|llll|l|}
  \hline
  \multicolumn{6}{|c|}{ $M_G/M_{0^+}$} \\ \hline
  $J^P$ 	& $SO(\infty)$ & $c_1/c_0 \sim$ & $\bar{\chi}^2_\text{dof}$& $N\geq$ & $SU(\infty)$ \\	\hline
  
$0^{+*}$	& 1.585(12) & -0.27 & 1.62 & 3 & 1.533(4) \\
$0^{+**}$	& 1.966(24) & +0.27 & 1.60 & 3 & 1.906(5) \\
$0^{+***}$	& 1.983(24) & -0.10 & 2.80 & 3 & 1.979(6) \\
$0^{+****}$	& 2.189(31) & -0.02 & 1.91 & 3 & 2.174(7) \\
$2^{+}$		& 1.708(14) & -0.05 & 1.04 & 3 & 1.680(4) \\
$2^{-}$         & 1.704(13) & -0.09 & 0.49 & 3 & 1.684(4) \\
$2^{+\star }$	& 2.114(21) & -0.13 & 2.84 & 3 & 2.046(6)  \\
$2^{-\star }$	& 2.166(26) & -0.21 & 2.11 & 3 & 2.062(7)  \\
$0^{-}$		& 2.175(26) & +0.06 & 1.36 & 3 & 2.186(8) \\
$1^{+}$	        & 2.389(45) & +0.16 & 1.72 & 3 & 2.408(8) \\
$1^{-}$	        & 2.421(46) & +0.09 & 1.72 & 3 & 2.402(8) \\
  \hline
\end{tabular}
\caption{Continuum glueball masses in units of the scalar glueball mass extrapolated
  to $N=\infty$, with the $\chi^2$ per degree of freedom and range of the best linear
  fit $c_0+c_1/N$ shown. The values for $SU(N)$ are from \cite{AAMT_SUN}.
 Grades as in Table~\ref{table_MK_largeN}.}
	\label{table_glueball_largeN}
\end{table}

\begin{table}
\centering
\begin{tabular}{|c||ll|ll|l||ll|l|}
  \hline
  \multicolumn{9}{|c|}{ $M_G/M_{0^+}$} \\ \hline
  $J^P$ 	& $SO(3)$ & & $SO(4)$ & & $SU(2)$ &  $SO(6)$ & & $SU(4)$  \\	\hline
  
$0^{+*}$	& 1.455(27) & $\alpha$  & 1.485(22) & $\alpha$  & 1.449(4) & 1.531(10) & $\beta$  & 1.518(5) \\
$0^{+**}$	& 1.855(52) & $\beta$  & 1.941(41) & $\gamma$  & 1.770(4) & 1.966(17) & $\gamma$ & 1.869(7)  \\
$0^{+***}$	& 2.095(62) & $\gamma$  & 2.235(45) & $\delta$  & 1.959(4) & 2.048(17) & $\delta$ & 1.961(9)  \\
$0^{+****}$	& 2.149(51) & $\delta$  & 2.282(53) & $\phi$  & 2.050(5) & 2.151(16) & $\phi$ & 2.131(13)  \\
$2^{+}$		& 1.638(34) & $\alpha$  & 1.708(27) & $\beta$  & 1.639(3) & 1.693(12) & $\beta$  & 1.672(6)  \\
$2^{-}$         & 1.647(28) & $\alpha$  & 1.675(27) & $\beta$  & 1.646(3) & 1.679(12) & $\beta$  & 1.673(6)  \\
$2^{+ *}$       & 2.011(56) & $\gamma$  & 2.100(62) & $\delta$  & 1.923(5) & 2.048(30) & $\gamma$ & 2.027(8)  \\
$2^{- *}$       & 2.043(44) & $\gamma$  & 2.160(68) & $\delta$  & 1.926(6) & 2.040(27) & $\gamma$ & 2.011(9)  \\	
$0^{-}$	       & 2.168(51) & $\delta$  & 2.211(48) & $\gamma$  & 2.087(6) & 2.216(31) & $\delta$ & 2.170(9)  \\	
$1^{+}$	       & 2.379(81) & $\phi$  & 2.647(65) & $\phi$  & 2.228(7) & 2.432(47) & $\phi$  & 2.346(16)  \\
$1^{-}$	       & 2.430(93) & $\phi$  & 2.605(65) & $\phi$  & 2.225(7) & 2.402(47) & $\phi$  & 2.357(13)  \\ \hline
  \hline
\end{tabular}
\caption{Continuum glueball massesin units of the mass gap: comparing  the spectra of $SO(N)$ gauge theories
  with those of $SU(N^\prime)$ with the same Lie algebra. Grades, e.g.$\alpha$, explained in Section~\ref{subsubsection_grades}.}
	\label{table_glueball_SONvsSUM}
\end{table}

\clearpage

\begin{figure}[htb]
         \centering
\begin{subfigure}{0.3\textwidth}
  \centering
	\begin{tikzpicture}
	  [nodewitharrow/.style 2 args={decoration={markings,mark=at position {#1} with {\arrow{stealth},\node[transform shape,above] {#2};}},postaction={decorate,ultra thick}}]
          
	\foreach \x in {0,1,2,3,4}
	\draw[black!25,very thin] (\x,-0.3)--(\x,4.3);
	\foreach \y in {0,1,2,3,4}
	\draw[black!25,very thin] (-0.3,\y)--(4.3,\y);	

	\draw (1.0,2.7) arc (0:90:0.3);
	\fill[white] (-0.2+0.7,-0.2+3.0)--++(0.4,0.0)--++(0.0,0.4)--++(-0.4,0.0)--cycle;
	\draw (0.3,3.0) arc (-90:-180:0.3);
	\fill[white] (-0.2+0.3,-0.2+3.0)--++(0.4,0.0)--++(0.0,0.4)--++(-0.4,0.0)--cycle;
  	\draw (0.0,3.3)--(0.0,3.7);
	\draw (0,3.7) arc (180:90:0.3);
  	\draw[nodewitharrow={0.75}{}] (0.3,4.0)--(2.0,4.0);
  	\draw[dashed] (2.0,4.0)--(3.0,4.0);
  	\draw (3.0,4.0)--(3.7,4.0);
	\draw (3.7,4.0) arc (90:0:0.3);
  	\draw (4.0,3.7)--(4.0,3.3);
	\draw (4.0,3.3) arc (0:-90:0.3);
  	\draw[nodewitharrow={0.45}{}] (3.7,3.0)--(3.0,3.0);
  	\draw[dashed] (3.0,3.0)--(2.0,3.0);
  	\draw (2.0,3.0)--(0.3,3.0);
	\draw (0.3,3.0) arc (90:180:0.3);
  	\draw (0.0,2.7)--(0.0,2.0);
  	\draw[dashed] (0.0,2.0)--(0.0,1.0);
  	\draw[nodewitharrow={0.75}{}] (0.0,1.0)--(0.0,0.3);
	\draw (0.0,0.3) arc (180:270:0.3);
  	\draw (0.3,0.0)--(0.7,0.0);
	\draw (0.7,0.0) arc (-90:0:0.3);
  	\draw (1.0,0.3)--(1.0,1.0);
  	\draw[dashed] (1.0,1.0)--(1.0,2.0);
  	\draw (1.0,2.0)--(1.0,2.7);
	\end{tikzpicture}
	\caption{Curve 1}
	\end{subfigure}
\begin{subfigure}{0.3\textwidth}
  \centering
	\begin{tikzpicture}
	[nodewitharrow/.style 2 args={decoration={markings,mark=at position {#1} with {\arrow{stealth},\node[transform shape,above] {#2};}},postaction={decorate,ultra thick}}]
	\foreach \x in {0,1,2,3,4}
	\draw[black!25,very thin] (\x,-0.3)--(\x,4.3);
	\foreach \y in {0,1,2,3,4}
	\draw[black!25,very thin] (-0.3,\y)--(4.3,\y);		

  	\draw (1.3,3.0)--(2.0,3.0);
	\draw (1.0,2.7) arc (180:90:0.3);
  	\draw (0.0,2.0)--(0.0,3.7);
	\draw (0,3.7) arc (180:90:0.3);
  	\draw[nodewitharrow={0.35}{}] (2.0,4.0)--(0.3,4.0);
  	\draw[dashed] (2.0,4.0)--(3.0,4.0);
  	\draw (3.0,4.0)--(3.7,4.0);
	\draw (3.7,4.0) arc (90:0:0.3);
  	\draw (4.0,3.7)--(4.0,3.3);
	\draw (4.0,3.3) arc (0:-90:0.3);
  	\draw[nodewitharrow={0.85}{}] (3.0,3.0)--(3.7,3.0);
  	\draw[dashed] (3.0,3.0)--(2.0,3.0);
  	\draw (0.0,2.7)--(0.0,2.0);
  	\draw[dashed] (0.0,2.0)--(0.0,1.0);
  	\draw[nodewitharrow={0.75}{}] (0.0,1.0)--(0.0,0.3);
	\draw (0.0,0.3) arc (180:270:0.3);
  	\draw (0.3,0.0)--(0.7,0.0);
	\draw (0.7,0.0) arc (-90:0:0.3);
  	\draw (1.0,0.3)--(1.0,1.0);
  	\draw[dashed] (1.0,1.0)--(1.0,2.0);
  	\draw (1.0,2.0)--(1.0,2.7);
	\end{tikzpicture}
	\caption{Curve 2}
	\end{subfigure}
	
	\vspace{5 mm}
	\begin{subfigure}{0.3\textwidth}
  \centering
	\begin{tikzpicture}
	[nodewitharrow/.style 2 args={decoration={markings,mark=at position {#1} with {\arrow{stealth},\node[transform shape,above] {#2};}},postaction={decorate,ultra thick}}]
	\foreach \x in {0,1,2,3,4}
	\draw[black!25,very thin] (\x,-0.3)--(\x,4.3);
	\foreach \y in {0,1,2,3,4}
	\draw[black!25,very thin] (-0.3,\y)--(4.3,\y);	

  	\draw (0.3,3.0)--(0.7,3.0);
	\draw (0.7,3.0) arc (-90:0:0.3);
	\draw (1.0,2.7) arc (180:90:0.3);
  	\draw (1.0,3.3)--(1.0,3.7);
	\draw (1,3.7) arc (180:90:0.3);
  	\draw[nodewitharrow={0.85}{}] (2.0,4.0)--(1.3,4.0);
  	\draw[dashed] (2.0,4.0)--(3.0,4.0);
  	\draw (3.0,4.0)--(3.7,4.0);
	\draw (3.7,4.0) arc (90:0:0.3);
  	\draw (4.0,3.7)--(4.0,3.3);
	\draw (4.0,3.3) arc (0:-90:0.3);
  	\draw[nodewitharrow={0.85}{}] (3.0,3.0)--(3.7,3.0);
  	\draw[dashed] (3.0,3.0)--(2.0,3.0);
  	\draw (2.0,3.0)--(1.3,3.0);
	\draw (0.3,3.0) arc (90:180:0.3);
  	\draw (0.0,2.7)--(0.0,2.0);
  	\draw[dashed] (0.0,2.0)--(0.0,1.0);
  	\draw[nodewitharrow={0.85}{}] (0.0,1.0)--(0.0,0.3);
	\draw (0.0,0.3) arc (180:270:0.3);
  	\draw (0.3,0.0)--(0.7,0.0);
	\draw (0.7,0.0) arc (-90:0:0.3);
  	\draw (1.0,0.3)--(1.0,1.0);
  	\draw[dashed] (1.0,1.0)--(1.0,2.0);
  	\draw (1.0,2.0)--(1.0,2.7);
	\end{tikzpicture}
	\caption{Curve 3}
	\end{subfigure}
\begin{subfigure}{0.3\textwidth}
  \centering
	\begin{tikzpicture}
	[nodewitharrow/.style 2 args={decoration={markings,mark=at position {#1} with {\arrow{stealth},\node[transform shape,above] {#2};}},postaction={decorate,ultra thick}}]
	\foreach \x in {0,1,2,3,4}
	\draw[black!25,very thin] (\x,-0.3)--(\x,4.3);
	\foreach \y in {0,1,2,3,4}
	\draw[black!25,very thin] (-0.3,\y)--(4.3,\y);		
	
  	\draw (1.0,2.0)--(1.0,3.7);
	\fill[white] (-0.2+1.0,-0.2+3.0)--++(0.4,0.0)--++(0.0,0.4)--++(-0.4,0.0)--cycle;
	\draw (1,3.7) arc (180:90:0.3);
  	\draw[nodewitharrow={0.85}{}] (2.0,4.0)--(1.3,4.0);
  	\draw[dashed] (2.0,4.0)--(3.0,4.0);
  	\draw (3.0,4.0)--(3.7,4.0);
	\draw (3.7,4.0) arc (90:0:0.3);
  	\draw (4.0,3.7)--(4.0,3.3);
	\draw (4.0,3.3) arc (0:-90:0.3);
  	\draw[nodewitharrow={0.85}{}] (3.0,3.0)--(3.7,3.0);
  	\draw[dashed] (3.0,3.0)--(2.0,3.0);
  	\draw (2.0,3.0)--(0.3,3.0);
	\draw (0.3,3.0) arc (90:180:0.3);
  	\draw (0.0,2.7)--(0.0,2.0);
  	\draw[dashed] (0.0,2.0)--(0.0,1.0);
  	\draw[nodewitharrow={0.55}{}] (0.0,0.3)--(0.0,1.0);
	\draw (0.0,0.3) arc (180:270:0.3);
  	\draw (0.3,0.0)--(0.7,0.0);
	\draw (0.7,0.0) arc (-90:0:0.3);
  	\draw (1.0,0.3)--(1.0,1.0);
  	\draw[dashed] (1.0,1.0)--(1.0,2.0);
	\end{tikzpicture}
	\caption{Curve 4}
	\end{subfigure}
\caption{Some typical paths for glueball operators built from rectangles.
  Each rectangle can be extended arbitrarily, so providing some curves with
  no overall rotational or reflection symmetry.}
	\label{fig_glueball:curves}
\end{figure}

\clearpage

\begin{figure}[htb]
\begin	{center}
\leavevmode
\input	{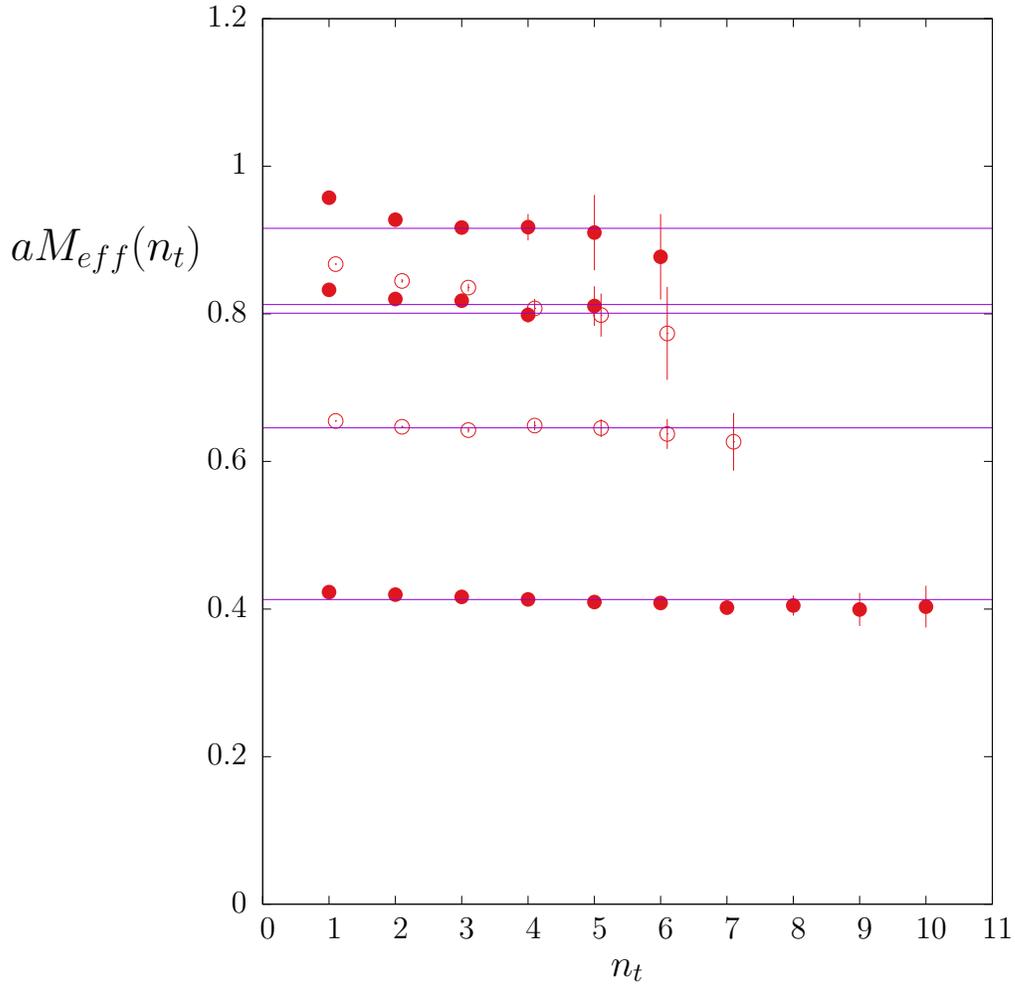}
\end	{center}
\caption{
  Effective masses of the lightest five $J^P=0^+$
  glueballs in $SO(12)$ at $\beta=250$. Lines are mass estimates.}
\label{fig_Eeff_M0p_so12_b250}
\end{figure}

\begin{figure}[htb]
\begin	{center}
\leavevmode
\input	{plot_Eeff_M2mp0m1mp_so12_b250.tex}
\end	{center}
\caption{
  Effective masses of the lightest and first excited $J^P=2^+$ ($\bullet$) and 
  $J^P=2^-$ ($\circ$) glueballs and the ground state $J^P=0^-$ ($\blacksquare$),
  $J^P=1^-$ ($\square$) and
  $J^P=1^+$ ($\vartriangle$) glueballs, all in $SO(12)$ at $\beta=250$. The
  $0^-$ and $1^\pm$ values have been shifted by $+0.4$ for clarity.
Lines are mass estimates}
\label{fig_Eeff_M2mp0m1mp_so12_b250}
\end{figure}

\begin{figure}[htb]
\begin	{center}
\leavevmode
\input	{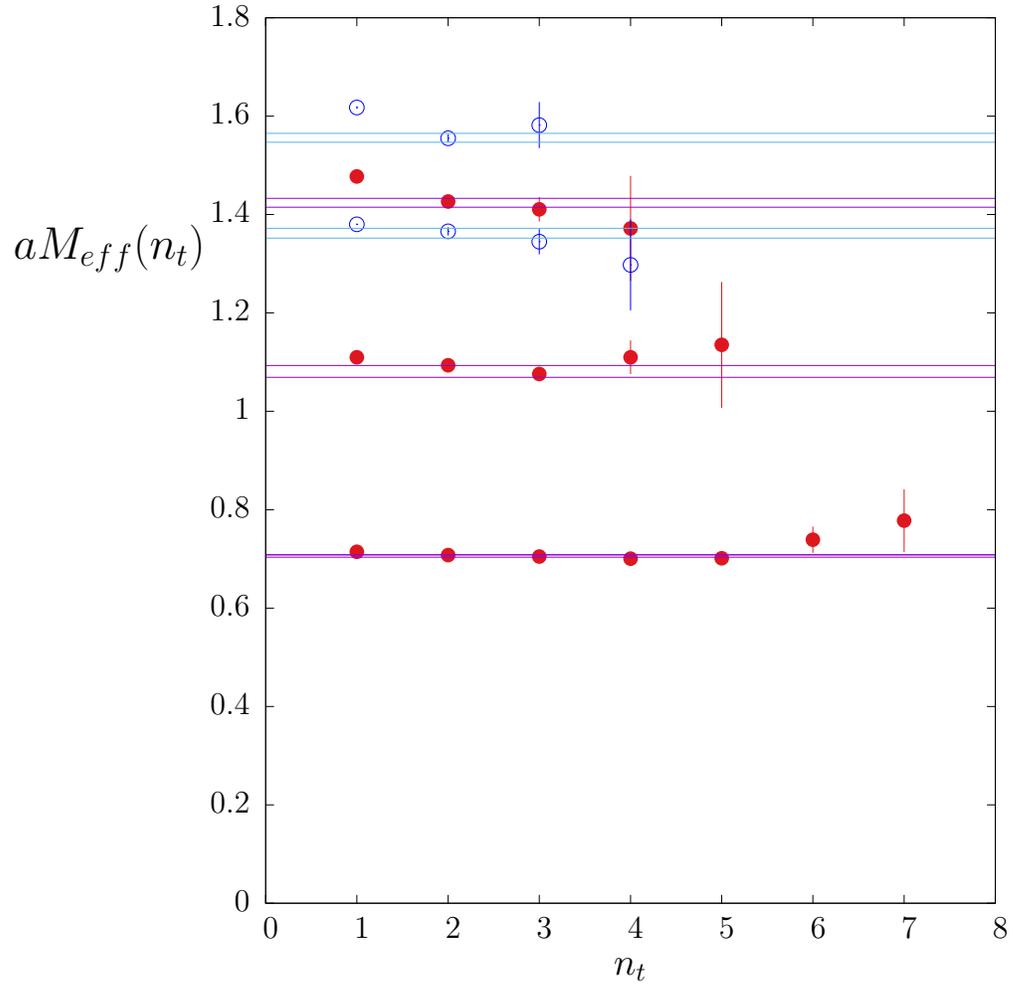}
\end	{center}
\caption{
  Effective masses of the lightest five $J^P=0^+$
  glueballs in $SO(12)$ at $\beta=155$. Lines are $\pm 1\sigma$ mass estimates.}
\label{fig_Eeff_M0p_so12_b155}
\end{figure}

\begin{figure}[htb]
\begin	{center}
\leavevmode
\input	{plot_Eeff_M2mp0m1mp_so12_b155.tex}
\end	{center}
\caption{
  Effective masses of the lightest and first excited $J^P=2^+$ ($\bullet$) and 
  $J^P=2^-$ ($\circ$) glueballs and the ground state $J^P=0^-$ ($\blacksquare$),
  $J^P=1^-$ ($\square$) and
  $J^P=1^+$ ($\vartriangle$) glueballs, all in $SO(12)$ at $\beta=155$. The
  $0^-$ and $1^\pm$ values have been shifted by $+0.4$ for clarity.
Lines are $\pm 1\sigma$ mass estimates}
\label{fig_Eeff_M2mp0m1mp_so12_b155}
\end{figure}



\begin{figure}[htb]
\begin	{center}
\leavevmode
\input	{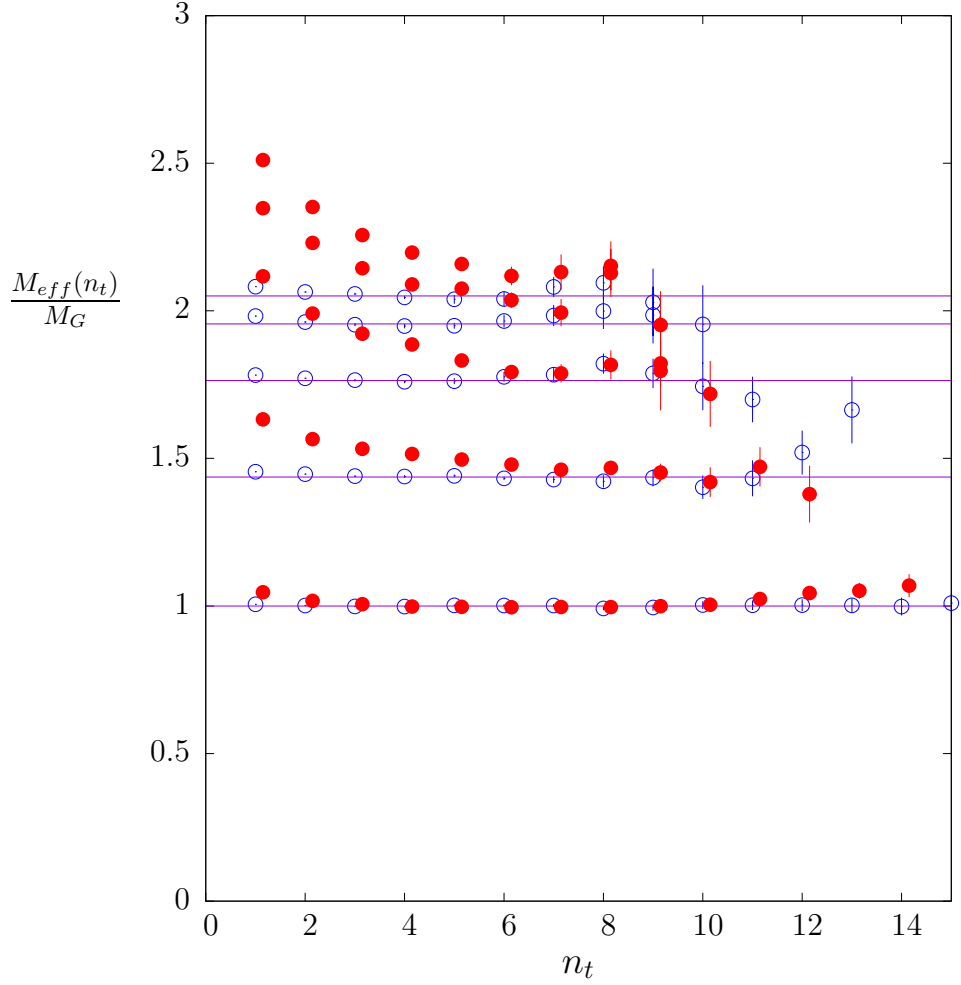}
\end	{center}
\caption{Effective masses of the lightest five scalar glueballs in
  $SO(3)$ ($\bullet$) at $\beta=11$ on a $100^2 80$ lattice and in $SU(2)$ ($\circ$)
  at $\beta=26.5$ on a $104^2 80$ lattice, in units of their respective mass gaps.
  Lines are $SU(2)$ mass estimates.}
\label{fig_Meff_0p_so3su2}
\end{figure}

\begin{figure}[htb]
\begin	{center}
\leavevmode
\input	{plot_Meff_Jm_so3su2.tex}
\end	{center}
\caption{Effective $SO(3)$ masses of the lightest two $2^-$ glueballs
  ($\bullet,\blacksquare$),
  the lightest $0^-$ glueball ($\blacklozenge$) and the lightest $1^-$ glueball
  ($\blacktriangledown$), with open points being the
  corresponding $SU(2)$ effective masses. Also the  $SO(3)$ mass of the first
  excited  $2^+$ ($\star$). All in units of their respective mass gaps and
  on the same lattices as in Fig~\ref{fig_Meff_0p_so3su2}. 
  Lines are $SU(2)$ mass estimates. The $0^-$ and $1^-$
  values have been shifted upwards by 0.5 for clarity.}
\label{fig_Meff_Jm_so3su2}
\end{figure}

\begin{figure}[htb]
\begin	{center}
\leavevmode
\input	{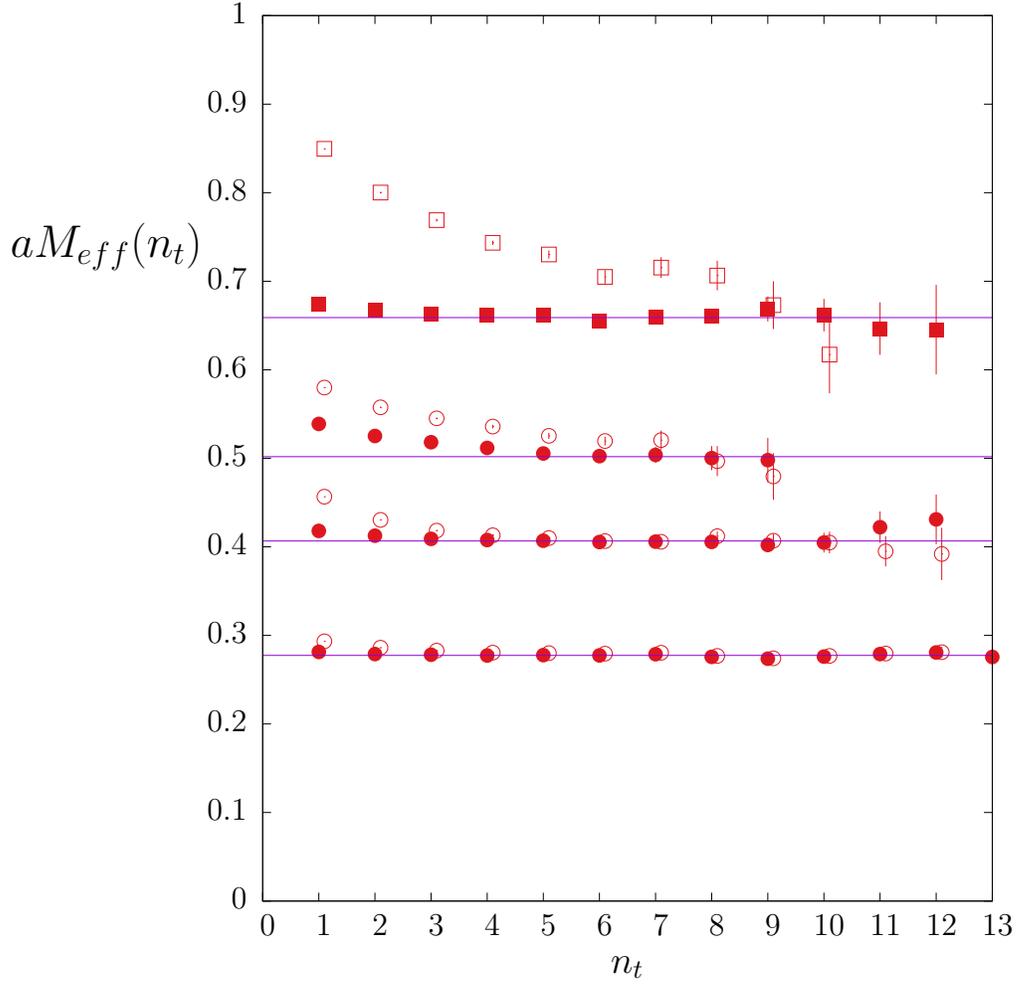}
\end	{center}
\caption{Effective masses of the lightest three $0^+$  glueballs ($\bullet, \circ $)
  and the lightest  $2^+$  glueball ($\blacksquare, \square $) in SU(2)
  on a $96^264$ lattice at $\beta=23.5$. The $2^+$ values have been shifted by $+0.2$ for clarity.
  Filled points use operators in the fundamental
  and open points in the adjoint representations. Lines are mass estimates.}
\label{fig_Eeff_M0p2pFA_su2_b23.5}
\end{figure}

\clearpage

\begin{figure}[htb]
\begin	{center}
\leavevmode
\input	{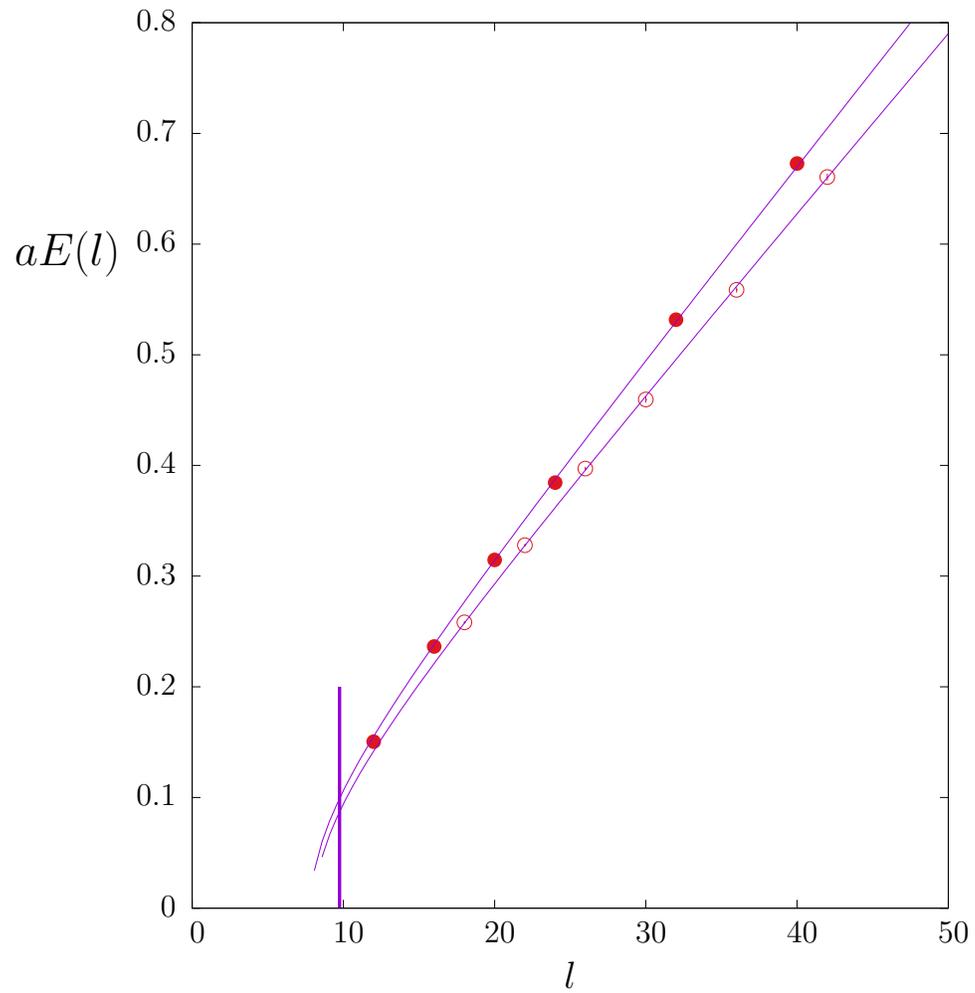}
\end	{center}
\caption{Ground state energy of a flux tube of length $l$ in $SO(8)$ at $\beta=86$,
  $\bullet$, and in $SO(6)$ at $\beta=46$, $\circ$.
Nambu-Goto fits shown. Vertical line is location of $SO(6)$ deconfining transition.}
\label{fig_Ek1_so6_so8}
\end{figure}

\begin{figure}[htb]
\begin	{center}
\leavevmode
\input	{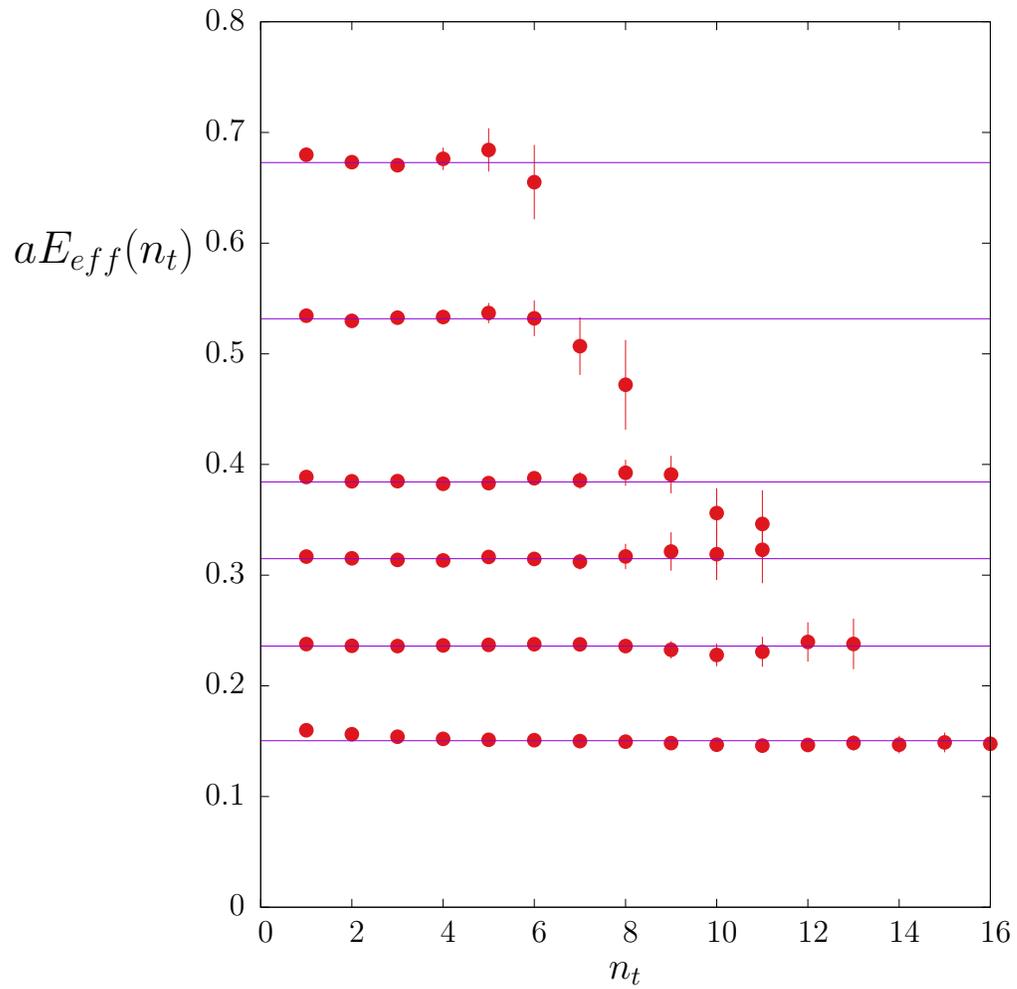}
\end	{center}
\caption{Effective energy of a flux tube of length $l$ in $SO(8)$ at $\beta=86$,
  as a function of $n_t=t/a$  for $l=12,16,20,24,32,40$. Lines are our
  energy estimates. }
\label{fig_Eeffk1_so8_b86}
\end{figure}

\begin{figure}[htb]
\begin	{center}
\leavevmode
\input	{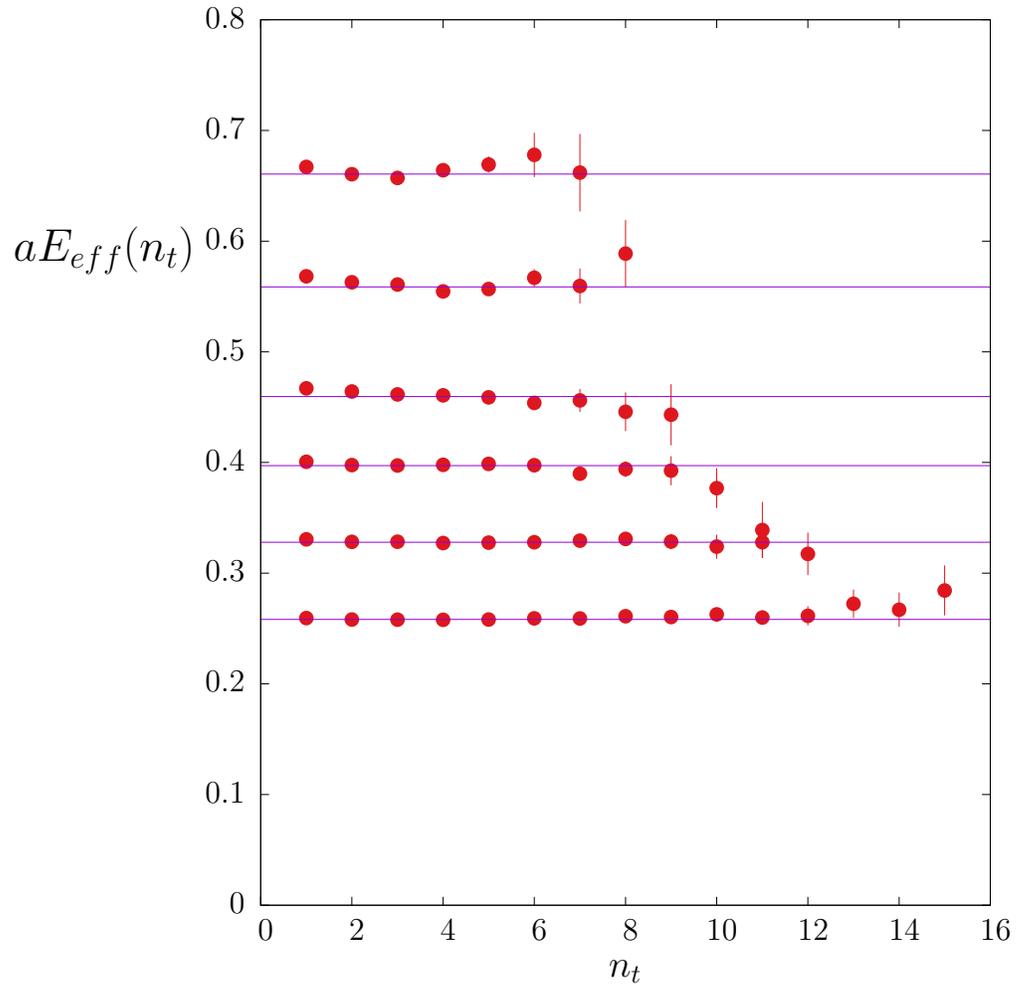}
\end	{center}
\caption{Effective energy of a flux tube of length $l$ in $SO(6)$ at $\beta=46$,
  as a function of $n_t=t/a$  for $l=18,22,26,30,36,42$. Lines are our
  energy estimates. }
\label{fig_Eeffk1_so6_b46}
\end{figure}

\begin{figure}[htb]
\begin	{center}
\leavevmode
\input	{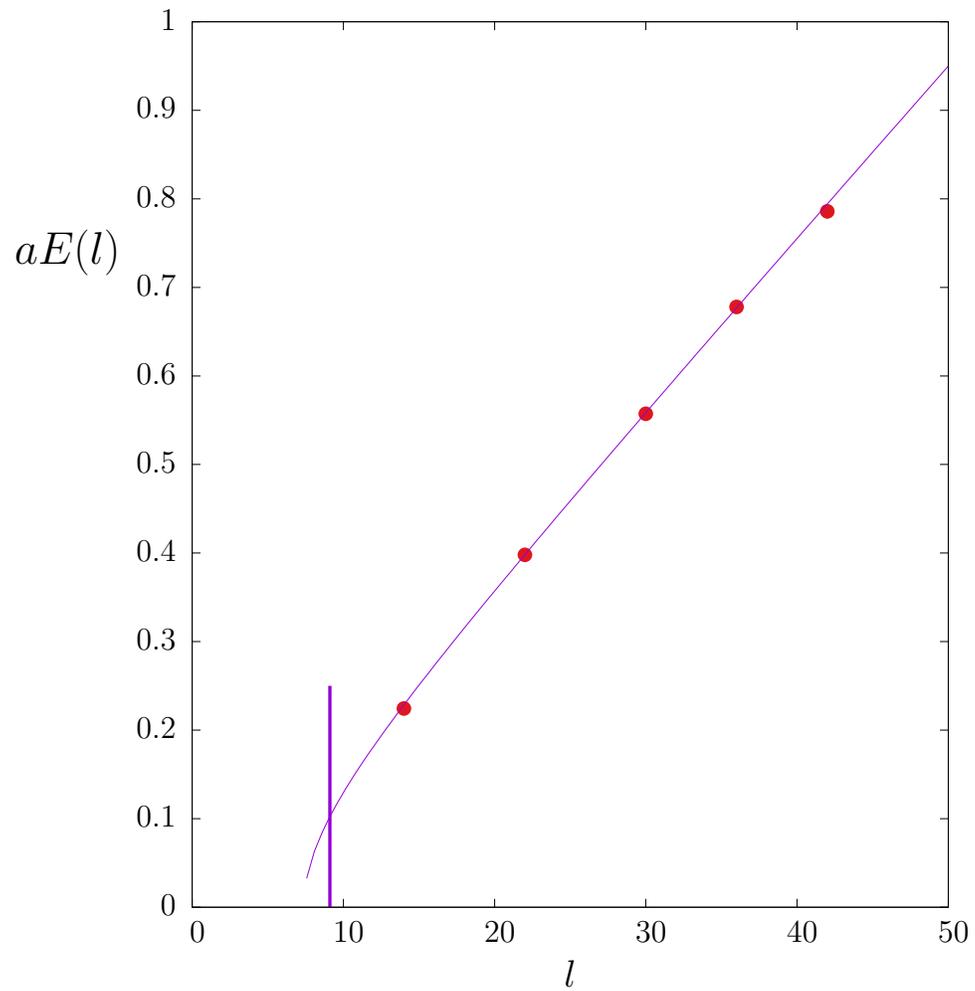}
\end	{center}
\caption{Ground state energy of a flux tube of length $l$ in $SO(5)$ at $\beta=27.5$.
Nambu-Goto fit shown. Vertical line gives location of the deconfining trensition.}
\label{fig_Ek1_so5_b27.5}
\end{figure}

\begin{figure}[htb]
\begin	{center}
\leavevmode
\input	{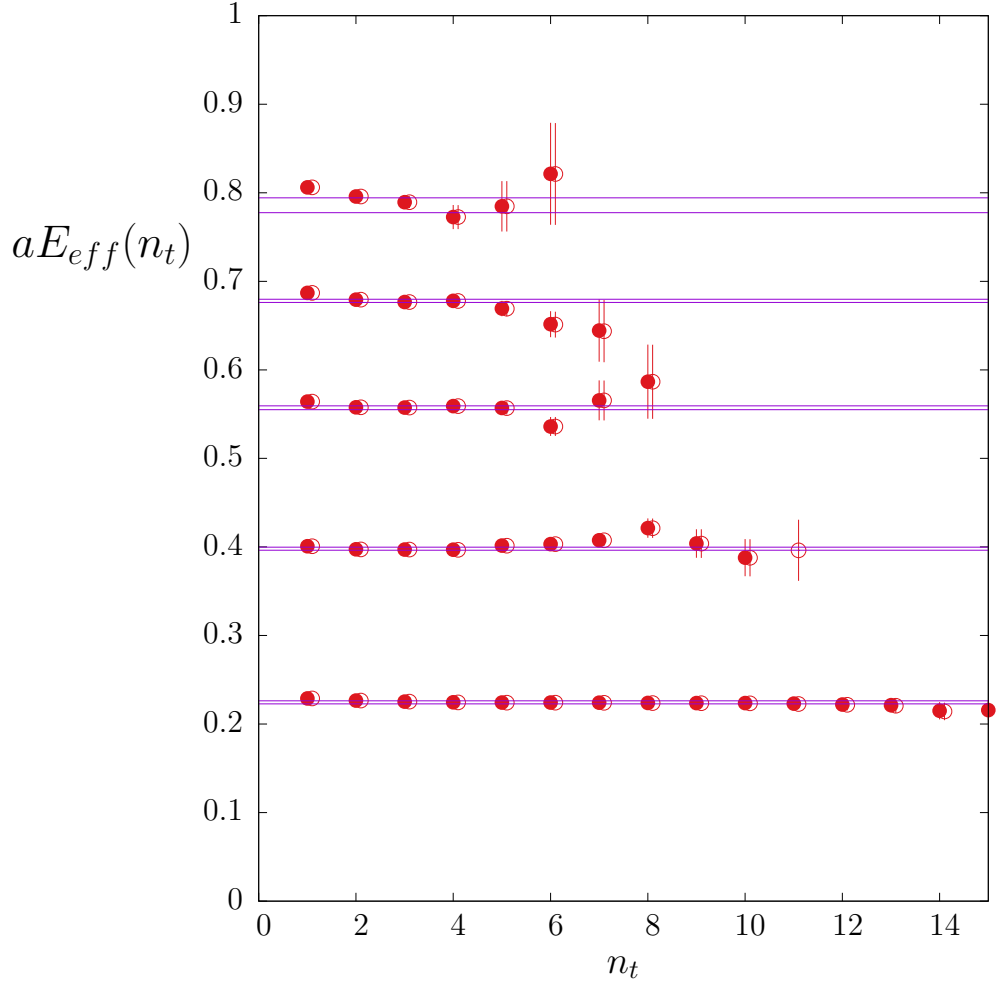}
\end	{center}
\caption{Effective energy of a flux tube of length $l$ in $SO(5)$ at $\beta=27.5$,
  as a function of $n_t=t/a$  for $l=14,22,30,36,42$. Bands show our
  energy estimates with errors. Closed (open) points are with (without) vacuum
subtraction. (Points shifted for clarity.)}
\label{fig_Eeffk1_so5_b27.5}
\end{figure}

\begin{figure}[htb]
\begin	{center}
\leavevmode
\input	{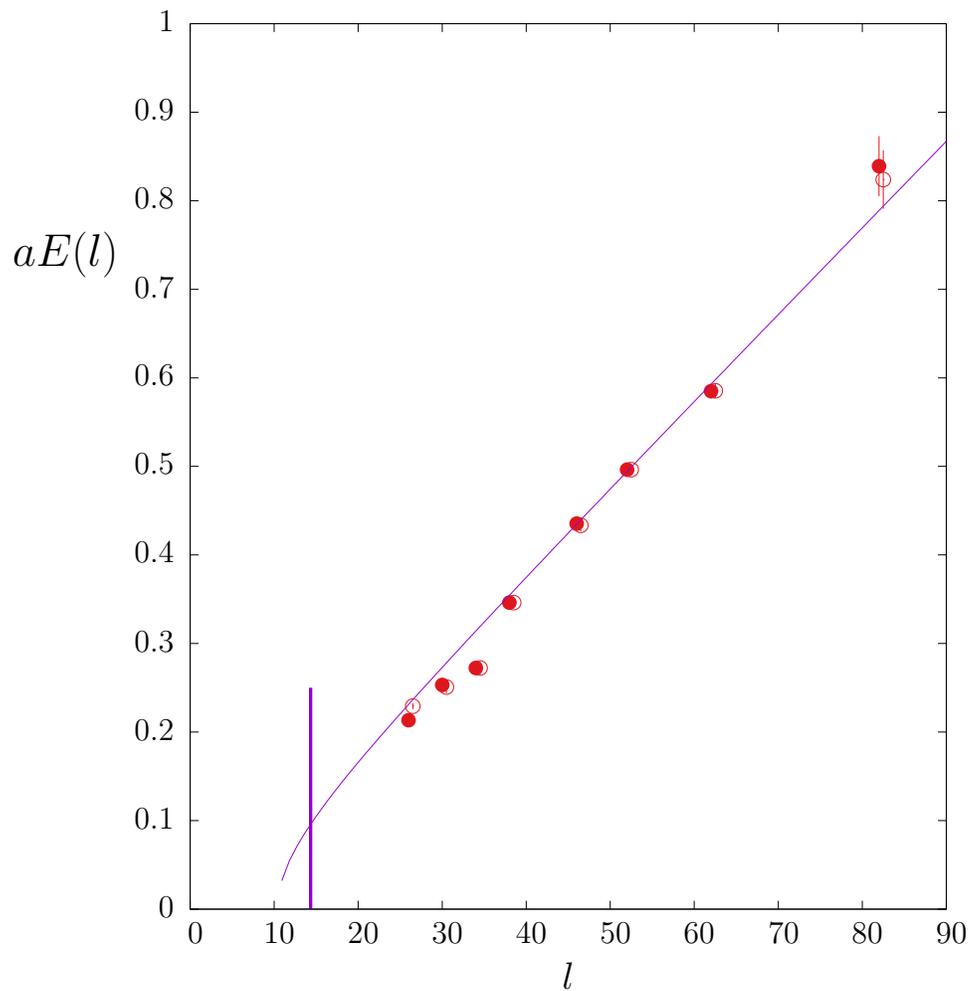}
\end	{center}
\caption{Ground state energy of a flux tube of length $l$ in $SO(3)$ at $\beta=9.0$
  with ($\bullet$) and without ($\circ$) vacuum subtraction. Nambu-Goto fit shown.
  Vertical line gives location of deconfining trensition.}
\label{fig_Ek1_so3_b9.0}
\end{figure}

\begin{figure}[htb]
\begin	{center}
\leavevmode
\input	{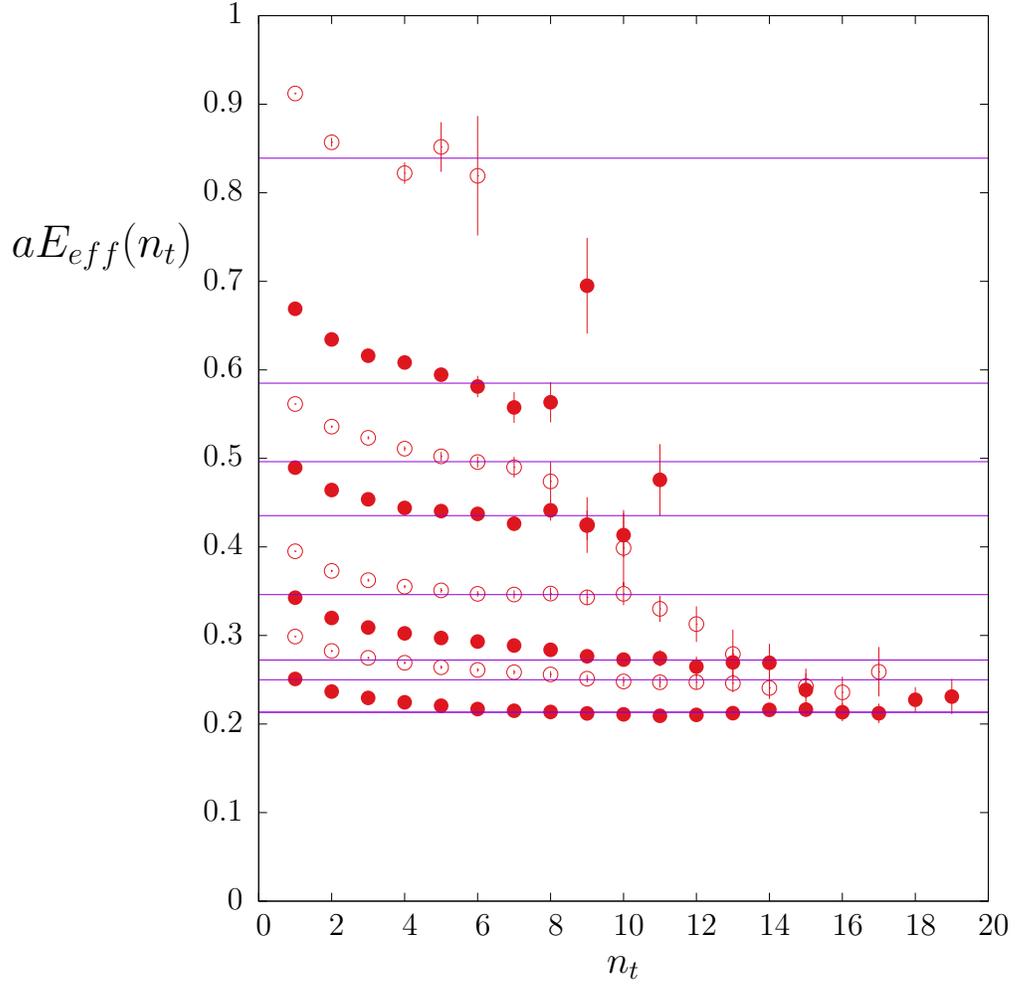}
\end	{center}
\caption{Effective energy of a flux tube of length $l$ in $SO(3)$ at $\beta=9.0$,
  as a function of $n_t=t/a$  for $l=26,30,34,38,46,52,62,82$. Correlators not vacuum subtracted.
  Lines 
  represent our energy estimates.}
\label{fig_Eeffk1_so3_b9.0}
\end{figure}

\begin{figure}[htb]
\begin	{center}
\leavevmode
\input	{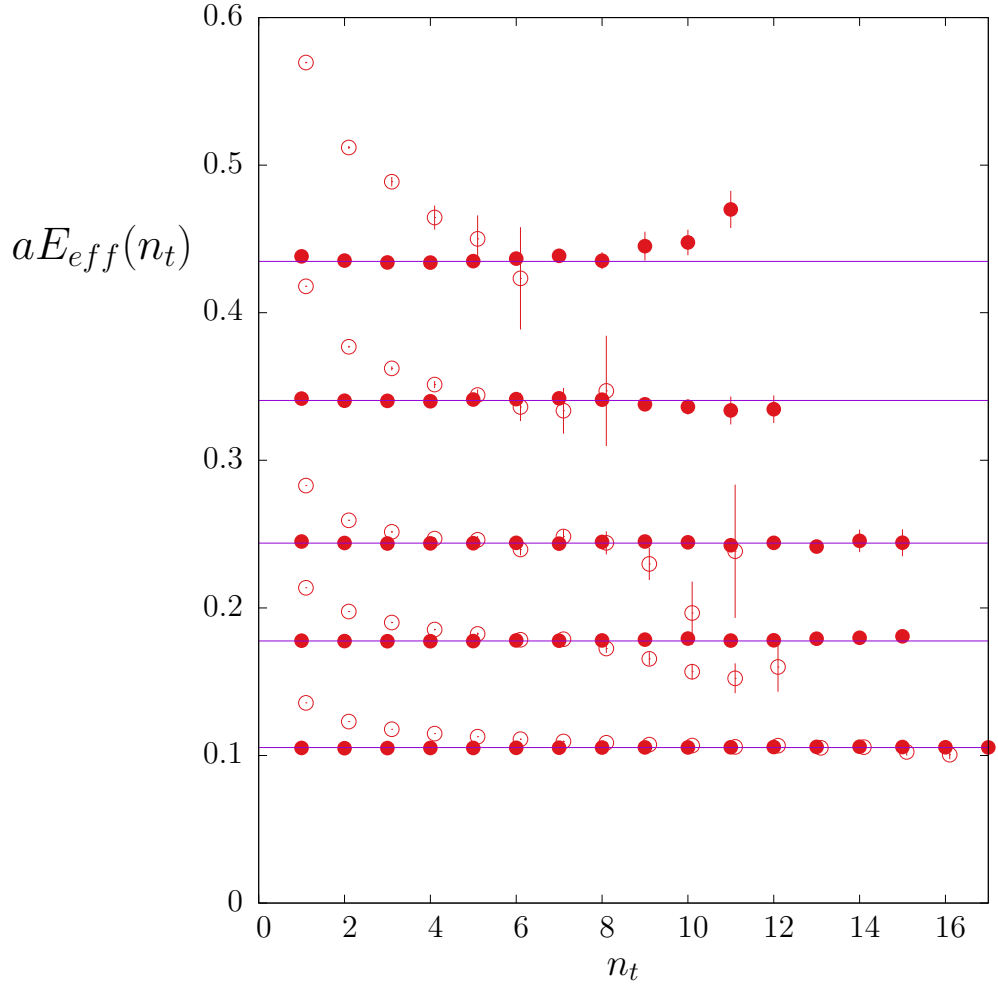}
\end	{center}
\caption{Effective energies of flux tubes of length $l=18,26,34,46,58$ in $SU(2)$ at $\beta=16.0$,
  as a function of $n_t=t/a$  for flux in the fundamental, $\bullet$, and in the adjoint, $\circ$.
  Adjoint values have been rescaled to asymptote to the fundamental energies. Lines shown are 
  (fundamental) energy estimates.}
\label{fig_Eeff_k1FA_su2_b16.0}
\end{figure}

\clearpage

\begin{figure}[htb]
\begin	{center}
\leavevmode
\input	{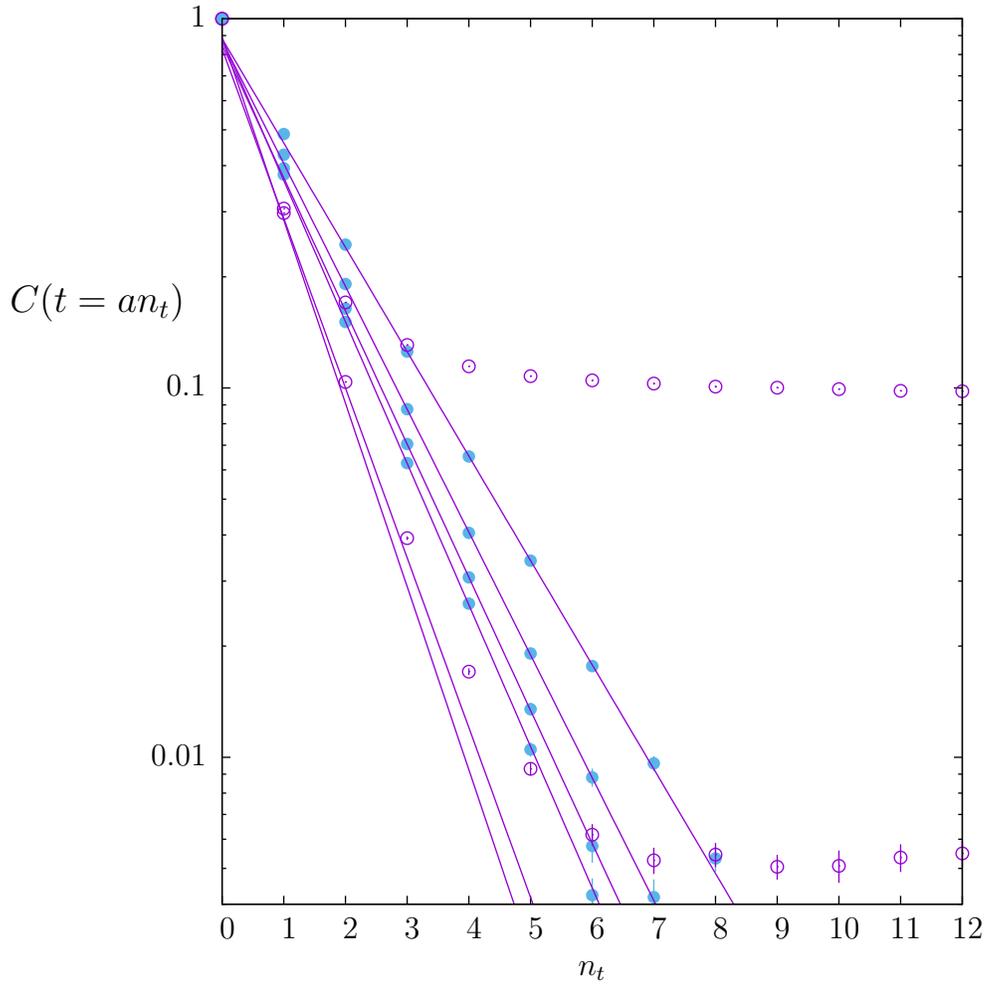}
\end	{center}
\caption{Correlation functions of the ground and first 3 excited flux tube states ($\bullet$)
  and also the 8'th and 10'th excited states ($\circ$). Exponential fits also shown,
   as described in text. In $SO(3)$ at $\beta=11$.}
\label{fig_Cork1_so3_b11}
\end{figure}

\begin{figure}[htb]
\begin	{center}
\leavevmode
\input	{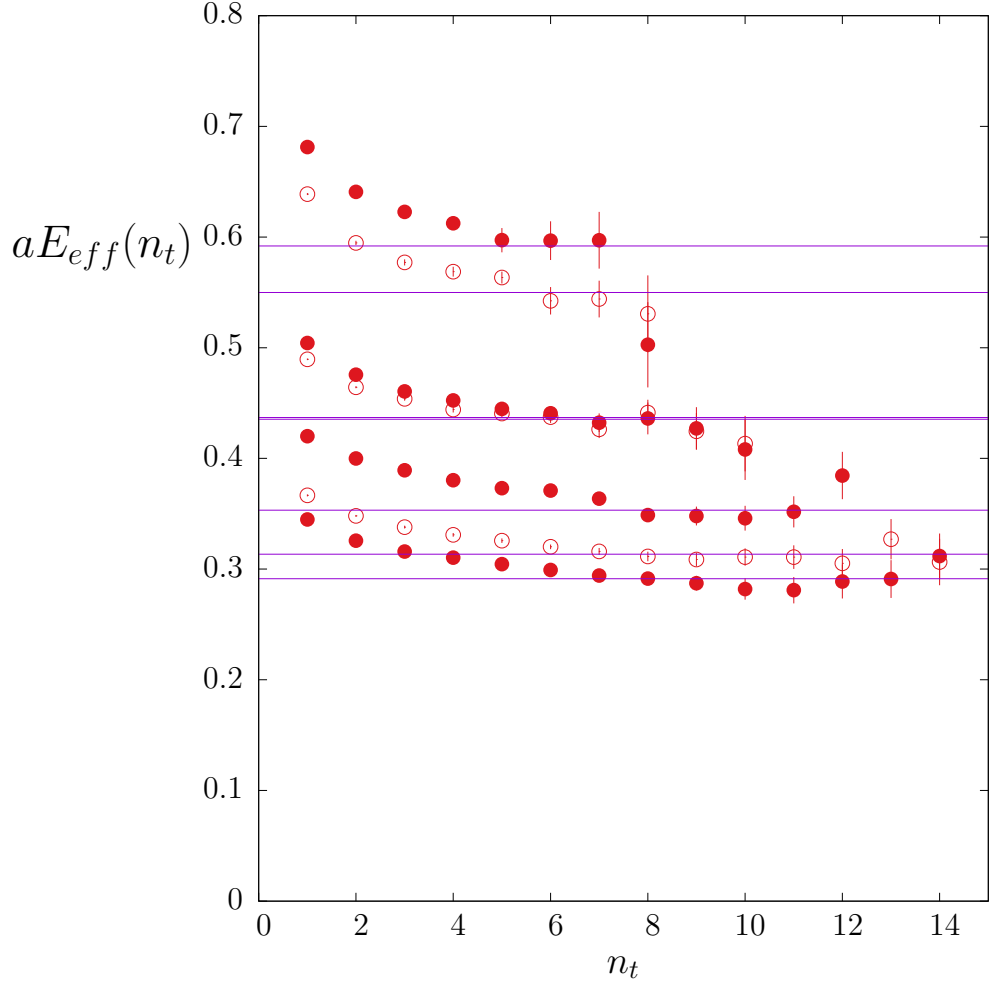}
\end	{center}
\caption{Effective energies of flux tubes of approximately the same physical 
  length, $l\surd\sigma \sim 4.2$, in $SO(3)$ at $\beta=6.5,7.0,8.5,9.0,10.0,11.0,12.0$,
  as a function of $n_t=t/a$. Correlators vacuum subtracted.
  Lines 
  represent our energy estimates.}
\label{fig_Eeffk1_vsm_so3}
\end{figure}

\begin{figure}[htb]
\begin	{center}
\leavevmode
\input	{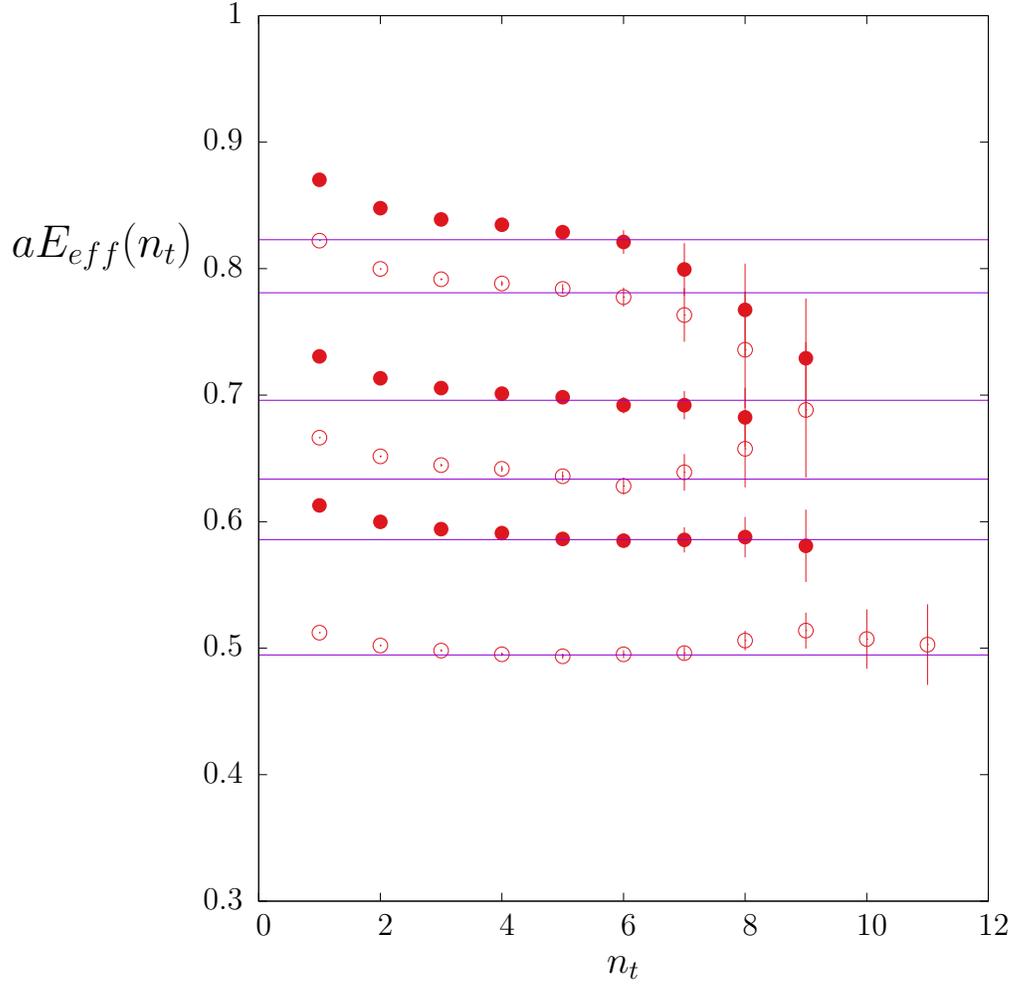}
\end	{center}
\caption{Effective energies of flux tubes of approximately the same physical 
  length, $l\surd\sigma \sim 4.4$, in $SO(4)$ at $\beta=11.0, 12.2, 13.7, 15.1, 16.5, 18.7$
  as a function of $n_t=t/a$.  Lines 
  represent our energy estimates.}
\label{fig_Eeffk1_sm_so4}
\end{figure}

\begin{figure}[htb]
\begin	{center}
\leavevmode
\input	{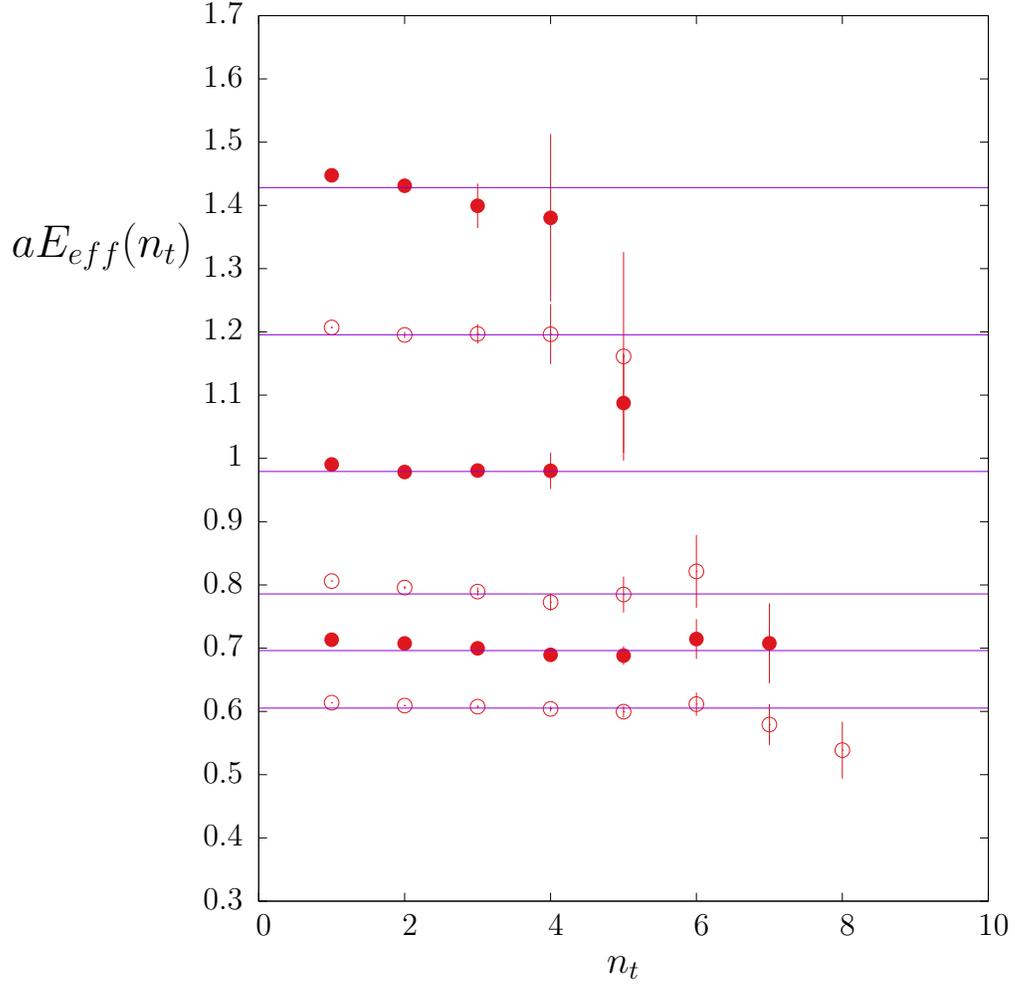}
\end	{center}
\caption{Effective energies of flux tubes of approximately the same physical 
  length, $l\surd\sigma \sim 6.0$, in $SO(5)$ at $\beta=17.5, 20.0, 23.5, 27.5, 32.0, 36.0$
  as a function of $n_t=t/a$.  Lines 
  represent our energy estimates.}
\label{fig_Eeffk1_so5}
\end{figure}

\begin{figure}[htb]
\begin	{center}
\leavevmode
\input	{plot_kg_soNd3.tex}
\end	{center}
\caption{Lattice values of $\surd\sigma/g^2_I N$ for
  SO(3), SO(4), SO(5), SO(6), SO(7), SO(8), SO(12), SO(16) in ascending order,
  with continuum extrapolations. }
\label{fig_kg_soNd3}
\end{figure}

\clearpage

\begin{figure}[htb]
\begin	{center}
\leavevmode
\input	{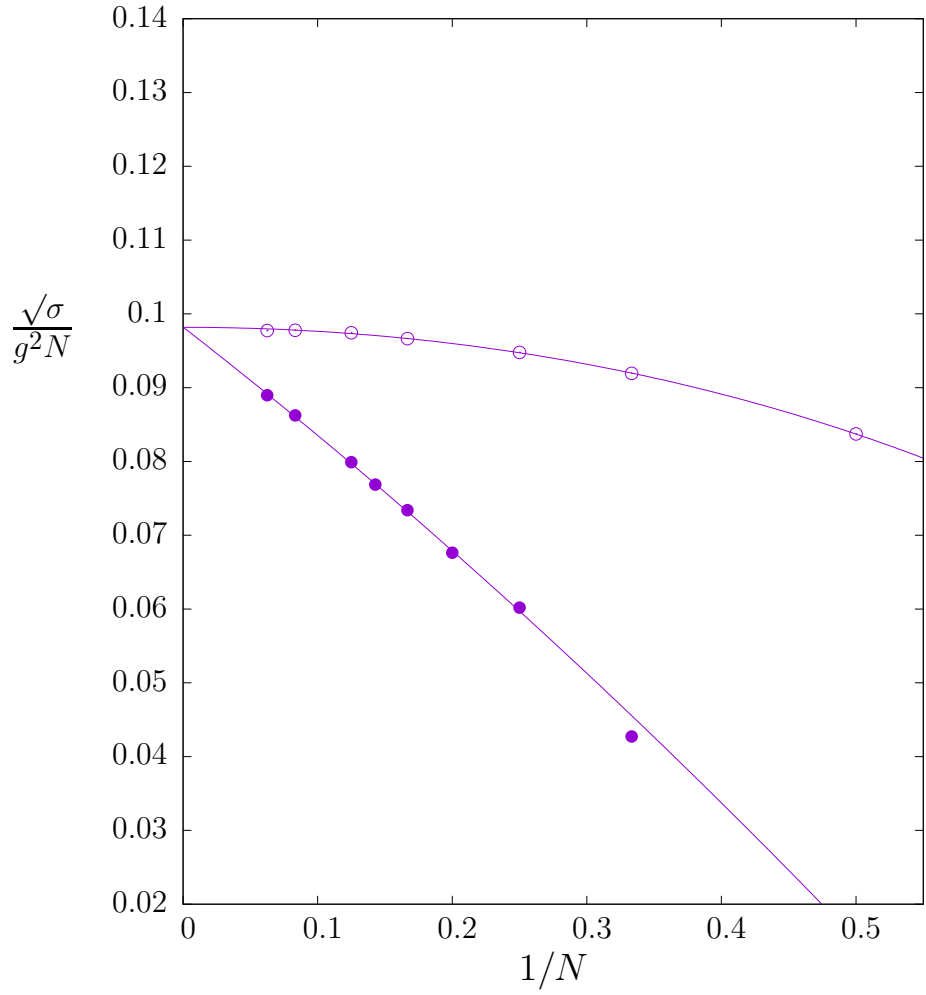}
\end	{center}
\caption{Continuum values of $\surd\sigma/g^2N$ in $SO(N)$, $\bullet$, with 
  extrapolation to $N=\infty$ as in eqn(\ref{eqn_kgN_N}). Also shown are $SU(N)$
  values of $\surd\sigma/2g^2N$, $\circ$, with the large $N$ fit given
  in eqn(\ref{eqn_kgN_suN}).}
\label{fig_kgNcont_soNd3}
\end{figure}

\begin{figure}[htb]
\begin	{center}
\leavevmode
\input	{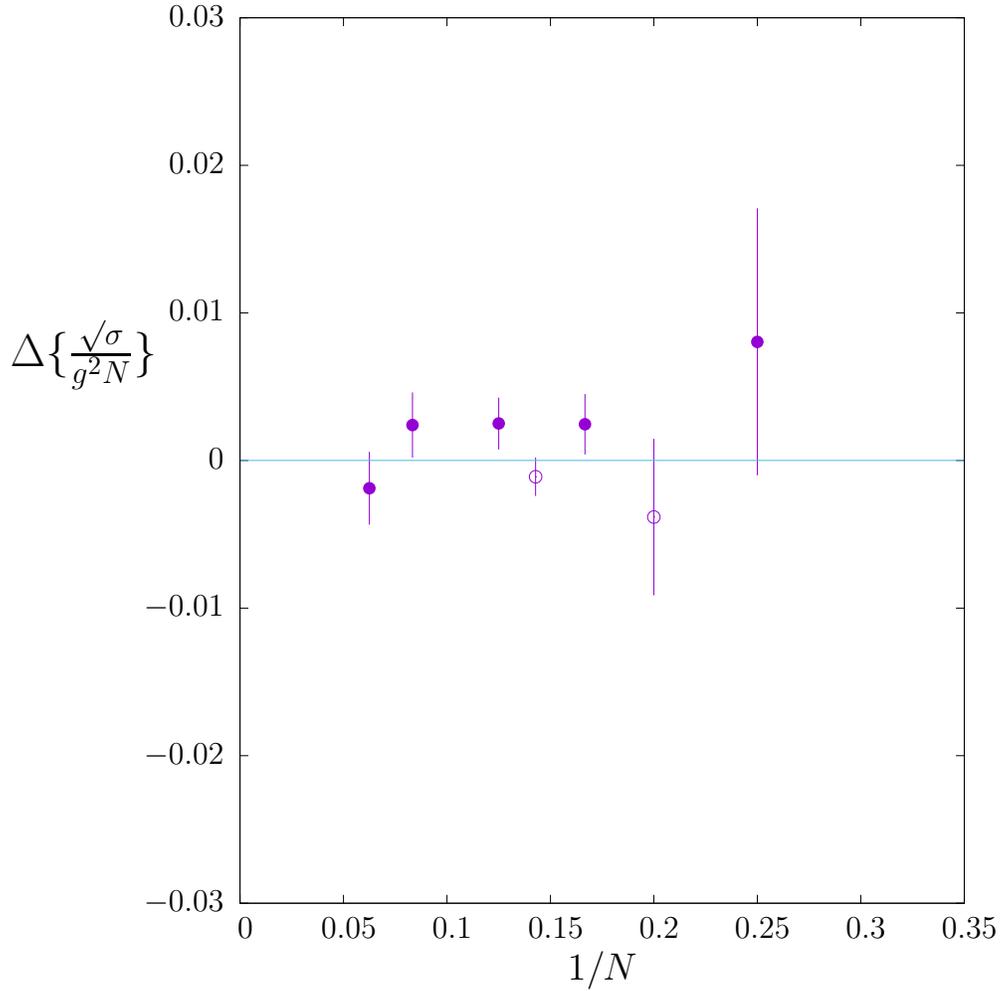}
\end	{center}
\caption{Difference between value of $\surd\sigma/g^2N$ and the best fit in 
  eqn(\ref{eqn_kgN_N}), normalised by its value. Open points are $N$ odd,
  full points are $N$ even. ($SO(3)$ is off the plot.)}
\label{fig_kgNdiff_soNd3}
\end{figure}

\begin{figure}[htb]
\begin	{center}
\leavevmode
\input	{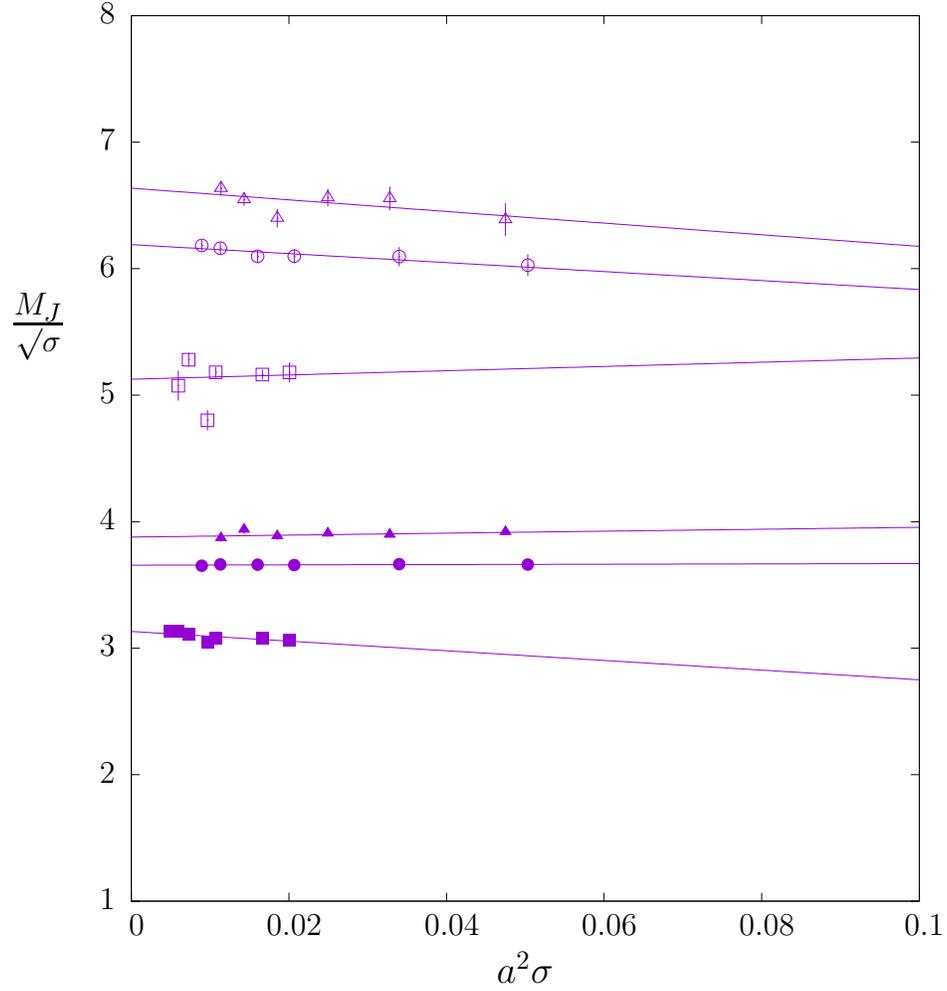}
\end	{center}
\caption{Lightest $0^+$ glueball 
  in $SO(3)$ ($\blacksquare$),   $SO(6)$ ($\bullet$),  $SO(12)$ ($\blacktriangle$), 
  and lightest $2^+$ glueball 
  in $SO(3)$ ($\square$),   $SO(6)$ ($\circ$),  $SO(12)$ ($\vartriangle$),
  all in units of the string tension and plotted versus the string tension, 
  with linear extrapolations to the continuum limit.}
\label{fig__M0p2pK_contd3}
\end{figure}

\begin{figure}[htb]
\begin	{center}
\leavevmode
\input	{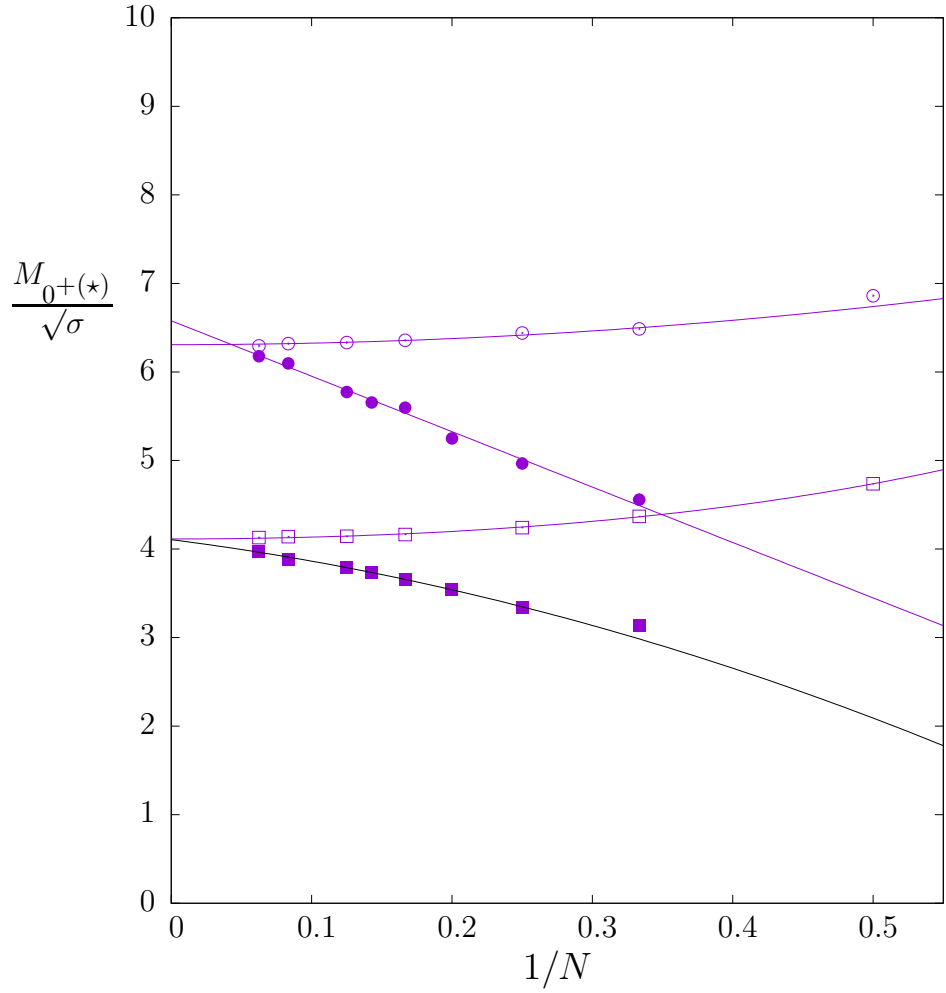}
\end	{center}
\caption{Lightest two $0^+$ glueballs in units of the string tension in $SO(N)$
  ($\blacksquare , \bullet$)
  and in $SU(N)$ ($\square , \circ$) with typical extrapolations to $N=\infty$.}
\label{fig__M0pgsex1K_soNsuN}
\end{figure}


\clearpage

\begin{figure}[htb]
\begin	{center}
\leavevmode
\input	{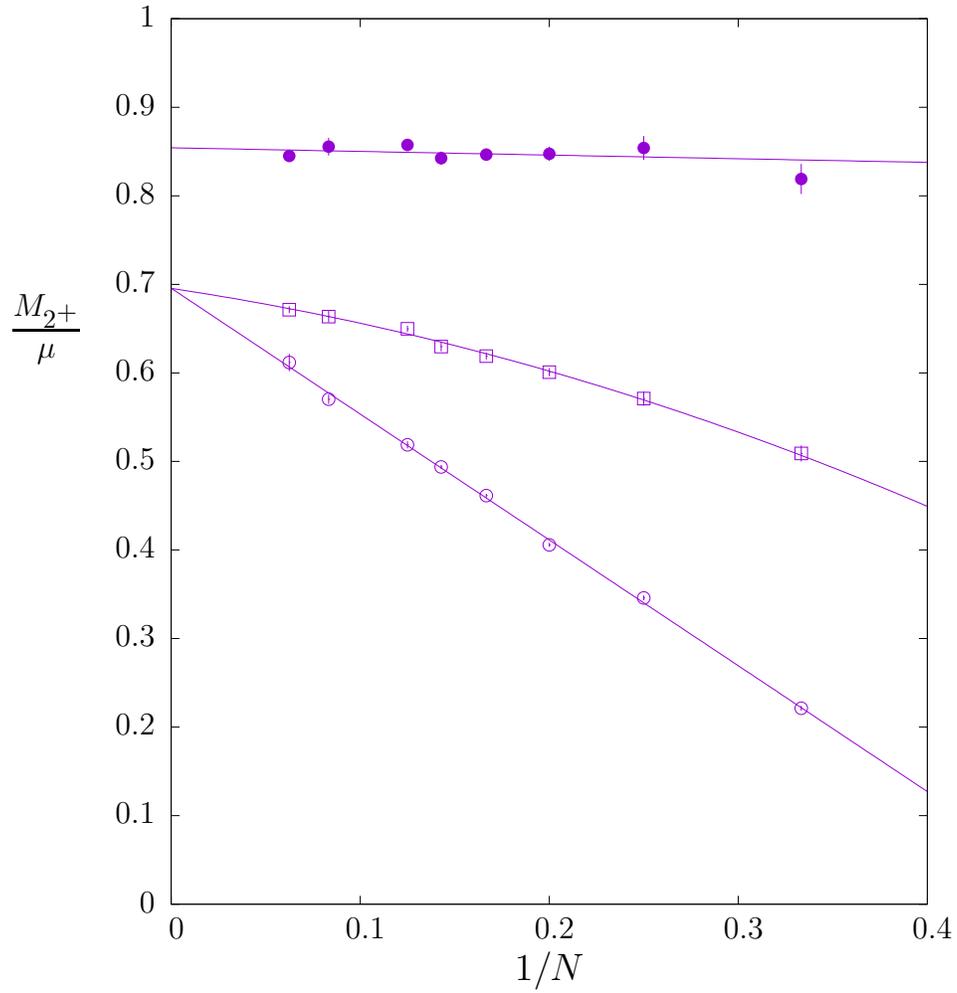}
\end	{center}
\caption{Mass ratios $M_{2^+}/\mu$ versus $1/N$ for $\mu=2M_{0^+}$ ($\bullet$), $\mu=g^2N$ ($\circ$)  and $\mu=10\sqrt{\sigma}$ ($\square$). Fits shown to guide the eye.}
\label{fig_m2pmu_soNd3}
\end{figure}

\begin{figure}[htb]
\begin	{center}
\leavevmode
\input	{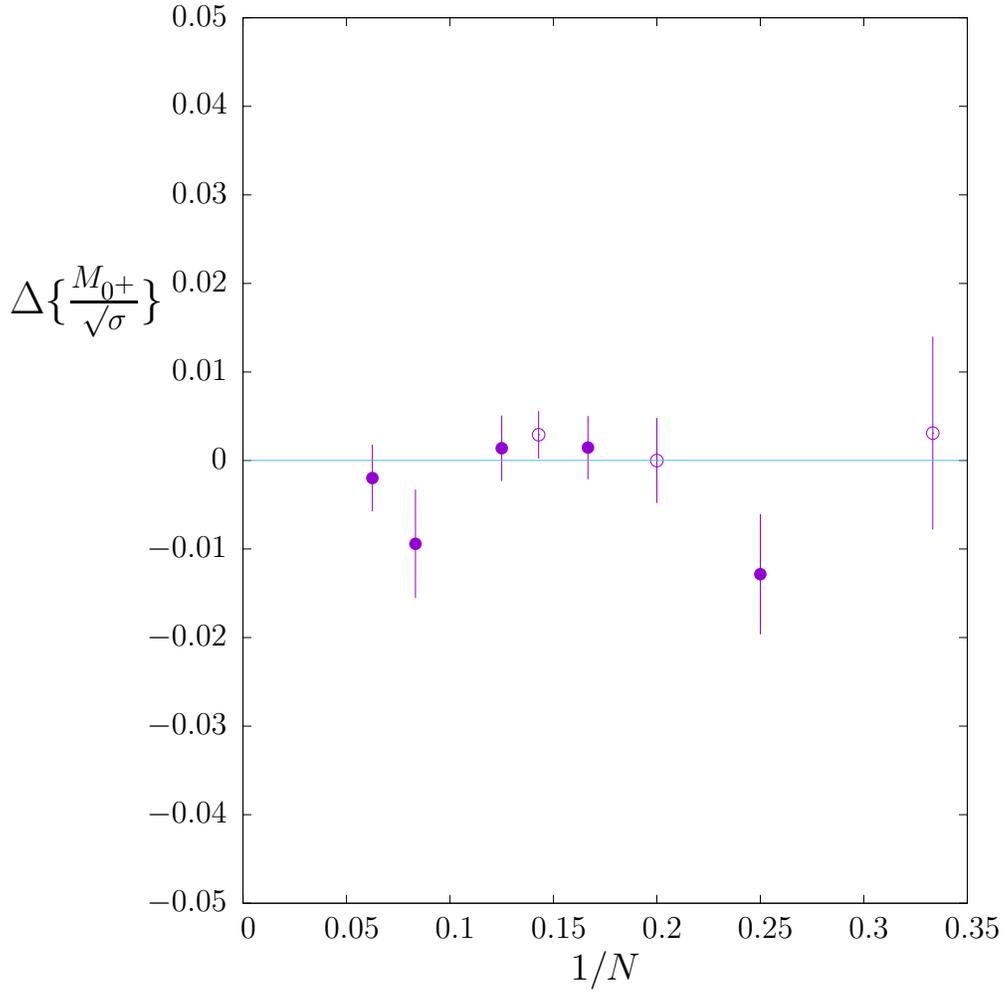}
\end	{center}
\caption{Difference between value of $M_{0^+}/\surd\sigma$ and the best fit
  eqn(\ref{eqn_M0pK_N}), normalised by value of ratio. Open points are $N$ odd,
  filled points are $N$ even. }
\label{fig_M0pgsKdiff_soN}
\end{figure}

\begin{figure}[htb]
\begin	{center}
\leavevmode
\input	{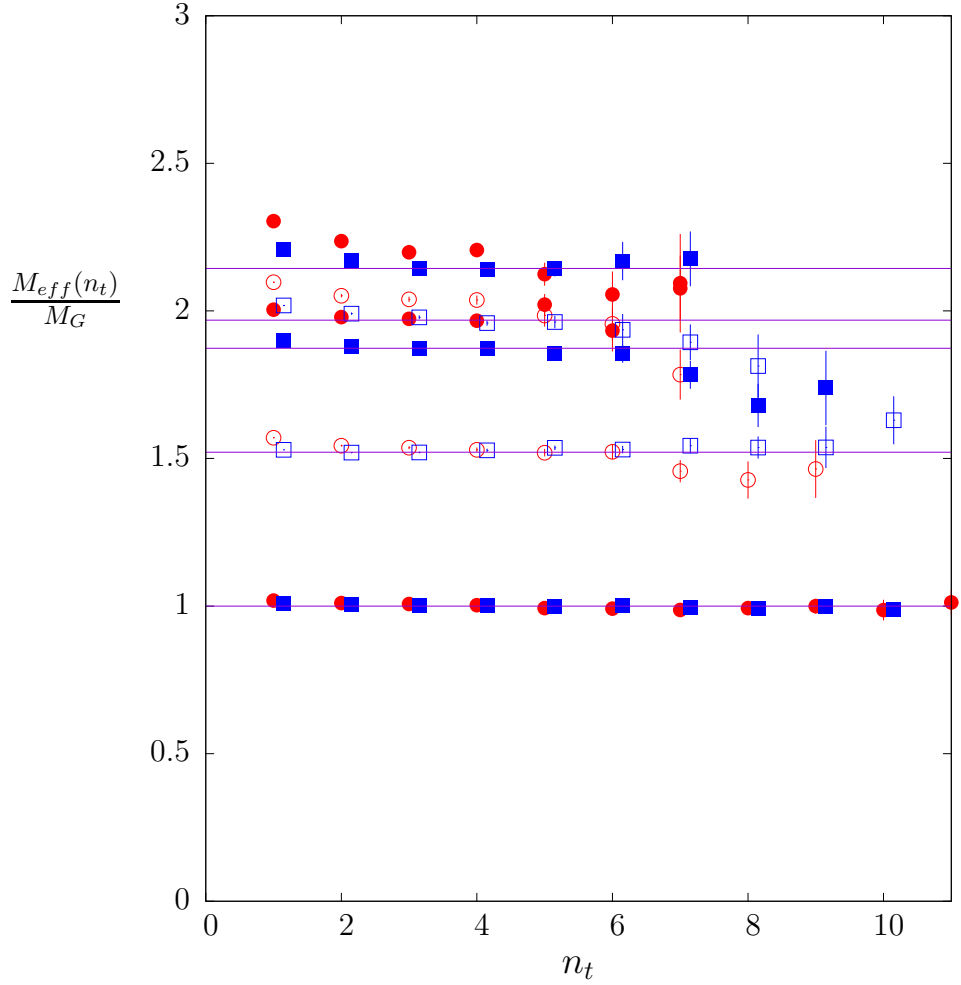}
\end	{center}
\caption{Lightest five $0^+$  glueball effective masses in $SU(4)$ ($\square$,$\blacksquare$) and
  $SO(6)$ ($\circ$,$\bullet$), in units of their respective mass gaps, and at
  the smallest values of $a$ in each case.}
\label{fig_Meff_0p_so6su4}
\end{figure}

\begin{figure}[htb]
\begin	{center}
\leavevmode
\input	{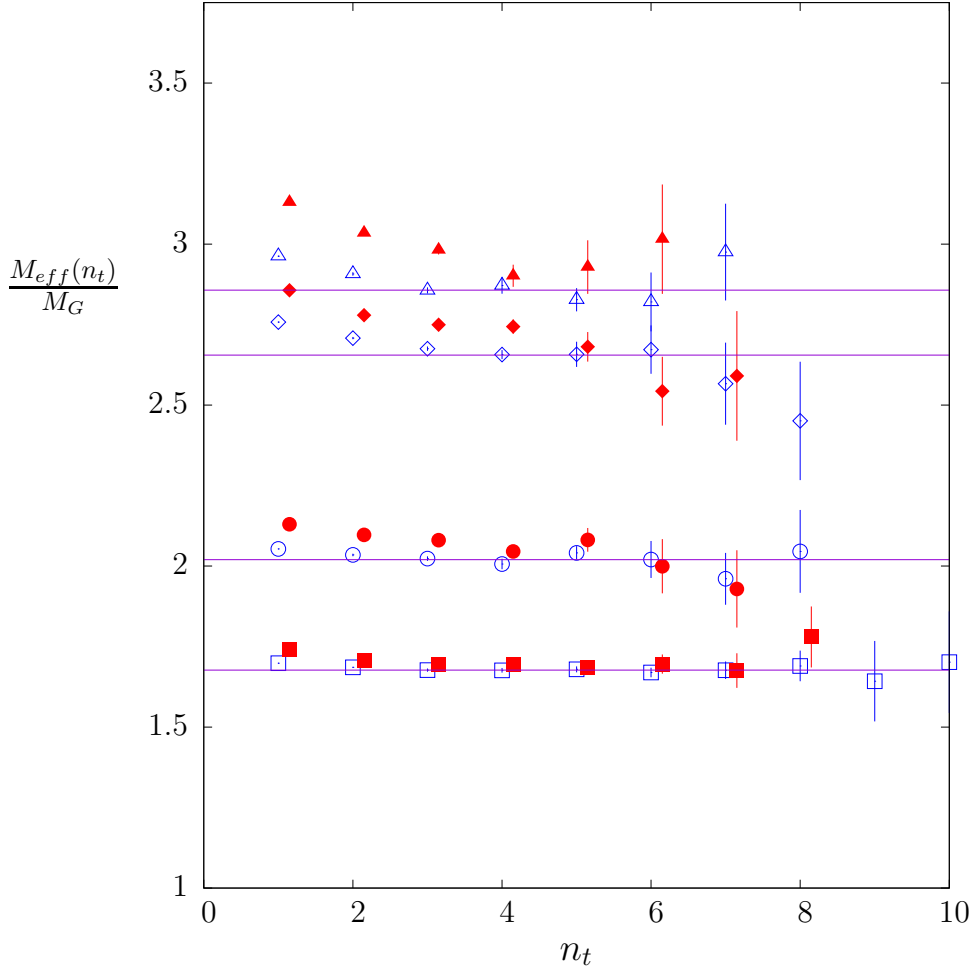}
\end	{center}
\caption{Some glueball effective masses in $SU(4)$ (open points) and
  $SO(6)$ (filled points), in units of their respective mass gaps, and at
  our smallest value of $a$:  lightest $2^+$, $\square$, first excited $2^+$, $\circ$,
  lightest $0^-$, $\lozenge$, and lightest $1^+$, $\vartriangle$. The $1^+$
  and $0^-$ have been shifted up by +0.5 for clarity.}
\label{fig_Meff_J_so6su4}
\end{figure}



\begin{figure}[htb]
\begin	{center}
\leavevmode
\input	{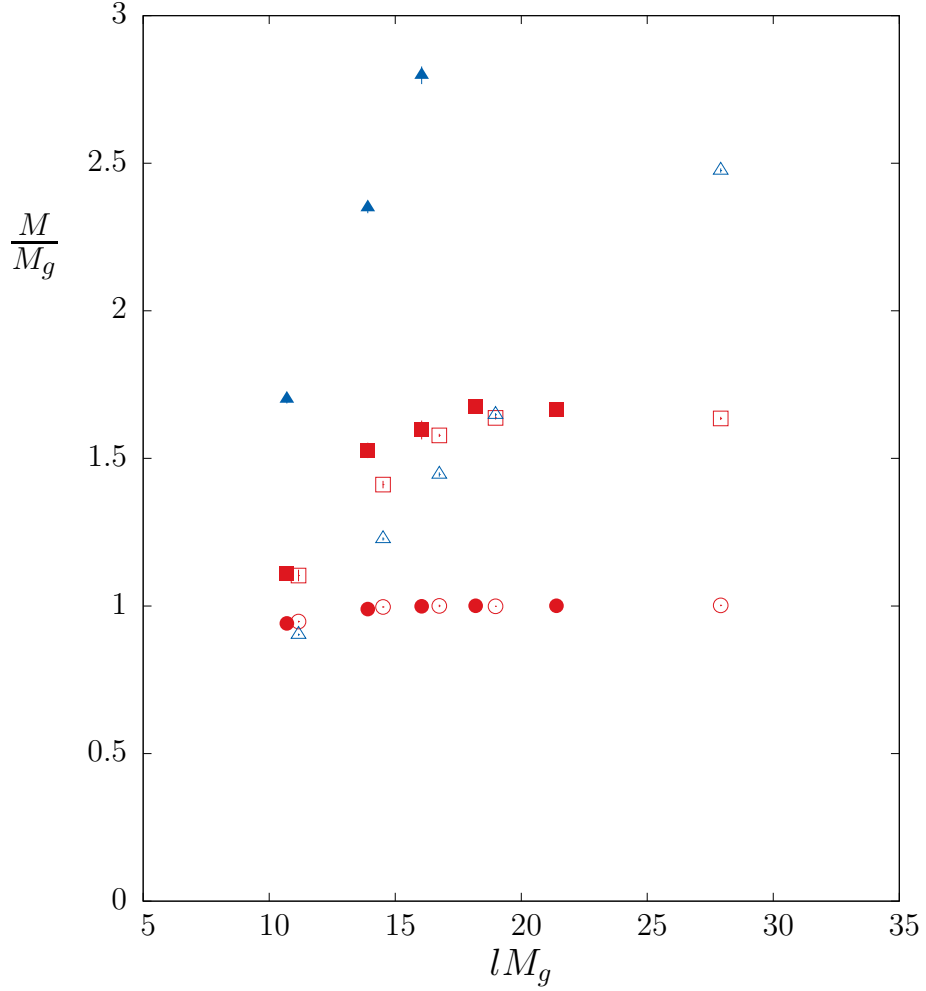}
\end	{center}
\caption{Lightest scalar ($\circ$) and tensor ($\Box$) glueballs
in SU(2) on various spatial volumes, $l^2$, all in units of the infinite 
volume mass gap $M_g$. And same for SO(4) ($\bullet,\blacksquare$ respectively).
Also shown is twice the fundamental flux loop mass in SU(2) ($\triangle$)
and SO(4) ($\triangle$).}
\label{fig_Vso4su2}
\end{figure}

\begin{figure}[htb]
\begin	{center}
\leavevmode
\input	{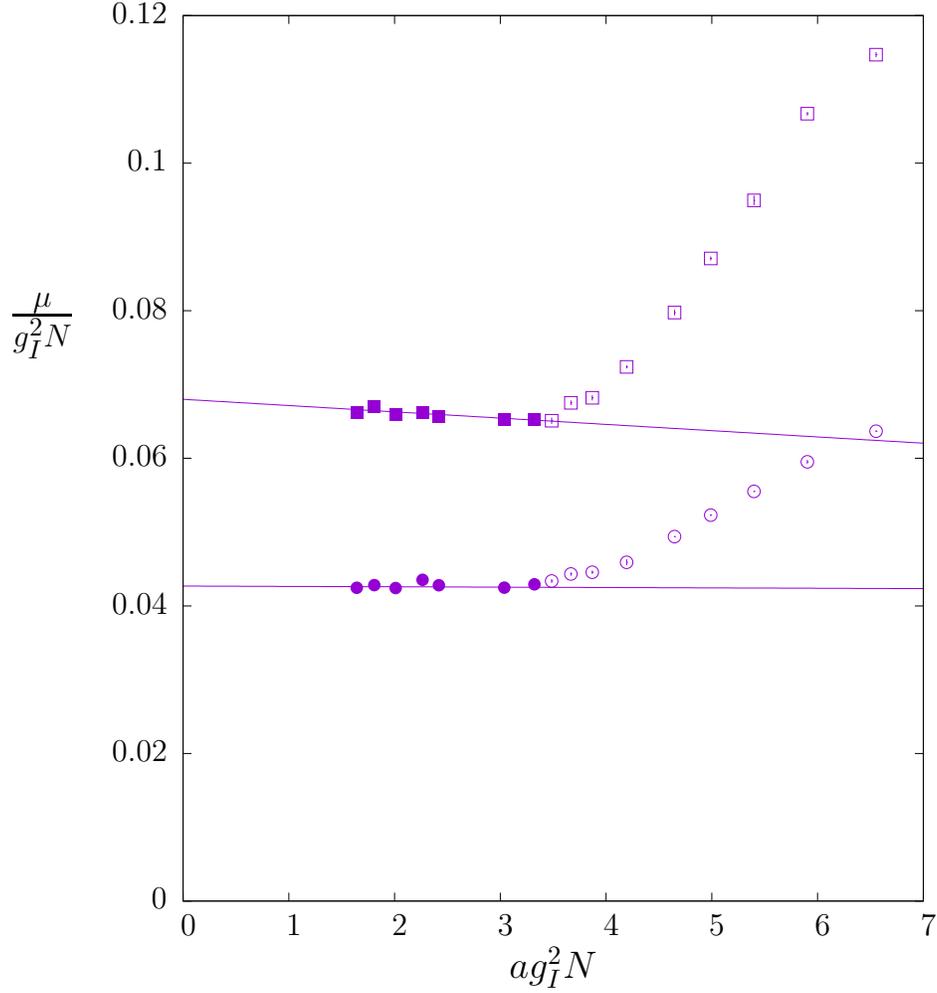}
\end	{center}
\caption{String tension, $\surd\sigma/g_I^2N$ ($\circ$,$\bullet$), and lightest scalar glueball, 
  $0.5\times M_{0^+}/g_I^2N$ ($\square$,$\blacksquare$), both in units of the (improved)
  't Hooft coupling and plotted against $ag_I^2N$; all in $SO(3)$. Open points are in the  strong coupling
  and cross-over region, and filled points are on the weak coupling side.}
\label{fig_kg_mg_so3d3}
\end{figure}

\begin{figure}[htb]
\begin	{center}
\leavevmode
\input	{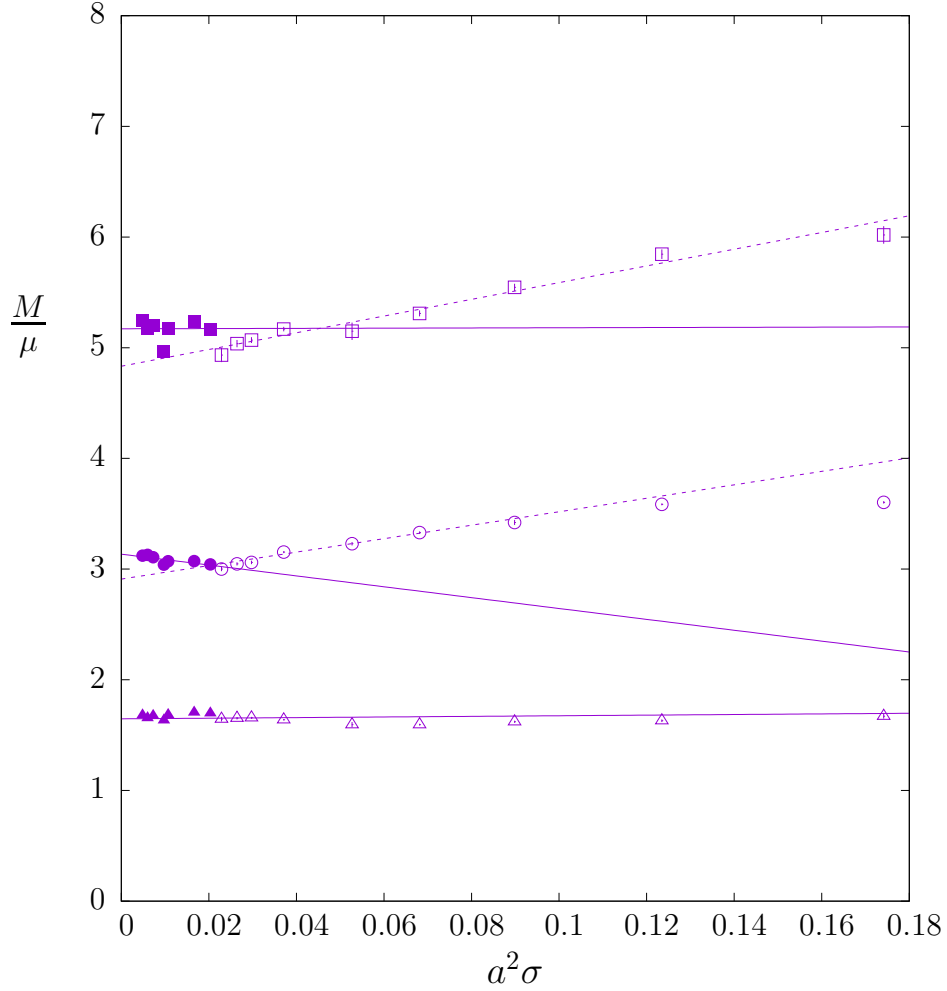}
\end	{center}
\caption{Mass ratios $M_{0^+}/\surd\sigma$ ($\circ$,$\bullet$),  $M_{2^-}/\surd\sigma$
  ($\square$,$\blacksquare$),
  and $M_{2^-}/M_{0^+}$ ($\vartriangle$,$\blacktriangle$) plotted against $a^2\sigma$,
  with open and filled
  at strong and weak coupling respectively; all in $SO(3)$. Linear fits to strong and
  weak coupling values shown as dashed and solid lines repectively.}
\label{fig_Mm_scwc_so3d3}
\end{figure}

\end{document}